\documentclass[twocolumn]{svjour3}

\usepackage[latin1]{inputenc}
\usepackage{times}

\usepackage{psfrag, amsmath, amssymb, amscd, amsfonts, latexsym, epsf, graphicx}
\usepackage{natbib}

\def\bbbr{{\mathbb R}} 
\def\bbbz{{\mathbb Z}} 

\newcommand{\htransf}{H}

\journalname{Journal of Mathematical Imaging and Vision}
\journalname{Biological Cybernetics}

\begin{document}

\title{\bf A time-causal and time-recursive scale-covariant
  scale-space representation of temporal signals and past time%
   \thanks{The support from the Swedish Research Council 
     (Contract 2018-03586) is gratefully acknowledged.}
   \thanks{A shortened version of this article is published Open Access
     in {\em Biological
   Cybernetics\/} under the Creative Commons Attribution 4.0
   International License,
   see https://doi.org/10.1007/s00422-022-00953-6.}}

\titlerunning{A time-causal and time-recursive scale-covariant
  temporal scale-space representation}

\author{Tony Lindeberg}

\institute{Tony Lindeberg,
                Computational Brain Science Lab,
                Division of Computational Science and Technology,
                KTH Royal Institute of Technology,
                SE-100 44 Stockholm, Sweden. 
                \email{tony@kth.se}
              ORCID: 0000-0002-9081-2170}
\date{}

\maketitle

\begin{abstract}
\noindent
This article presents an overview of a theory for performing temporal
smoothing of temporal signals in such a way that:
(i)~temporally smoothed signals at coarser temporal scales are guaranteed
to constitute simplifications of corresponding temporally smoothed
signals at any finer temporal scale (including the original signal) and
(ii)~the temporal smoothing process is both time-causal and
time-recursive, in the sense that it does not require access to future
information and can be performed with no other temporal memory buffer of the
past than the resulting smoothed temporal scale-space representations
themselves.

For specific subsets of parameter settings for the classes of linear
and shift-invariant temporal smoothing operators that obey this
property, it is shown how temporal scale covariance can be additionally 
obtained, guaranteeing that if the temporal input signal is rescaled by a
uniform temporal scaling factor, then also the resulting temporal scale-space
representations of the rescaled temporal signal will constitute mere
rescalings of the temporal scale-space representations of the original
input signal, complemented by a shift along the temporal scale
dimension.
The resulting time-causal limit kernel that obeys this property
constitutes a canonical temporal kernel for processing temporal signals
in real-time scenarios when the regular Gaussian kernel cannot be
used, because of its non-causal access to information from the future,
and we cannot additionally require the temporal smoothing process to
comprise a complementary memory of the past beyond the information
contained in the temporal smoothing process itself, which in this way
also serves as a multi-scale temporal memory of the past.

We describe how the time-causal limit kernel relates to previously used
temporal models, such as Koenderink's scale-time kernels and the
ex-Gaussian kernel. We do also give an overview of how the time-causal
limit kernel can be used for modelling the temporal processing in
models for spatio-temporal and spectro-temporal receptive fields, and
how it more generally has a high potential for modelling neural
temporal response functions in a purely time-causal and time-recursive
way, that can also handle phenomena at multiple temporal scales in a
theoretically well-founded manner.

We detail how this theory can be efficiently implemented for
discrete data, in terms of a set of recursive filters coupled in
cascade.
Hence, the theory is generally applicable for both: (i)~modelling
continuous temporal phenomena over multiple temporal scales and
(ii)~digital processing of measured temporal signals in real time.

We conclude by stating implications of the theory for modelling
temporal phenomena in biological, perceptual, neural and memory
processes by mathematical models, as well as implications regarding the philosophy of
time and perceptual agents. Specifically, we propose that for A-type theories of
time, as well as for perceptual agents, the notion of a
non-infinitesimal inner temporal scale of the temporal receptive
fields has to be included in 
representations of the present, where the inherent non-zero temporal
delay of such time-causal receptive fields implies a need for
incorporating predictions from the actual time-delayed present in the 
layers of a perceptual hierarchy, to make it possible for a
representation of the perceptual present to constitute a representation
of the environment with timing properties closer to the actual present.

\end{abstract}

\section{Introduction}
\label{sec-intro}

When processing time-dependent measurement signals, there is often a
need to perform temporal smoothing prior to more refined data analysis.
A commonly stated general motivation for this need is to suppress
measurement noise, often based on the assumption that there is a
well-defined underlying noise free signal that has been corrupted with
some amount of measurement noise.

A more fundamental approach to take on the need for performing temporal
smoothing of temporal signals is to follow a multi-scale approach,
based on the observation that measurements performed on real-world
data may reflect different types of temporal structures at different
temporal scales. In other words, even for the underlying noise free signal
in the above signal+noise model, it may hold that the data reflect
different types of underlying physical or biological processes at different
temporal scales. The measurement process itself, by which a
non-infinitesimal amount of energy needs to be integrated over some
non-infinitesimal temporal duration on the physical sensor, does in
this respect define an inner temporal scale of the measurements, beyond
which there is no way to resolve temporal phenomena that occur faster
than this inner temporal scale. Any real-world physical measurement
does in this respect involve an inherent notion of temporal scale.%
\footnote{For a popular overview over the wide range of temporal
  scales in physics and how the choice of temporal scale of
  observation thus will influence our modelling and understanding
  of the world, see
  't Hooft and Vandoren (\citeyear{HooVan14-book}).}

Specifically, in the areas of image processing, computer vision, 
machine listening%
\footnote{Other names for this field, which develops
  methods for audio understanding by machines, are machine hearing
  (Lyon \citeyear{Lyo10-SignProcMag}, \citeyear{Lyo17-book})
and computer audition.}
and computational modelling of visual and auditory perception,
this need is well understood, and has lead to
multi-scale approaches for spatial, spatio-temporal and
spectro-temporal receptive fields expressed in terms of multi-scale
representations over the spatial, spectral and temporal domains,
where specifically the theoretical framework known as scale-space
theory is based upon solid theory in terms of axiomatic derivations
concerning how the multi-scale processing operations should be performed
(Iijima \citeyear{Iij62};
 Witkin \citeyear{Wit83};
 Koenderink \citeyear{Koe84};
 Koenderink and van Doorn \citeyear{KoeDoo87-BC,KoeDoo92-PAMI};
 Lindeberg \citeyear{Lin93-Dis,Lin94-SI,Lin10-JMIV,Lin13-ImPhys};
 Florack \citeyear{Flo97-book};
 Sporring {\em et al.\/}\ \citeyear{SpoNieFloJoh96-SCSPTH};
 Weickert {\em et al.\/}\ \citeyear{WeiIshImi99-JMIV};
 ter Haar Romeny \citeyear{Haa04-book}).
It has also been found
that biological perception, memory and cognition has developed
biological processes at multiple temporal scales
(DeAngelis {\em et al.\/}\
\citeyear{DeAngOhzFre95-TINS,deAngAnz04-VisNeuroSci};
G{\"u}tig and Sompolinsky \citeyear{GulSom08-NatNeuroSci};
Gentner \citeyear{Gen08-JASA});
Holcombe \citeyear{Hol09-TICS};
Goldman \citeyear{Gol09-Neur};
Gauthier {\em et al.\/} \citeyear{GauEgeHesGIrKle12-JNeuroSci};
Atencio and Schreiner \citeyear{AteSch12-PONE};
Chait {\em et al.\/} \citeyear{ChaGreAraSimPoe15-FrontNeurSci};
Teng {\em et al.\/} \citeyear{TenTiaPoe16-SciRep};
Buzs{\'a}ki and Llin{\'a}s \citeyear{BuzLli17-Sci};
Tsao {\em et al.\/} \citeyear{TsaSugLuWanKniMosMos18-Nature};
Osman {\em et al.\/} \citeyear{OsmLeeEscRea18-JNeuroSci};
Latimer {\em et al.\/} \citeyear{LatBarSokAwwKatNelLamFaiPri19-NeuroSci};
Bright {\em et al.\/} \citeyear{BriMeiCruTigBufHow20-PNAS};
Cavanagh {\em et al.\/} \citeyear{CavHunKen20-FronNeurCirc};
Monsa {\em et al.\/} \citeyear{MonPeeArz20-JCognNeuroSci};
Spitmaan {\em et al.\/} \citeyear{SpiSeoLeeSol20-PNAS};
Howard and Hasselmo \citeyear{HowHas20-arXiv};
Howard \citeyear{How21-HandBookHumMem};
Guo {\em et al.\/} \citeyear{GuoHusMacReg21-Nature};
Miri {\em et al.\/} \citeyear{MirBhaAksTanGol22-JPhys});
see Section~\ref{sec-temp-sc-neur-sign} for a more detailed retrospective review.

The subject of this article is to describe a theoretical framework
for representing temporal signals at multiple temporal scales,
intended for a more
general audience without background in these areas and with the focus
on the temporal domain only, thus without the
complementary spatial or spectral domains that this theory has
previously been combined with for expressing spatio-temporal and
spectro-temporal receptive fields
(Lindeberg and Fagerstr{\"o}m \citeyear{LF96-ECCV};
Lindeberg \citeyear{Lin97-AFPAC,Lin16-JMIV,Lin17-JMIV,Lin18-JMIV,Lin18-SIIMS,Lin21-Heliyon};
Lindeberg and Friberg \citeyear{LinFri15-SSVM,LinFri15-PONE}).
This theoretical framework, referred to as {\em temporal scale-space
theory\/}, guarantees {\em non-creation of the temporal structures
with increasing temporal scales,\/} in the sense that it ensures that
a temporal representation at any coarser temporal
scale constitutes a simplification of a temporal representation at
any finer temporal scale, in the respect that the number of local
temporal extrema, alternatively the number of temporal zero-crossings, is guaranteed to
not increase from finer to coarser temporal scales.

Additionally, these temporal scale-space representations are
{\em time-causal,\/} in the sense that they do not require access to
future data, and are {\em time-recursive,\/} in the respect that the temporal
representation at the next temporal moment can be computed with no
other additional {\em memory of the past\/} than the temporal scale-space
representation itself. For a specific choice of temporal scale-space
kernel, referred to as the {\em time-causal limit kernel\/}, the temporal
scale-space representations are also {\em scale covariant\/},
meaning that the set of temporal scale-space representations is closed
under temporal rescalings of the input. A rescaling of the input signal by a
uniform scaling factor merely corresponds to a rescaling of the temporal scale-space
representations complemented by a shift of the temporal scale levels in the
temporal scale-space representation.
In this way, the temporal scale-space representation ensures an
internally consistent way of processing temporal signals that may be
subject to temporal scaling transformations, by phenomena or events
that may occur faster or slower in the world.

A main purpose of this article is to describe this theory in a
self-contained manner, without need for the reader to digest the
original references, where the information is distributed over several
papers, and may require a substantial effort for a reader not
previously familiar with this framework, to get an updated
view of the latest version of this theory.%
\footnote{For the reader interested in an overview of the developments
  of the different parts of temporal scale-space theory that this
  paper is based on, follows and extends,
  see the treatment in Section~\ref{sec-hist-devel-temp-scsp-theory}.}
Furthermore, we will describe
explicit relations to other previously used temporal models, such as
Koenderink's scale-time kernels (Koenderink \citeyear{Koe88-BC})
and the ex-Gaussian model
(Grushka \citeyear{Gru72-AnalChem};
Bright {\em et al.\/} \citeyear{BriMeiCruTigBufHow20-PNAS}), making it
possible to transfer modelling results from those temporal models to the
time-causal limit kernel described in this article.

We will also relate the presented temporal scale-space theory to other
approaches for processing signals at multiple temporal scales, such as
wavelet analysis and time-frequency analysis. Specifically, we will outline
how the temporal derivatives of the proposed time-causal limit kernel
described and analyzed in this article allow for fully time-causal
and time-recursive wavelet analysis methods, without need for additional
temporal buffering, and thus enabling minimal temporal response
times in a time-critical context. We will also outline how a
complex-valued extension of the proposed time-causal limit kernel can
be seen as a time-causal analogue of Gabor functions, thus allowing
for capturing essentially similar transformations of temporal signals
as for the family of Gabor functions, and thereby providing a way to define a
scale-covariant time-frequency representation over a time-causal
temporal domain, which by a slight modification can also be 
extended to additionally being implemented in terms strictly time-recursive operations.

Additionally, we will describe implications of using this theory for
modelling perceptual, neural and memory processes in biological
systems by mathematical models, as well as implications of the
theory with regard to the philosophy of time and perceptual agents.
Specifically, we will argue that when modelling a perceptual representation of the
present, it is essential to include the inner temporal scales of the
perceptual processes that lead to any percept, where the inherent
temporal delays of such time-causal operations imply that a
representation of the present will {\em de facto\/} constitute a representation of
some temporal intervals in the past, unless complemented by prediction
processes to enable better timing properties of a perceptual agent that interacts
with a dynamic world.

\subsection{Structure of this article}

This paper is organized as follows:
Section~\ref{sec-time-caus-time-rec-temp-scsp-model} introduces the
problem of constructing a temporal scale-space representation, as
constituting a multi-scale representation of temporal signals, with
the property that a measure of the amount of structure in the signal,
quantified as the number of local extrema over time, must not increase
from any finer to any coarser temporal scale.
A complete classification of the time-causal convolution kernels that
enable this property is given, and it is shown that the only possible
time-causal scale-space kernels over a continuous temporal domain
consist of truncated exponential kernels coupled in cascade.

Section~\ref{sec-sc-cov-temp-scsp} then adds a complementary condition
on  this structure, in terms of temporal scale covariance, and meaning
that if the temporal input signal is rescaled by a uniform temporal
scaling factor, then the result of temporal scale-space filtering of
this kernel should also be a mere rescaling of the result of
performing temporal scale-space filtering on the input signal,
complemented by a shift in the temporal scale channels
and a possibly complementary shift
in the magnitude of the signal.
It is shown that a specific kernel, the time-causal limit kernel, defined from an infinite
convolution of truncated exponential kernels in cascade, with specially chosen
time constants, obeys temporal scale covariance.
We do also show how this time-causal limit kernel relates to
previously used temporal models, such as Koenderink's scale-time
kernels and the ex-Gaussian kernel.

In Section~\ref{sec-comp-impl-disc-signals}, we complement the above
treatment for continuous signals with a corresponding discrete theory,
ensuring that the number of local extrema in a discrete signal is also
guaranteed to not increase from any finer to any coarser temporal
scale. The discrete analogue of the truncated exponential kernels are
first-order recursive filters, which are then again coupled in cascade.
Section~\ref{sec-temp-scsp-ders} furthermore generalizes the
above theory from temporal smoothing of a raw temporal signal, to the
computation of temporal scale-space derivatives, which measure the
amount of change in the signal with respect to any level of temporal
scale.
Section~\ref{sec-rel-wavelets-time-frequency-repr} outlines how the
proposed temporal scale-space representation is related to other
approaches for handling temporal signals at multiple temporal scales,
specifically wavelet analysis and time-frequency analysis, with
conceptual extensions of these notions with respect to strictly time-causal and
time-recursive operations for real-time applications.

Section~\ref{sec-temp-model-biol-perc-neur} describes how this general
theory can be used for modelling time-dependent processes and
mechanisms in perceptual and neural systems, with emphasis on
spatio-temp\-oral and spectro-temporal receptive fields as well as temporal
memory processes. Section~\ref{sec-impl-phil-time-perc-agent} outlines
more general implications of the theory with regard to the philosophy
of time and how time is handled by a perceptual agent. Specifically,
we develop how the inner temporal scale associated with any
biophysical measurement of time-dependent phenomena implies that a non-infinitesimal
inner temporal scale needs to be included in a representation of the
perceptual present, and also that the non-zero temporal delay of such
time-causal kernels implies that a biophysical representation of the
present will {\em de facto\/} constitute a representation of what has occurred over some temporal
intervals in the past, in turn implying a need for prediction
mechanisms to extra\-polate the {\em de facto\/} time-delayed representation of
the present into a better predicted representation of the actual present.

Section~\ref{sec-hist-devel-temp-scsp-theory} gives a retrospective
historic overview of the different parts of temporal scale-space
theory that this paper is based on, follows and extends, as well as a
conceptual overview of some of the main contributions to temporal
scale-space theory made in this article.
Finally, Section~\ref{sec-summ-concl} summarizes some of the main results.

\begin{figure}[hbt]
    \begin{center}
      \includegraphics[width=0.48\textwidth]{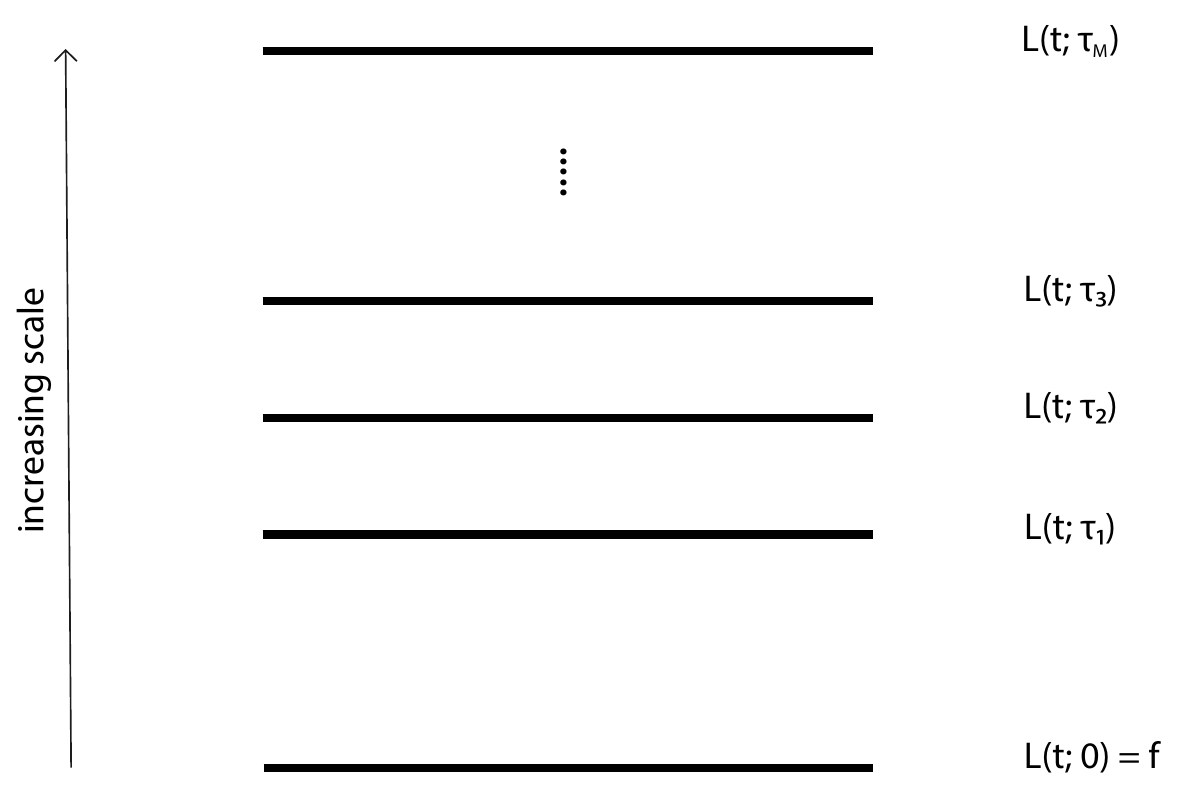}
    \end{center}
  \caption{The main idea of a scale-space representation is to, given
    any input signal $f(t)$, create a set of derived signals $L(t;\;
    \tau)$ intended to represent the information in the original
    signal at a set of coarser levels of scale $\tau$, with $L(t;\; 0) = f(t)$.
    These derived signals
    should preferably constitute true simplifications of each other,
    in such a way that the signal at a coarser level of scale does not
    contain more structures or information than any signal at any finer level of
    scale. Over spatial image domains, the notion of scale-space
    representation has been extensively studied, where several
    axiomatic derivations have shown that the Gaussian kernel and its
    corresponding Gaussian derivatives constitute a canonical class of
    convolution kernels for generating a spatial scale-space
    representation and have also been demonstrated to constitute a
    suitable basis of image primitives for computing different types of
    features from spatial image data. In this paper, we develop the
    associated notion of temporal scale-space theory, based on the
    additional constraints that (i)~the temporal scale-space kernels are not
    allowed to access information from the future in relation to any
    time moment and that (ii)~the computations should be possible to
    perform in a purely time-recursive manner, implying no other need
    for a temporal memory of the past than the temporal scale-space
    representation itself. Furthermore, we add a complementary requirement of
    (iii)~temporal scale covariance, meaning that under temporal scaling
    variations of the input, the temporal scale-space representations
    should also constitute mere temporal rescalings of the temporal
    scale-space representation computed from the original temporal signal before
    the temporal rescaling operation, complemented by a shift along
    the temporal scale axis.}
  \label{fig-schem-ill-temp-scsp}
\end{figure}

\section{Time-causal and time-recursive scale-space model for temporal
  signals}
\label{sec-time-caus-time-rec-temp-scsp-model}

The problem that we consider is that we are given a temporal signal $f(t)$
and want to define a set of successively smoothed temporal scale-space
representations $L(t;\; \tau)$ for different values of a temporal
scale parameter $\tau \geq 0$,
as schematically illustrated in Figure~\ref{fig-schem-ill-temp-scsp}.
We will throughout this treatment assume linearity and translational shift
covariance, implying that the transformation from the original signal
$f \colon \bbbr \rightarrow \bbbr$ to the temporal scale-space
representation $L \colon \bbbr \times \bbbr_+ \rightarrow \bbbr$
is given by convolution with some one-parameter family of
scale-dependent convolution kernels
$h \colon \bbbr \times \bbbr_+ \rightarrow \bbbr$
\begin{equation}
  L(t;\; \tau)
  = (h(\cdot;\; \tau) * f(\cdot))(t;\; \tau)
  = \int_{\xi \in \bbbr} h(\xi;\; \tau) \, f(t - \xi) \, d\xi.
\end{equation}
A crucial condition on this family of temporal scale-space
representations is that the temporal scale-space representation
$L(t;\; \tau_2)$ at any coarser temporal scale $t_2$ should
correspond to a simplification of the temporal scale-space
representation $L(t;\; \tau_1)$ at any finer temporal scale $t_1$.

Following Lindeberg (\citeyear{Lin90-PAMI}), we shall measure this
simplification property in terms of the number of local extrema in the
signal at any temporal scale, and define a {\em scale-space kernel\/}
as a kernel that obeys the property that the number of local extrema
in the signal after convolution is guaranteed to not exceed the number of local extrema
prior to the convolution operation, with the important qualifier that this
property should hold {\em for any input signal\/}.
Equivalently, this property can also be expressed by measuring the 
number of zero-crossings before and after the convolution operation.
A scale-space kernel $h(t;\; \tau)$ is referred to as a
{\em temporal scale-space kernel\/}
(Lindeberg and Fagerstr{\"o}m \citeyear{LF96-ECCV})
if it additionally satisfies $h(t;\; \tau) = 0$ for $t < 0$, meaning that it does not
require access to the future relative to any time moment.

To make the scale simplification property from finer to coarser temporal
scales hold, we will assume that the family of temporal smoothing
kernels $h(u;\ \tau)$ should obey the following cascade smoothing
property%
\footnote{Note that in contrast to some other temporal scale-space formulations
  (Lindeberg \citeyear{Lin97-AFPAC,Lin10-JMIV};
  Fagerstr{\"o}m \citeyear{Fag05-IJCV,Fag07-ScSp}),
  we do not here assume a semi-group
  property over temporal scales, since such an assumption leads to
  poor temporal dynamics, {\em e.g.\/}, longer temporal delays given a
  variance-based measure of the temporal duration of the kernel, as
  explained in more detail in (Lindeberg \citeyear[Appendix~1]{Lin17-JMIV}).}
\begin{equation}
  \label{eq-casc-rel-temp-scsp}
   h(\cdot;\; \tau_2) = (\Delta h)(\cdot;\; \tau_1 \mapsto \tau_2) * h(\cdot;\; \tau_1)
\end{equation}
for any pair of temporal scales $(\tau_1, \tau_2)$ with $\tau_2 > \tau_1$
and for some family of transformation kernels $(\Delta h)(t;\; \tau_1 \mapsto \tau_2)$.
We can then obtain a temporal scale-space representation if and
only if the transformation kernel $(\Delta h)(t;\; \tau_1 \mapsto \tau_2)$
between adjacent temporal scale levels $t_1$ and $t_2$ is always a temporal scale-space
kernel.

\subsection{Classification of scale-space kernels for continuous
  signals}
\label{sec-class-cont-scsp-kernels}

A fundamental question with regard to smoothing of temporal signals
concerns what convolution kernels satisfy the
conditions of being scale-space kernels.

\subsubsection{Complete classification of continuous scale-space kernels}

Interestingly, the class of one-dimensional scale-space kernels can be completely
classified based on classical results by
Schoenberg (\citeyear{Sch30,Sch46,Sch47,Sch48,Sch50,Sch53,Sch88-book}),
see also the excellent monograph by Karlin (\citeyear{Kar68}).
Summarizing the treatment in (Lindeberg
\citeyear[Section~3.5]{Lin93-Dis}; \citeyear[Section~3.2]{Lin16-JMIV}), a
continuous smoothing kernel is a scale-space kernel if and only if it has
a bilateral Laplace-Stieltjes transform
of the form (Schoenberg \citeyear{Sch50})
  \begin{equation}
    \label{eq-char-var-dim-kernels-cont-case-Laplace}
    \int_{\xi = - \infty}^{\infty} e^{-s \xi} \, h(\xi) \, d\xi =
    C \, e^{\gamma s^2 + \delta s}
    \prod_{i = 1}^{\infty} \frac{e^{a_i s}}{1 + a_i s}
      \quad
  \end{equation}
for $-c < \mbox{Re}(s) < c$ and some $c > 0$,
where $C \neq 0$, $\gamma \geq 0$, $\delta$ and $a_i$ are real and
$\sum_{i=1}^{\infty} a_i^2$ is convergent.

\subsubsection{Basic classes of primitive scale-space kernels over a continuous
  signal domain}

Interpreted over the temporal domain,%
\footnote{In the general expression
  (\ref{eq-char-var-dim-kernels-cont-case-Laplace})
  for the bilateral
  Laplace-Stieltjes transform of a continuous scale-space kernel, the factor
  $e^{\gamma s^2}$ is the Laplace-Stieltjes transform of the Gaussian
  kernel $e^{-\gamma \xi^2}$, the factor $1/(1 + a_i s)$ is the
  Laplace-Stieltjes transform of a truncated exponential function
  $e^{-a_i \xi}/a_i$ with time constant $a_i$, whereas the factors $e^{\delta s}$ and
  $e^{a_i s}$ correspond to translations in the temporal
  domain. Furthermore, the general product form of this expression in the
  Laplace-Stieltjes domain corresponds to a convolution of the corresponding
  primitives over the original temporal domain.}
this result means that there,
beyond trivial rescaling and translation, are two main classes of
one-dimensional scale-space kernels:
\begin{itemize}
\item
  convolution with {\em Gaussian kernels\/}
  \begin{equation}
  \label{eq-Gauss-polya-comp}
  h(\xi) = e^{-\gamma \xi^2},
\end{equation}
\item
  convolution with {\em truncated exponential functions\/}
  \begin{equation}
    \label{eq-truncexp-polya-comp}
    h(\xi) =
    \left\{
      \begin{array}{lcl}
        e^{- |\lambda| \xi} & & \xi \geq 0, \\
        0                 & & \xi < 0,
      \end{array}
    \right.
      \quad\quad
    h(\xi) =
    \left\{
      \begin{array}{lcl}
        e^{|\lambda| \xi} & & \xi \leq 0, \\
        0                 & & \xi > 0,
      \end{array}
    \right.
  \end{equation}
   for some strictly positive $|\lambda|$.
\end{itemize}
Moreover, the result means that a continuous smoothing kernel is a
scale-space kernel {\em if and only if\/} it can be decomposed into a
cascaded convolution of these primitives.

\subsection{Time-causal temporal scale-space kernels over continuous
  temporal domain}
\label{sec-cont-temp-scsp-kern}

Among the above primitive smoothing kernels, we recognize the Gaussian
kernel,
which is a good and natural temporal smoothing kernel to use
when analysing pre-recorded signals in offline scenarios.
When analysing temporal signals in a real-time situation, or when
modelling biological processes that operate in real time, we
cannot, however, use a temporal smoothing kernel that requires access
to information in the future relative to any time moment.

For building a time-causal temporal scale-space representation, the
truncated exponential kernels are therefore the only possible
primitive time-causal temporal smoothing kernels
(Lindeberg and Fagerstr{\"o}m \citeyear{LF96-ECCV})
\begin{equation}
    h_{\mbox{\scriptsize exp}}(t;\; \mu_k) 
    = \left\{
        \begin{array}{ll}
          \frac{1}{\mu_k} e^{-t/\mu_k} & t \geq 0, \\
          0         & t < 0,
        \end{array}
      \right.
\end{equation}
where we will throughout this treatment adopt the convention of
normalizing these kernels to unit $L_1$-norm.
The Laplace transform of such a kernel is given by
\begin{equation}
  \label{eq-FT-composed-kern-casc-truncexp}
    H_{\mbox{\scriptsize exp}}(q;\; \mu_k) 
    = \int_{t = - \infty}^{\infty} h_{\mbox{\scriptsize exp}}(t;\; \mu_k) \, e^{-qt} \, dt
    = \frac{1}{1 + \mu_k q}.
 \end{equation}
Coupling $K$ such kernels in cascade leads to a composed kernel
\begin{equation}
    \label{eq-comp-trunc-exp-cascade}
    h_{\mbox{\scriptsize composed}}(\cdot;\; \mu) 
    = *_{k=1}^{K} h_{\mbox{\scriptsize exp}}(\cdot;\; \mu_k)
\end{equation}
having a Laplace transform of the form
\begin{align}
    \begin{split}
       \label{eq-expr-comp-kern-trunc-exp-filters}
       H_{\mbox{\scriptsize composed}}(q;\; \mu) 
       & = \int_{t = - \infty}^{\infty} 
              *_{k=1}^{K} h_{\mbox{\scriptsize exp}}(\cdot;\; \mu_k)(t) 
                 \, e^{-qt} \, dt
   \end{split}\nonumber\\
   \begin{split}
       & =  \prod_{k=1}^{K} \frac{1}{1 + \mu_k q}.
    \end{split}
\end{align}
The temporal mean and variance of the composed kernel is
\begin{equation}
     \label{eq-mean-var-trunc-exp-filters}
     m_K = \sum_{k=1}^{K} \mu_k, \quad\quad \tau_K = \sum_{k=1}^{K} \mu_k^2.
\end{equation}
The temporal mean $m_K$ is a coarse measure of the temporal delay of the
time-causal temporal scale-space kernel, and the temporal variance $\tau_K$ 
is a measure of the temporal duration, also referred to as the temporal scale.
   
In terms of physical models, repeated convolution with this class of
temporal scale-space kernels
corresponds to coupling a series
of {\em first-order integrators\/} with time constants $\mu_k$ in cascade
\begin{equation}
  \label{eq-first-ord-int}
   \partial_t L(t;\; \tau_k) 
   = \frac{1}{\mu_k} \left( L(t;\; \tau_{k-1}) - L(t;\; \tau_k) \right)
\end{equation}
with $L(t;\; 0) = f(t)$, where the temporal scale-space
representations for larger values of the scale parameter $t_k$
constitute successively temporally smoothed representations of each
other. An important property of this type of temporal scale-space
representation is that it is also {\em time-recursive\/}. The temporal
scale-space representations $L(t;\; \tau_k)$ constitute a {\em sufficient
temporal memory of the past\/} to compute the temporal scale-space
representation and the next temporal moment, given a new input in the
input signal $f(t)$.

An important consequence of the above necessity result, is that
this type of scale-space representation constitutes the {\em only\/} way
to compute a time-causal temporal scale-space representation, given the
requirement that the number of local extrema, or equivalently the
number of zero-crossings, in the signal must not increase from finer
to coarser temporal scales. In this respect, the temporal scale-space
representations can be seen as gradual simplifications of each other
from finer to coarser temporal scales.

\begin{figure}[!h]
   \begin{center}
      \includegraphics[width=0.45\textwidth]{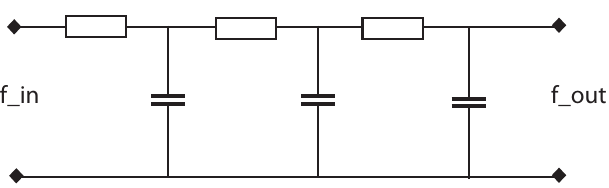}
   \end{center}
\caption{Electric wiring diagram consisting of a set of resistors and
  capacitors that emulate a series of first-order integrators coupled in
  cascade, if we regard the time-varying voltage $f_{\mbox{\scriptsize in}}$ as
  representing the time varying input signal and the resulting output
  voltage and $f_{\mbox{\scriptsize out}}$ as representing the time varying output signal at a
  coarser temporal scale.
 Such first-order temporal integration can be used as a straightforward
  computational model for temporal processing in biological neurons;
  see also Koch (\protect\citeyear[Chapters~11--12]{Koch99-book}) regarding 
  physical modelling of the information transfer in the dendrites of neurons.}
  \label{fig-first-order-integrators-electric}
\end{figure}

Figure~\ref{fig-first-order-integrators-electric} shows an
illustration of this model in terms of an electric wiring diagram for
transforming an input signal $f_{\mbox{\scriptsize in}}$ to an output signal $f_{\mbox{\scriptsize out}}$
using a set of first-order integrators coupled in cascade.

\subsection{Logarithmic distribution of the temporal scale levels}
\label{sec-log-distr-temp-sc-levels}

When implementing this temporal scale-space concept in practice,
a set of intermediate temporal scale levels $\tau_k$ has to be distributed
between some minimum and maximum temporal scale levels $\tau_{\mbox{\scriptsize min}} = \tau_1$
and $\tau_{\mbox{\scriptsize max}} = \tau_K$.
Then, it is natural to choose these temporal scale levels according to
a geometric series,
corresponding to a uniform distribution in units of
{\em effective temporal scale\/} $\tau_{\scriptsize{\mbox{eff}}} = \log \tau$
(Lindeberg \citeyear{Lin92-PAMI}).

If we have a free choice of what minimum temporal
scale level $\tau_{\mbox{\scriptsize min}}$ to use, a natural way
of parameterizing these temporal scale levels is by using a distribution
parameter $c > 1$ such that
\begin{equation}
  \label{eq-distr-tau-values}
  \tau_k = c^{2(k-K)} \tau_{\mbox{\scriptsize max}} \quad\quad (1 \leq k \leq K),
\end{equation}
which by equation~(\ref{eq-mean-var-trunc-exp-filters}) implies that
the time
constants of the individual first-order integrators should be given by
(Lindeberg \citeyear[Equations~(19)--(20)]{Lin16-JMIV})
\begin{align}
  \begin{split}
     \label{eq-mu1-log-distr}
     \mu_1 & = c^{1-K} \sqrt{\tau_{\mbox{\scriptsize max}}}
  \end{split}\\
  \begin{split}
     \label{eq-muk-log-distr}
     \mu_k & = \sqrt{\tau_k - \tau_{k-1}} = c^{k-K-1} \sqrt{c^2-1} \sqrt{\tau_{\mbox{\scriptsize max}}} \quad (2 \leq k \leq K).
  \end{split}
\end{align}
If the temporal signal is on the other hand given at some minimum
temporal scale $\tau_{\mbox{\scriptsize min}}$, corresponding to an {\em a priori\/} given
inner temporal scale of the measurement device,
we can instead determine
\begin{equation}
  c = \left( \frac{\tau_{\mbox{\scriptsize max}}}{\tau_{\mbox{\scriptsize min}}} \right)^{\frac{1}{2(K-1)}}
\end{equation}
 in (\ref{eq-distr-tau-values})
such that $\tau_1 = \tau_{\mbox{\scriptsize min}}$ 
and add $K - 1$ temporal scales with $\mu_k$ according to (\ref{eq-muk-log-distr}).

Temporal smoothing kernels of this form, combined with
temporal differentiation for different orders of differentiation, to obtain ripples of
opposite contrast in the resulting temporal receptive fields, have
been used for modelling the temporal part of the processing in models
for spatio-temporal receptive fields
(Lindeberg and Fagerstr{\"o}m \citeyear{LF96-ECCV};
Lindeberg \citeyear{Lin15-SSVM,Lin16-JMIV,Lin21-Heliyon})
and spectro-temporal receptive fields
(Lindeberg and Friberg \citeyear{LinFri15-SSVM,LinFri15-PONE}).

\subsection{Logarithmic memory of the past}
\label{sec-log-mem-of-past}

When using a logarithmic distribution of the temporal scale levels
according to either of these methods,
the different levels in the temporal scale-space representation at
increasing temporal scales will serve as a logarithmic memory of the
past, with qualitative similarity to the mapping of the
past onto a logarithmic time axis in the scale-time model by
Koenderink (\citeyear{Koe88-BC}).
Such a logarithmic memory of the past can also be extended to later
stages in a visual, auditory or other form of neural hierarchy.

An alternative type of temporal memory structure can be obtained if
the different truncated exponential kernels are applied, not in a
cascade as above, but instead in parallel with a single temporal time
constant for each temporal memory channel,
\begin{equation}
  \label{eq-temp-channel-single-exp-fcn}
    h_{\mbox{\scriptsize composed}}(\cdot;\; \tau_k) 
    = h_{\mbox{\scriptsize exp}}(\cdot;\; \mu_k)
\end{equation}
for $\mu_k = \sqrt{\tau_k}$, again with a logarithmic
distribution of the temporal scale levels $\tau_k$. Such a model for temporal
memory has been studied by Howard and his co-workers
(Howard \citeyear{How21-HandBookHumMem};
 Bright {\em et al.\/} \citeyear{BriMeiCruTigBufHow20-PNAS}).
Then, each temporal memory channel is also a simplification of the
input signal $f(t)$, and a record of the past with a given temporal delay and
temporal duration. Inversion from the temporal memory channels to the
input signal is also more straightforward, from the conceptual similarity to
a real-valued Laplace transform
(Howard {\em et al.\/} \citeyear{HowLuzTig18-CompBrainBehav};
 Howard and Hasselmo \citeyear{HowHas20-arXiv}).
The different temporal memory channels are, however,
not guaranteed to constitute formal simplifications of each other, as
they are for the cascade model.

The theoretical framework for time-causal and time-recur\-sive temporal scale-space
representations presented earlier in
(Lindeberg and Fagerstr{\"o}m \citeyear{LF96-ECCV};
Lindeberg \citeyear{Lin16-JMIV})
and here can be seen as providing a theoretical foundation for such
time-recursive temporal memory models.

\subsection{Uniform distribution of the temporal scale levels}
\label{sec-uni-distr-temp-sc-levels}

An alternative approach to distributing the temporal scale levels is
to use a uniform distribution of the intermediate temporal scales
\begin{equation}
  \label{eq-distr-tau-values-uni}
  \tau_k = \frac{k}{K} \, \tau_{\mbox{\scriptsize max}},
\end{equation}
implying that  the time constants in the individual smoothing steps are
given by
\begin{equation}
  \label{eq-mu-k-uni}
   \mu_k = \mu = \sqrt{\frac{\tau_{\mbox{\scriptsize max}}}{K}}.
 \end{equation}
Then, a compact expression can be easily obtained for the composed
convolution kernel corresponding to a cascade of $K$ such kernels
\begin{equation}
 \label{eq-temp-scsp-kernel-uni-distr}
  h_{\mbox{\scriptsize composed}}(t;\; \mu, K) 
  = \frac{t^{K-1} \, e^{-t/\mu}}{\mu^K \, \Gamma(K)}.
\end{equation}
Such kernels have also been used in memory models
(Goldman \citeyear{Gol09-Neur}).
The temporal Poisson model studied in more detail in
(Lindeberg \citeyear{Lin97-AFPAC}) can be seen
as the limit case of such a uniform distribution of the temporal scale
levels in the time-discrete case, when the difference between adjacent temporal scales tends to
zero, a limit case that, however, only exists for discrete temporal
signals (Lindeberg and Fagerstr{\"o}m \citeyear{LF96-ECCV}),
and which also serves as a multi-scale temporal memory of the past
(see the illustrations of how the temporal scale-space representation evolves over time and
temporal scales in the time-scale diagrams in Figures~3--5 in
(Lindeberg \citeyear{Lin97-AFPAC}), which demonstrate the temporal
memory properties of such a temporal scale-space representation ---
specifically observe the property that an event that occurs at a
certain temporal moment first appears in the temporal scale-space
representation at the finest temporal scale, and then moves to
gradually coarser temporal scales as time passes by, and is thus also
after some short times gradually forgotten at the finer temporal
scales, being taken over temporal structures that appear after the
initial temporal event).

For constructing temporal memory processes that are to operate over wide ranges of
temporal scales, such models based on a uniform sampling of the
temporal scale levels do, however, require a larger number of
primitive temporal integrators, and thus more hardware or wetware, compared to a
temporal memory model based on a logarithmic distribution of the
temporal scale levels.

Combined with temporal differentiation of the smoothing kernel, such
temporal kernels have been used for modelling the temporal response
properties of neurons in the visual system
(den Brinker and Roufs \citeyear{BriRou92-BC})
and for computing spatio-temporal image features in
computer vision (Rivero-Moreno and Bres \citeyear{RivBre04-ImAnalRec};
Berg et al.\ \citeyear{BerReyRid14-SensMEMSElOptSyst}).

For a given value of the temporal scale (the temporal variance) of
such time-causal kernels, the temporal delay for a temporal kernel based on a
uniform distribution of the temporal scale levels will, however, also be
longer than for a temporal kernel constructed from a logarithmic distribution
of the intermediate temporal scale levels. Thus, for formulating computational
algorithms for expressing time-critical decision
processes in computer vision or machine listening, as well as for modelling
time-critical decision processes in biological perception or
cognition, we argue that a logarithmic distribution of the temporal
scale levels should be a much better choice.

For these reasons, we will henceforth in this treatment focus solely
on models based on a logarithmic distribution of the temporal scale levels.

\section{Time-causal temporal scale-space representations that also
  obey temporal scale covariance}
\label{sec-sc-cov-temp-scsp}

Beyond the task of representing temporal signals at multiple temporal scales,
a main requirement on a temporal scale-space representation should
also be the notion of {\em temporal scale covariance\/},%
\footnote{In certain literature, the property that we refer to as
  ``covariance'' is instead referred to as ``equivariance''. In this
  paper, we throughout use the terminology ``covariance'', to maintain
  consistency with the scale-space literature (Lindeberg
  \citeyear{Lin13-ImPhys}).}
so as to be able to
consistently handle temporal phenomena and events that
occur faster or slower in the world. Temporal scale covariance
means that if a
signal $f(t)$ is subject to a temporal scaling transformation
\begin{equation}
  f'(t') = f(t) \quad\quad \mbox{for}  \quad\quad  t' = S t
\end{equation}
and then
processed, here with a temporal convolution kernel $T(t';\; \tau')$
that depends on a temporal scale parameter $\tau'$,
\begin{equation}
   L'(t';\; \tau') = (T(\cdot;\; \tau') * f'(\cdot))(t';\; \tau'),
\end{equation}
the result should be essentially similar to the result of applying the same type of
processing to the original signal
\begin{equation}
   L(t;\; \tau) = (T(\cdot;\; \tau) * f(\cdot))(t;\; \tau)
\end{equation}
and then rescaling the processed
original signal 
\begin{equation}
   L'(t';\; \tau') = L(t;\; \tau) 
\end{equation}
(for other types of processes possibly also complemented with some minor modification,
such as a correction of the magnitude of the response).
For the task of temporal filtering in a temporal scale-space
representation, this implies that the temporal scale-space
kernel should commute with temporal scaling transformations, as
illustrated in the commutative dia\-gram in
Figure~\ref{fig-comm-diag-temp-sc-transf}.

\begin{figure}[hbt]
  \[
    \begin{CD}
       L(t;\; \tau)
       @>{\scriptsize\begin{array}{c} \scriptsize t' = S t  \\
                         \scriptsize \tau' = S^2 \tau \end{array}}>> L'(t';\; \tau') \\
       \Big\uparrow\vcenter{\rlap{$\scriptstyle{{*T(t;\; 
               \tau)}}$}} & &
       \Big\uparrow\vcenter{\rlap{$\scriptstyle{{*T(t';\; 
               \tau')}}$}} \\
       f(t) @>{\scriptsize t' = S t}>> f'(t')
    \end{CD}
 \]
\caption{Commutative diagram for temporal receptive field responses under
  {\em temporal scaling transformations\/} of the temporal domain. Such
  transformations describe the effect of events occurring slower or
  faster in the world.
(The commutative diagram should be read from the lower left corner to
the upper right corner, and means that irrespective of whether the
image is first convolved with a temporal
smoothing kernel and then subject to
temporal scaling transformation, or
whether the temporal signal is first subject to a
temporal scaling transformation and then convolved with
a temporal smoothing kernel,
we should get the same result provided that the 
temporal scale parameters $\tau$ and $\tau'$ are properly matched to the relative temporal
scaling factor $S$ between the
two temporal patterns.)}
\label{fig-comm-diag-temp-sc-transf}
\end{figure}

This algebraic closedness property under temporal scaling
transformations will imply that similar
temporal phenomena that occur faster or slower in the world will be treated in a
conceptually similar manner. Under variations caused by scaling
transformations in the input, the output of applying scale-covariant
processing to such temporally rescaled data will be mere temporal
rescalings of each other, thus without bias to any particular scales,
which would otherwise be a severe shortcoming, if the computational
model is not well-behaved under temporal scaling transformations.

In this section, we will describe a theory for how to obtain 
time-causal temporal scale-space representations that also obey such temporal scale
covariance, which in turn makes it possible to construct provably
scale-invariant temporal representations at higher levels in a
temporal processing hierarchy.
The way that we will reach this goal is by constructing a limit kernel that is the
convolution of an infinite number of truncated exponential kernels
in cascade, with specially chosen time constants that correspond to a geometric
distribution of the intermediate temporal scale levels.

Unfortunately, there is no known simple compact explicit expression for this
limit kernel in the temporal domain, implying that some of the
closed-form calculations using the limit kernel may be interpreted as
somewhat technical at the first encounter with this function.
Once these algebraic transformation properties have been established
for the limit kernel, however, this function can be handled and used
in a similar way as other standard functions in mathematics.

For practical implementations, the limit kernel can furthermore for
the purpose of computing the representation at a single temporal scale
often be very
well approximated by a moderate finite number of truncated exponential
kernels coupled in cascade, usually between 4 and 8 in our
implementations of this concept, because of its rapid convergence
properties for suitable values of its internal distribution parameter.
In turn, for the purpose of computing another temporal scale-space
representation at the next coarser temporal scale, applying a {\em single\/}
truncated exponential kernel to the nearest finer temporal scale is sufficient.

In this section, we will first define the limit kernel and derive 
its transformation properties. Then, we will turn to relating and
comparing the limit kernel to two other models used 
for expressing temporal variations over time.

\begin{figure*}[hbtp]
  \begin{center}
    \begin{tabular}{ccc}
       {\footnotesize $h(t;\; c = \sqrt{2})$} 
      & {\footnotesize $h_{t}(t;\; c = \sqrt{2})$} 
      & {\footnotesize $h_{tt}(t;\; c = \sqrt{2})$} \\
      \includegraphics[width=0.30\textwidth]{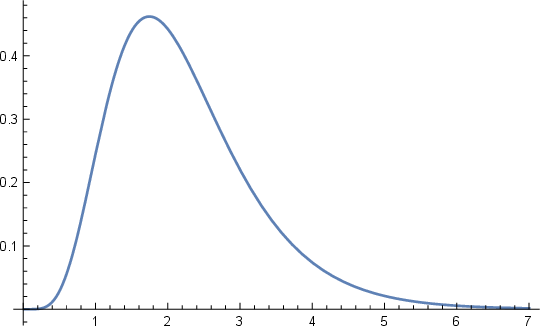} &
      \includegraphics[width=0.30\textwidth]{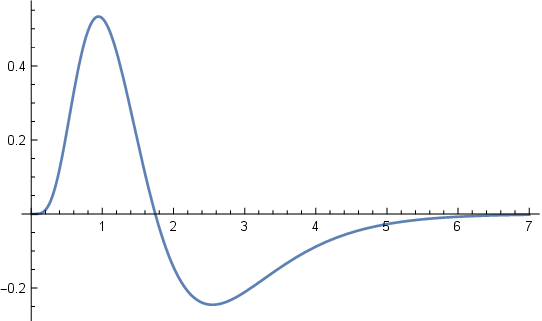} &
      \includegraphics[width=0.30\textwidth]{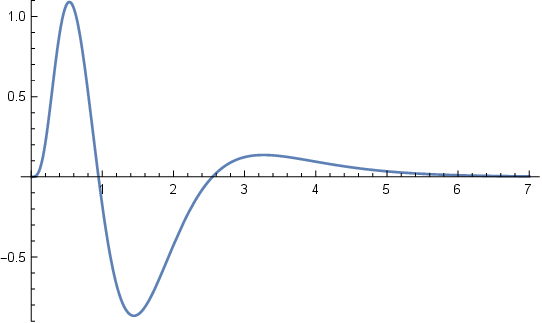} \\
\\
       {\footnotesize $h(t;\; c = 2)$} 
      & {\footnotesize $h_{t}(t;\; c = 2)$} 
      & {\footnotesize $h_{tt}(t;\; c = 2)$} \\
     \includegraphics[width=0.30\textwidth]{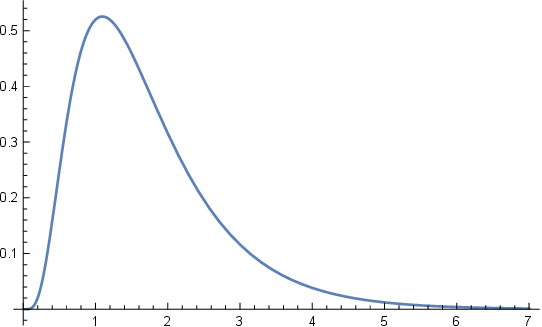} &
      \includegraphics[width=0.30\textwidth]{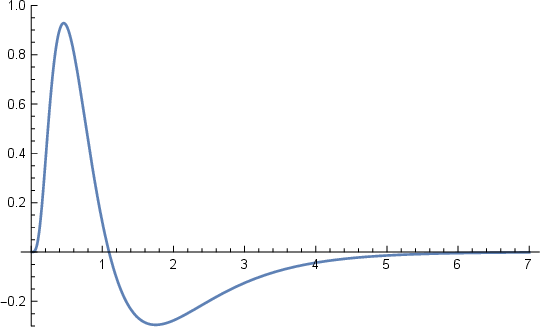} &
      \includegraphics[width=0.30\textwidth]{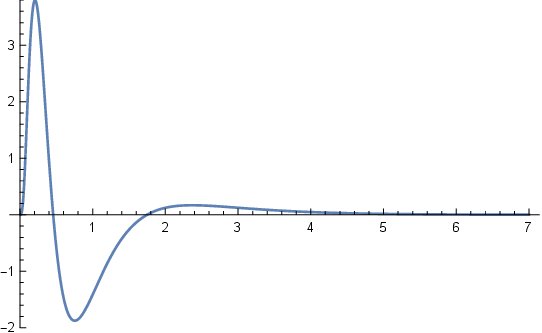} \\
     \end{tabular} 
  \end{center}
   \caption{Approximations of the time-causal limit kernel for $\tau =
     1$ using $K = 7$ {\em truncated exponential kernels\/} in cascade and
           their first- and second-order derivatives.
           (top row) Logarithmic distribution of the scale levels
           for $c = \sqrt{2}$.
          (bottom row) Logarithmic distribution 
          for $c = 2$.
          (Horizontal axes: time. Vertical axes: function values.)}
  \label{fig-trunc-exp-kernels-1D}
\end{figure*}

\subsection{The time-causal limit kernel}
\label{sec-time-caus-limit-kern}

Consider the Fourier transform of the composed convolution kernel that
we obtain by coupling $K$ truncated exponential kernels in cascade
with a logarithmic distribution of the temporal scale levels and thus
time constants according to (\ref{eq-mu1-log-distr}) and
(\ref{eq-muk-log-distr}) for some $c > 1$:
\begin{multline}
  \label{eq-FT-comp-kern-log-distr}
    \hat{h}_{\mbox{\scriptsize composed}}(\omega;\; \tau, c, K) 
       = \\ \frac{1}{1 + i \, c^{1-K} \sqrt{\tau} \, \omega}
               \prod_{k=2}^{K} \frac{1}{1 + i \, c^{k-K-1} \sqrt{c^2-1} \sqrt{\tau} \, \omega}.
\end{multline}
By formally letting the number of primitive smoothing steps $K$ tend
to infinity and renumbering the indices by a shift in terms of one unit, we
obtain a limit object of the form (Lindeberg \citeyear[Equation~(38)]{Lin16-JMIV})
\begin{align}
  \begin{split}
     \hat{\Psi}(\omega;\; \tau, c) 
     & = \lim_{K \rightarrow \infty} \hat{h}_{\mbox{\scriptsize composed}}(\omega;\; \tau, c, K) 
  \end{split}\nonumber\\
  \begin{split}
     \label{eq-FT-comp-kern-log-distr-limit}
     & = \prod_{k=1}^{\infty} \frac{1}{1 + i \, c^{-k} \sqrt{c^2-1} \sqrt{\tau} \, \omega}.
  \end{split}
\end{align}
By treating this limit kernel as an object by itself, which will be well-defined because of the rapid convergence by the summation of variances according to a geometric series, interesting relations can be expressed
between the temporal scale-space representations
\begin{equation}
  \label{eq-temp-scsp-conv-limit-kernel}
  L(t;\; \tau, c) = \int_{u = 0}^{\infty} \Psi(u;\; \tau, c) \, f(t-u) \, du
\end{equation}
obtained by convolution with this limit kernel.

\subsubsection{Self-similar recurrence relation for the time-causal limit kernel over temporal scales}

Using the limit kernel, an infinite number of discrete temporal scale levels
is implicitly defined given the specific choice of one temporal scale $\tau = \tau_0$:
\begin{equation}
  \label{eq-temp-scale-levels-limit-kernel}
  \dots \frac{\tau_0}{c^6}, \frac{\tau_0}{c^4}, \frac{\tau_0}{c^2}, \tau_0,
  c^2 \tau_0, c^4 \tau_0, c^6 \tau_0, \dots 
\end{equation}
Directly from the definition of the limit kernel, we obtain the following recurrence relation
between adjacent temporal scales:
\begin{equation}
  \label{eq-recur-rel-limit-kernel}
   \Psi(\cdot;\; \tau, c) = h_{\mbox{\scriptsize exp}}(\cdot;\; \tfrac{\sqrt{c^2-1}}{c} \sqrt{\tau}) * \Psi(\cdot;\; \tfrac{\tau}{c^2}, c)
\end{equation}
and in terms of the Fourier transform:
\begin{equation}
  \label{eq-recur-rel-limit-kernel-FT}
   \hat{\Psi}(\omega;\; \tau, c) 
   = \frac{1}{1 + i \, \tfrac{\sqrt{c^2-1}}{c} \sqrt{\tau} \, \omega} \,
       \hat{\Psi}(\omega;\; \tfrac{\tau}{c^2}, c).
\end{equation}

\subsubsection{Behaviour under temporal rescaling transformations}

From the Fourier transform of the limit kernel (\ref{eq-FT-comp-kern-log-distr-limit}),
we can observe that for any temporal scaling factor $S$ it holds that
\begin{equation}
  \label{eq-sc-transf-limit-kernel-FT}
   \hat{\Psi}(\tfrac{\omega}{S};\; S^2 \tau, c) = \hat{\Psi}(\omega;\; \tau, c).
\end{equation}
Thus, the limit kernel transforms as follows under a scaling
transformation of the temporal domain:
\begin{equation}
  \label{eq-sc-transf-limit-kernel}
   S \, \Psi(S \, t;\; S^2 \tau, c) = \Psi(t;\; \tau, c).
\end{equation}
If we, for a given choice of distribution parameter $c$, rescale the 
input signal $f$ by a temporal scaling factor $S = 1/c$ such that $t' = t/c$, it then follows that 
the scale-space representation of $f'$ at temporal scale $\tau' = \tau/c^2$
\begin{equation}
  L'(t';\; \tfrac{\tau}{c^2}, c) = (\Psi(\cdot;\; \tfrac{\tau}{c^2}, c) * f'(\cdot))(t';\; \tfrac{\tau}{c^2}, c)
\end{equation}
will be equal to the temporal scale-space representation of the
original signal $f$ at scale $\tau$ (Lindeberg \citeyear[Equation~(46)]{Lin16-JMIV})
\begin{equation}
  \label{eq-scale-cov-limit-kernel-scsp}
  L'(t';\; \tau', c) = L(t;\; \tau, c).
\end{equation}
Hence, under a rescaling of the original signal by a temporal scaling factor $c$,
a rescaled copy of the temporal scale-space representation of the 
original signal can be found at the next lower discrete temporal scale, relative 
to the temporal scale-space representation of the original signal.

\subsubsection{Provable temporal scale covariance}
\label{sec-prov-temp-sc-cov}

Applied recursively, the above result implies that 
the temporal scale-space representation
obtained by convolution with the limit kernel {\em obeys a 
closedness property over all temporal scaling transformations $t' = c^j t$ 
with temporal rescaling factors $S = c^{j}$ ($j \in \bbbz$) that are 
integer powers of the distribution parameter $c$\/}
(Lindeberg \citeyear[Equation~(47)]{Lin16-JMIV}),
\begin{equation}
  L'(t';\; \tau', c) = L(t;\; \tau, c) \quad\mbox{for}\quad t' = 
c^j t \quad\mbox{and} \quad \tau' = c^{2j} \tau,
\end{equation} 
thus allowing for perfect scale covariance over the restricted subset of scaling factors
$S = c^j$ that precisely matches the specific set
of discrete temporal scale levels that is defined by a specific choice of the
distribution parameter $c$.
Based on this desirable and highly useful property, it is natural to refer to the limit kernel 
as {\em the scale-covariant time-causal limit kernel\/}
(Lindeberg \citeyear[Section~5]{Lin16-JMIV}).

\begin{figure*}[hbtp]
  \begin{center}
    \begin{tabular}{ccc}
      {\footnotesize\em Temporal scale $\sqrt{\tau} = 16384$}
      &  $\quad\quad$ & {\footnotesize\em Temporal scale $\sqrt{\tau} = 16$} \\
      \includegraphics[width=0.30\textwidth]{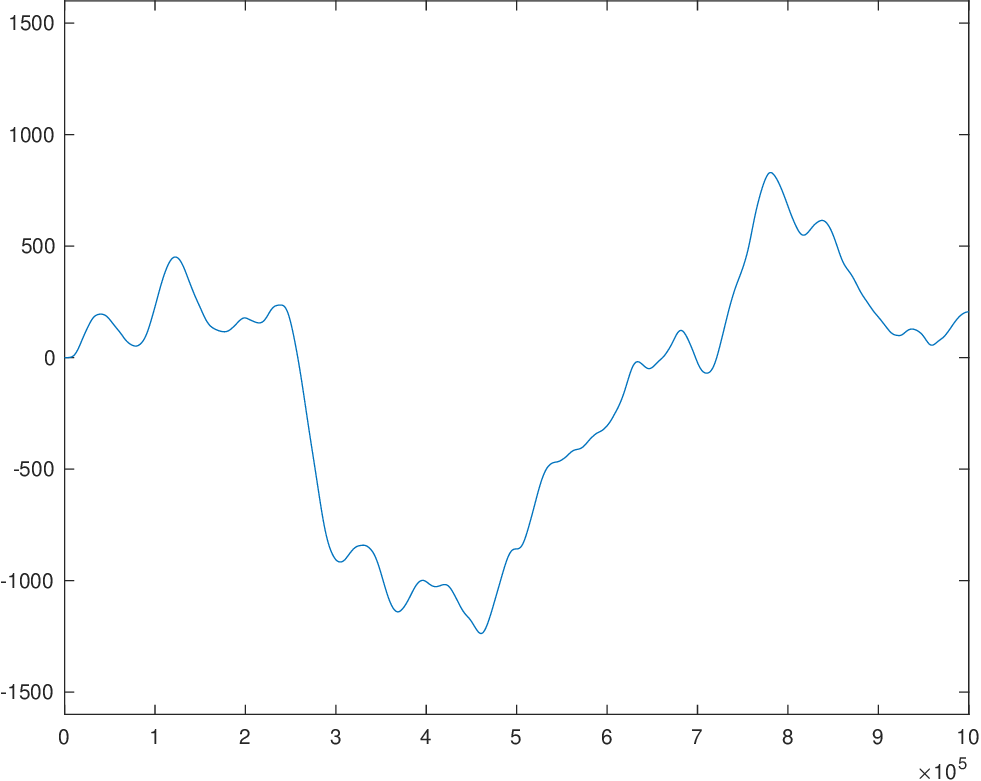}
      & $\quad\quad$ & \includegraphics[width=0.30\textwidth]{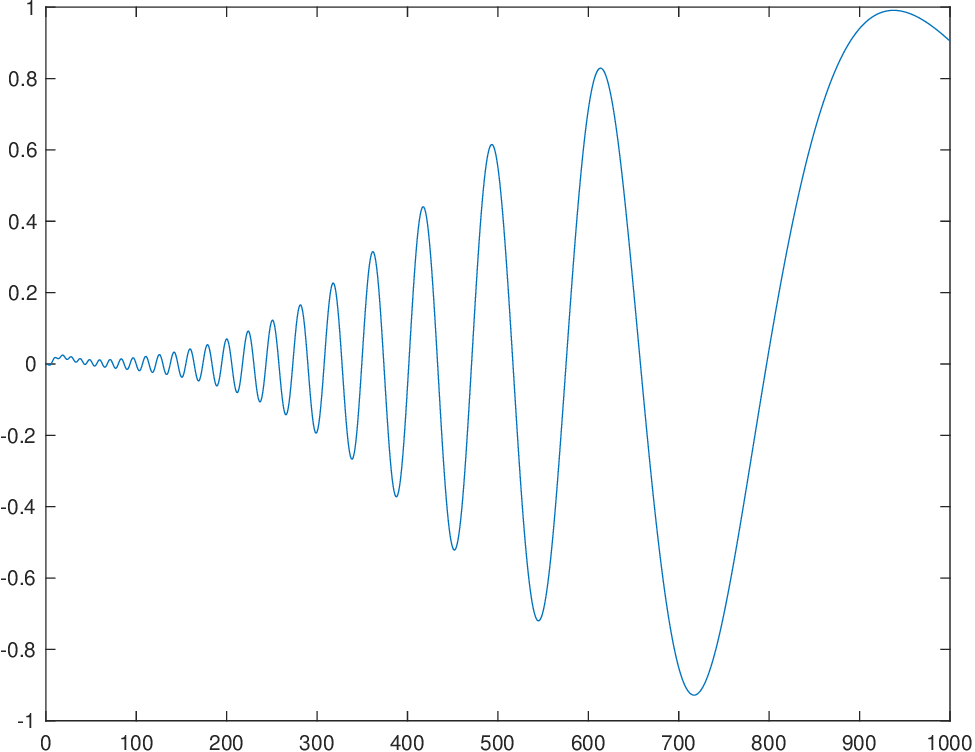}
      \\
      {\footnotesize\em Temporal scale $\sqrt{\tau} = 4096$}
      & $\quad\quad$ & {\footnotesize\em Temporal scale $\sqrt{\tau} = 4$} \\
      \includegraphics[width=0.30\textwidth]{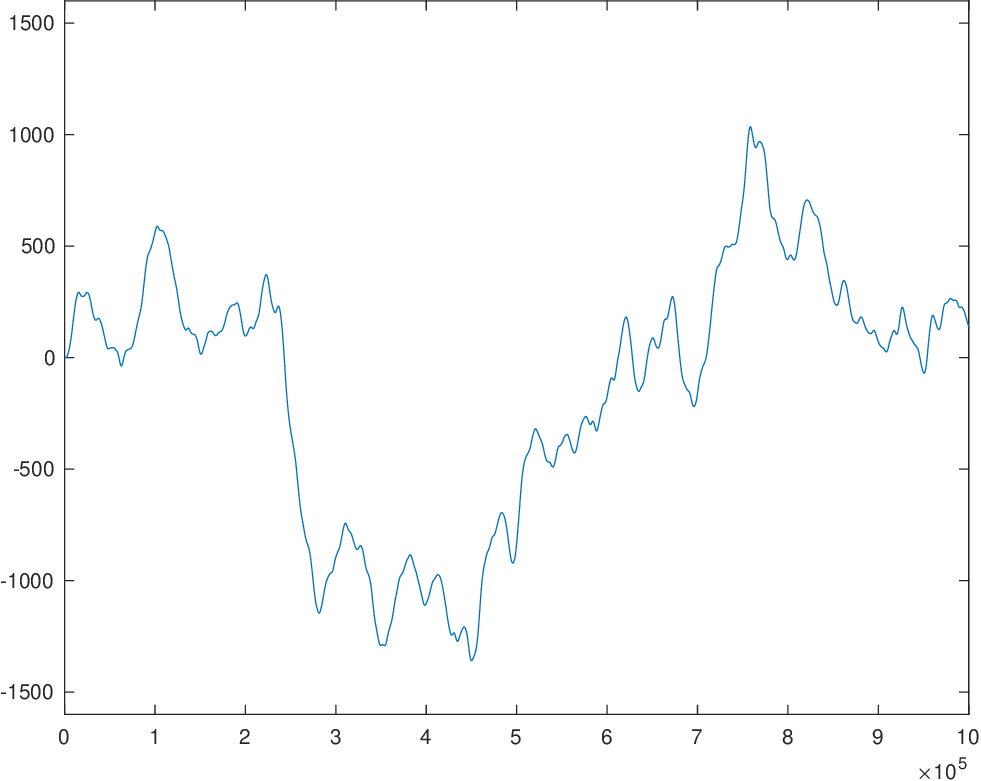}
      & $\quad\quad$ & \includegraphics[width=0.30\textwidth]{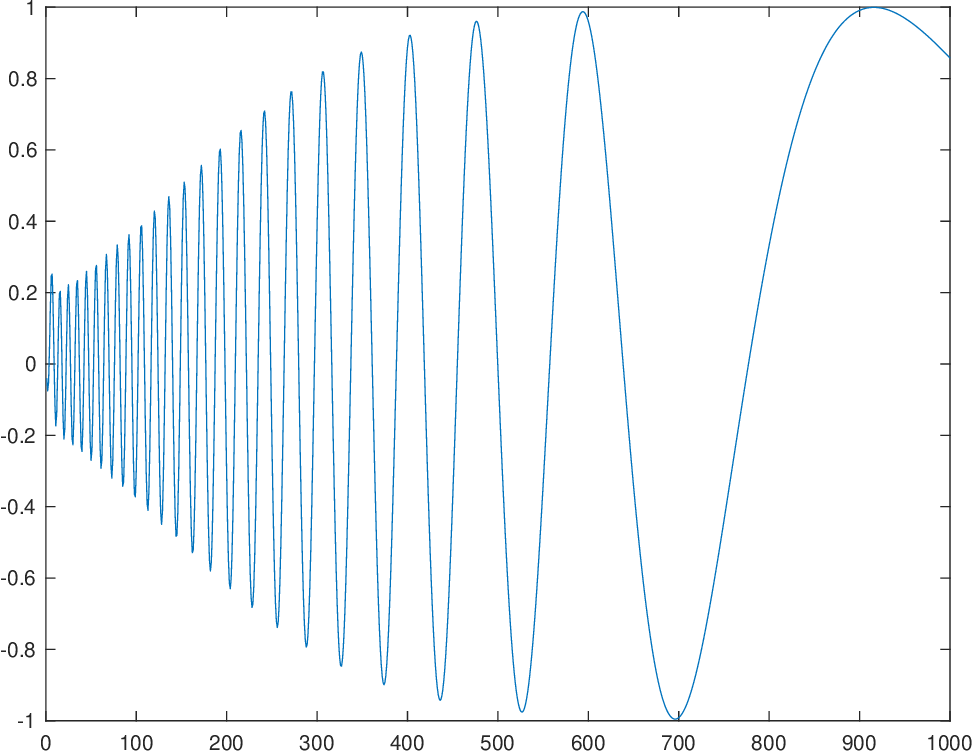} \\
      {\footnotesize\em Temporal scale $\sqrt{\tau} = 1024$}
      & $\quad\quad$ & {\footnotesize\em Temporal scale $\sqrt{\tau} = 1$} \\
      \includegraphics[width=0.30 \textwidth]{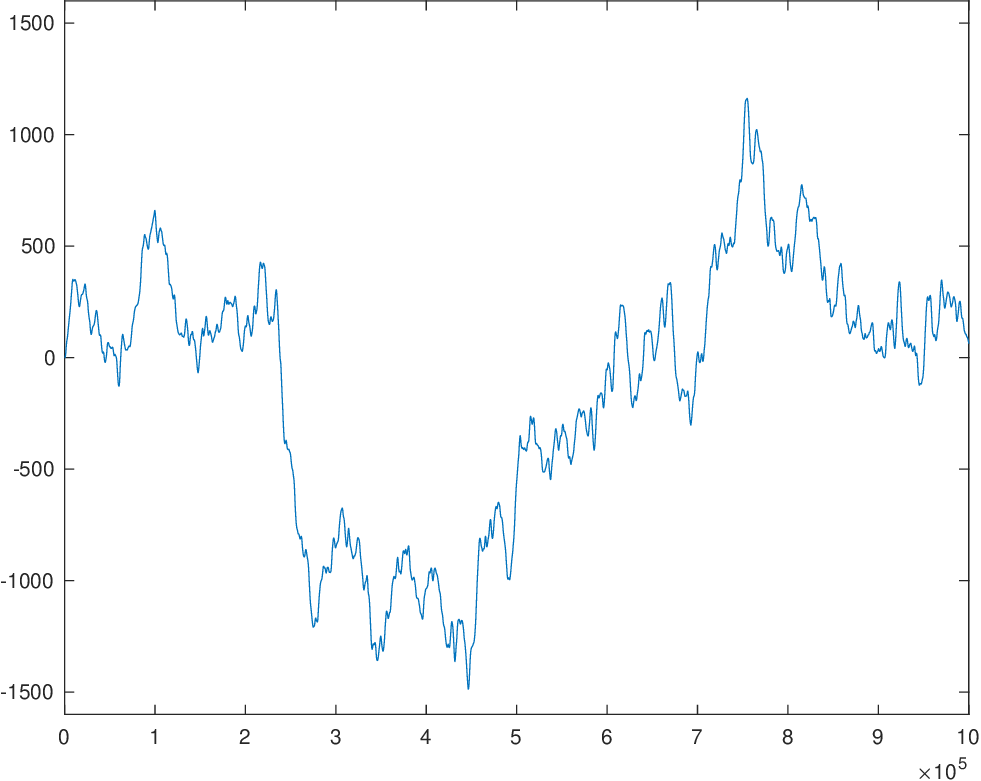}
      & $\quad\quad$ & \includegraphics[width=0.30\textwidth]{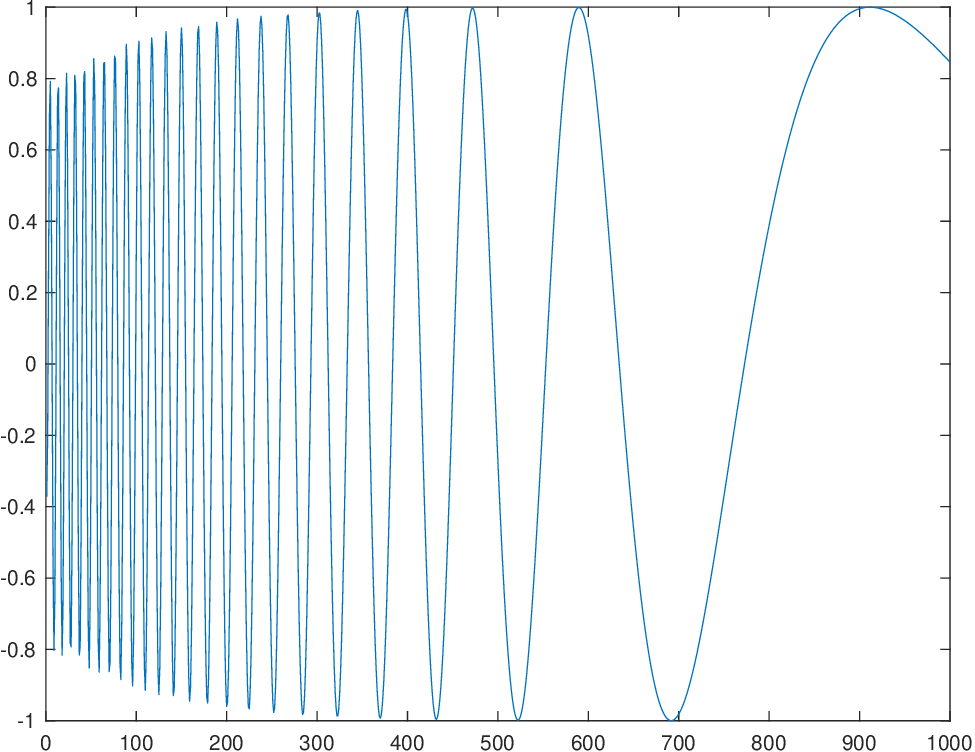} \\
      {\footnotesize\em Original signal}
      & $\quad\quad$ & {\footnotesize\em Original signal} \\
      \includegraphics[width=0.30\textwidth]{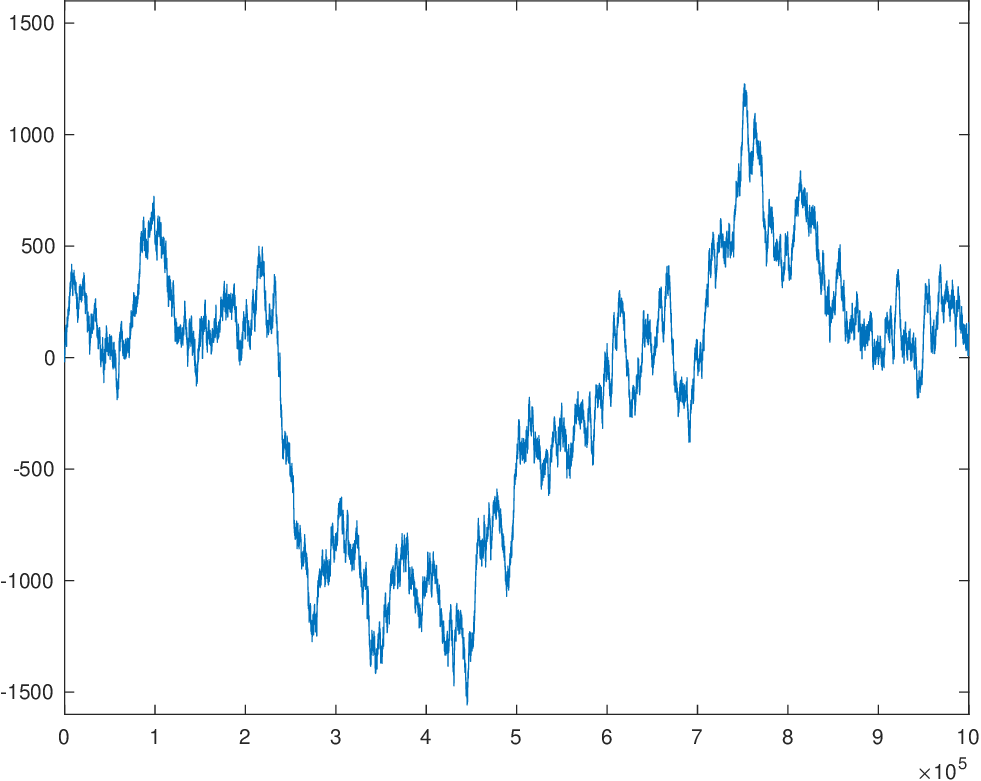}
      & $\quad\quad$ & \includegraphics[width=0.30\textwidth]{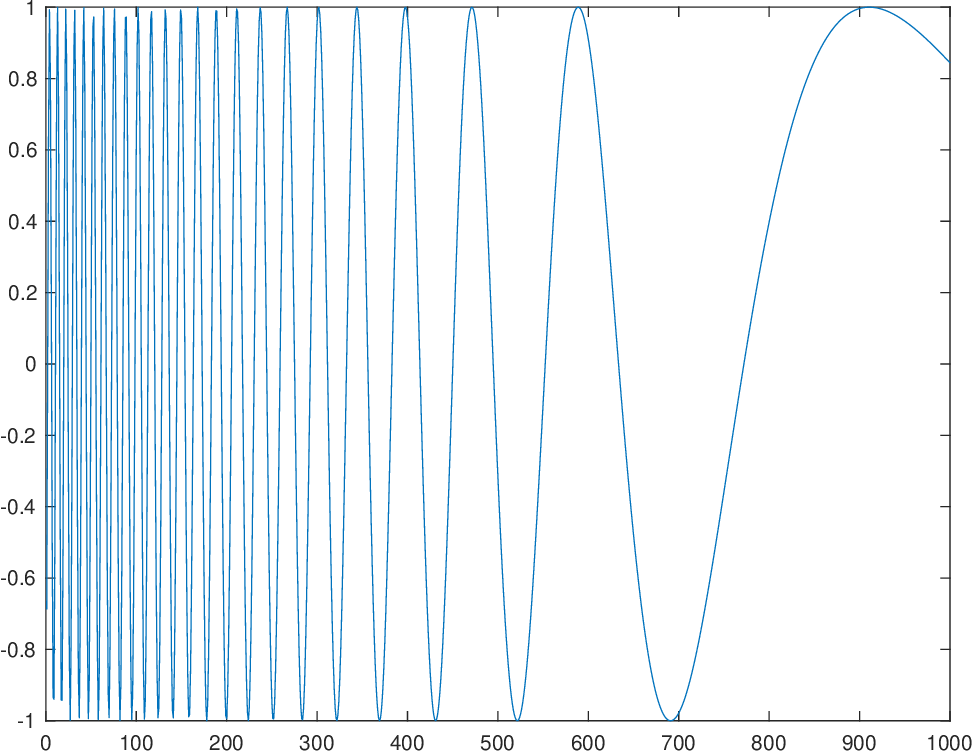} \\
      \end{tabular} 
  \end{center}
   \caption{Illustration of temporal smoothing of two 
     signals:
     (left) a Brownian noise signal generated from a simulated Wiener process and
     (right) a synthetic sine wave signal $f(t) = \sin(\exp((b-t)/a))$ for $a = 200$
     $b = 1000$ with temporally varying frequency so that the
     wavelength increases with time $t$,
    computed using a discrete approximation of the time-causal limit
    kernel for $c = 2$ in terms of a set of recursive filters coupled in
    cascade. Observe how fine-scale structures corresponding to higher
    frequencies are successively suppressed when going from finer to
    coarser temporal scales, and also that the
    temporal scale-space representations at coarser temporal scales
    are associated with longer
    temporal delays, in this figure seen as different offsets in the
    positions of the peaks in the temporal signal at different
    temporal scales.
    (Horizontal axes: time. Vertical axes: signal values.)}
  \label{fig-temp-brown-smooth}
\end{figure*}

\begin{figure*}[hbtp]
  \begin{center}
    \begin{tabular}{ccc}
      \includegraphics[width=0.30\textwidth]{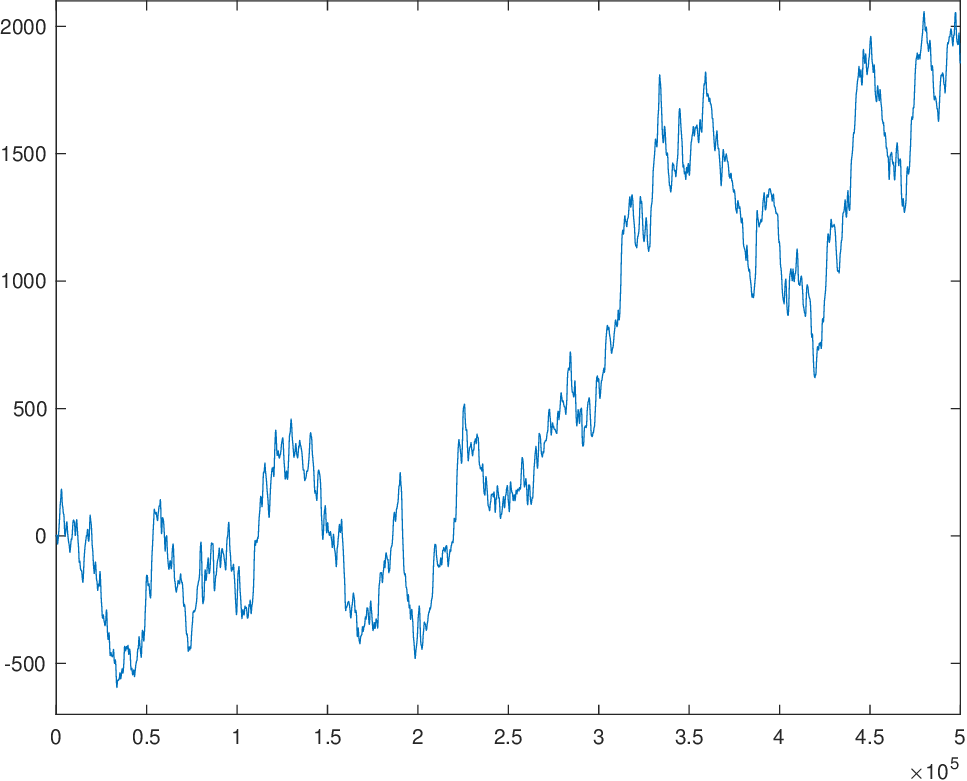}
      & {\footnotesize ${\begin{array}{c} \footnotesize t' = S t  \\
                           \footnotesize \tau' = S^2 \tau \\
                           \longrightarrow 
                         \end{array}}$}
      & \includegraphics[width=0.30\textwidth]{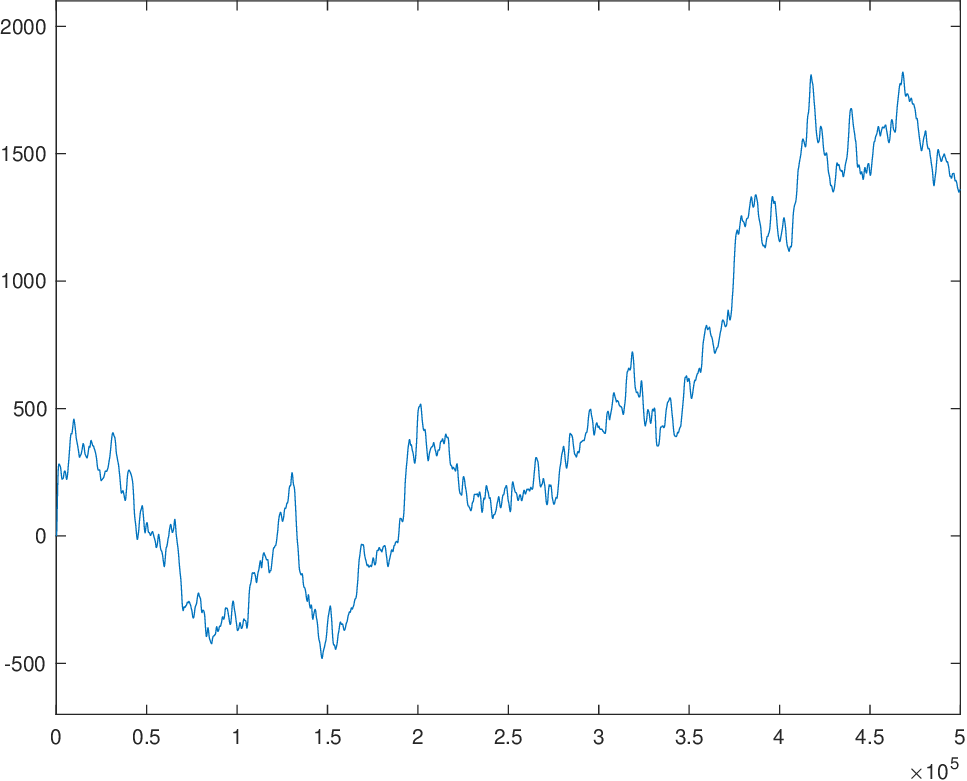} \\
      $\uparrow{\rlap{$\scriptstyle{{*T(t;\; 
               \tau)}}$}}$ & $\quad$ & $\uparrow{\rlap{$\scriptstyle{{*T(t';\; 
                                       \tau')}}$}}$ \\
      $\quad$ & $\quad$ & $\quad$ \\
      \includegraphics[width=0.30\textwidth]{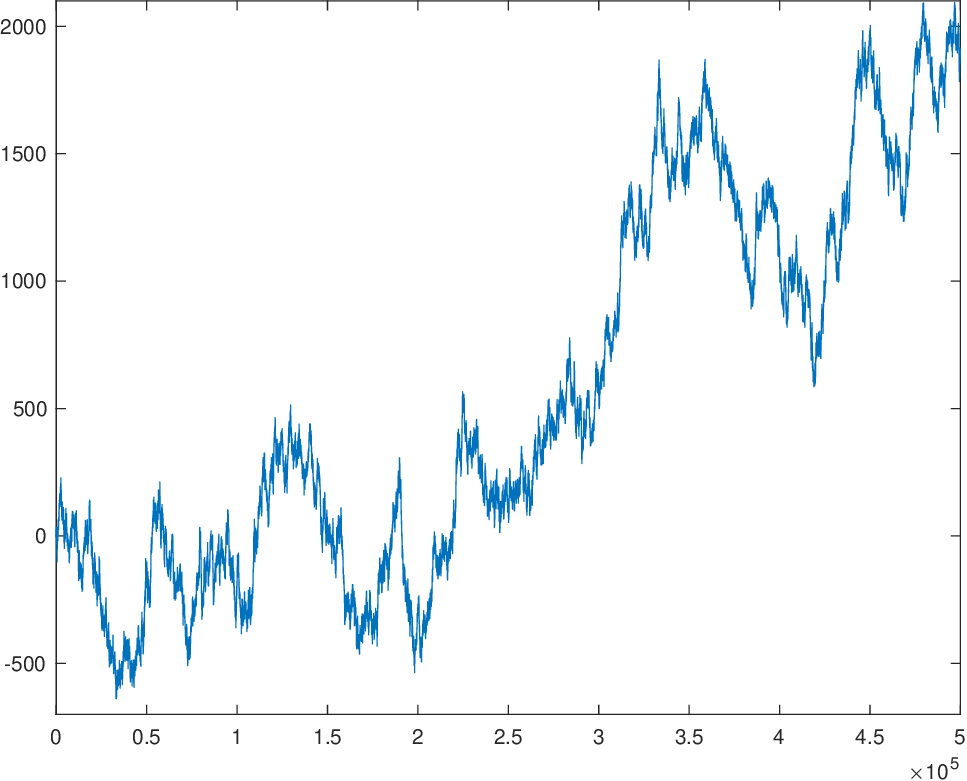}
      & {\footnotesize ${\begin{array}{c} \footnotesize t' = S t  \\
                          \longrightarrow \end{array}}$}
        &
          \includegraphics[width=0.30\textwidth]{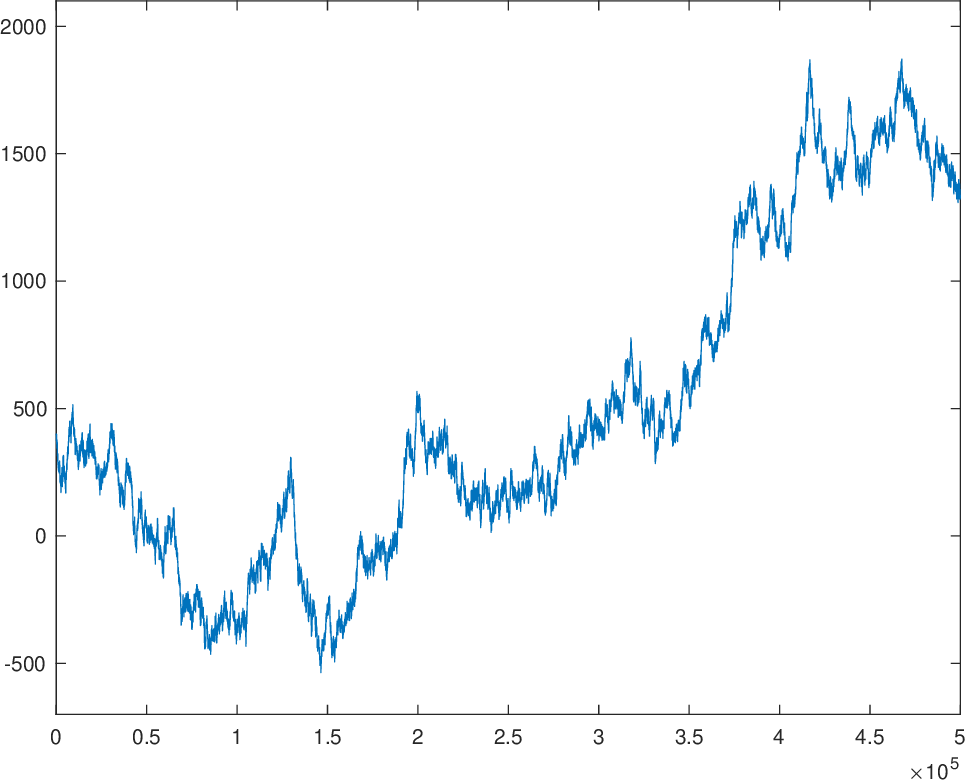} \\
    \end{tabular}
  \end{center}
  \caption{Illustration of the temporal scale covariance property of the
    temporal scale-space representation defined from convolutions with
  the time-causal limit kernel. In the bottom row, the signal in the
  right column is a rescaling of the signal in the left column by a temporal scaling factor $S = 2$
  (with the temporal
  rescaling performed relative to the center of the temporal interval).
  In the top row, the temporal
  scale-space representations at the matching temporal scale levels $\sqrt{\tau} = 128$
  and $\sqrt{\tau'} = 256$ have for distribution parameter $c = 2$
  been computed from the corresponding input signals in
  the bottom row. Due to the temporal scale-covariance property,
  these temporal scale-space representations are in the ideal
  continuous case related by a temporal scaling transformation with
  the same temporal scaling factor $S = 2$ as between the input
  signals. If one for experimental purposes compares a corresponding
  temporal rescaling of the output from
  the discrete implementation in terms of recursive filters
  (described in more detail in Section~\ref{sec-comp-impl-disc-signals}), one can
  see that the
  corresponding graphs are practically indistinguishable (see Figure~\ref{fig-ill-temp-sc-cov-diff}).
  In this way, this experiment verifies and visualizes the theoretical properties
  reflected in the commutative diagram in Figure~\ref{fig-comm-diag-temp-sc-transf}.
  (Horizontal axes: time. Vertical axes: signal values.)}
  \label{fig-ill-temp-sc-cov}

  \bigskip
  \bigskip
  
  \begin{center}
    \begin{tabular}{cc}
      \includegraphics[width=0.30\textwidth]{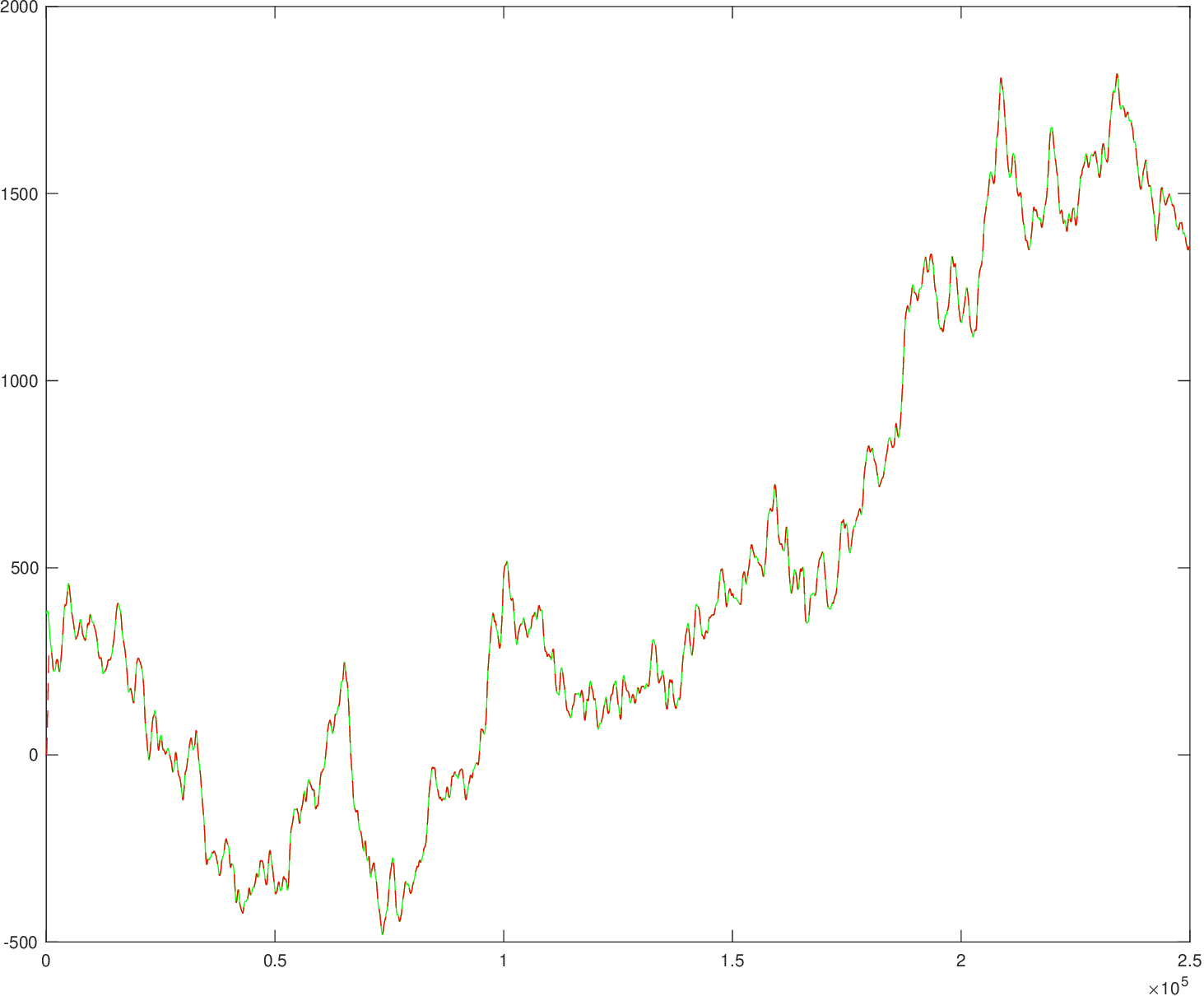}
          & \includegraphics[width=0.30\textwidth]{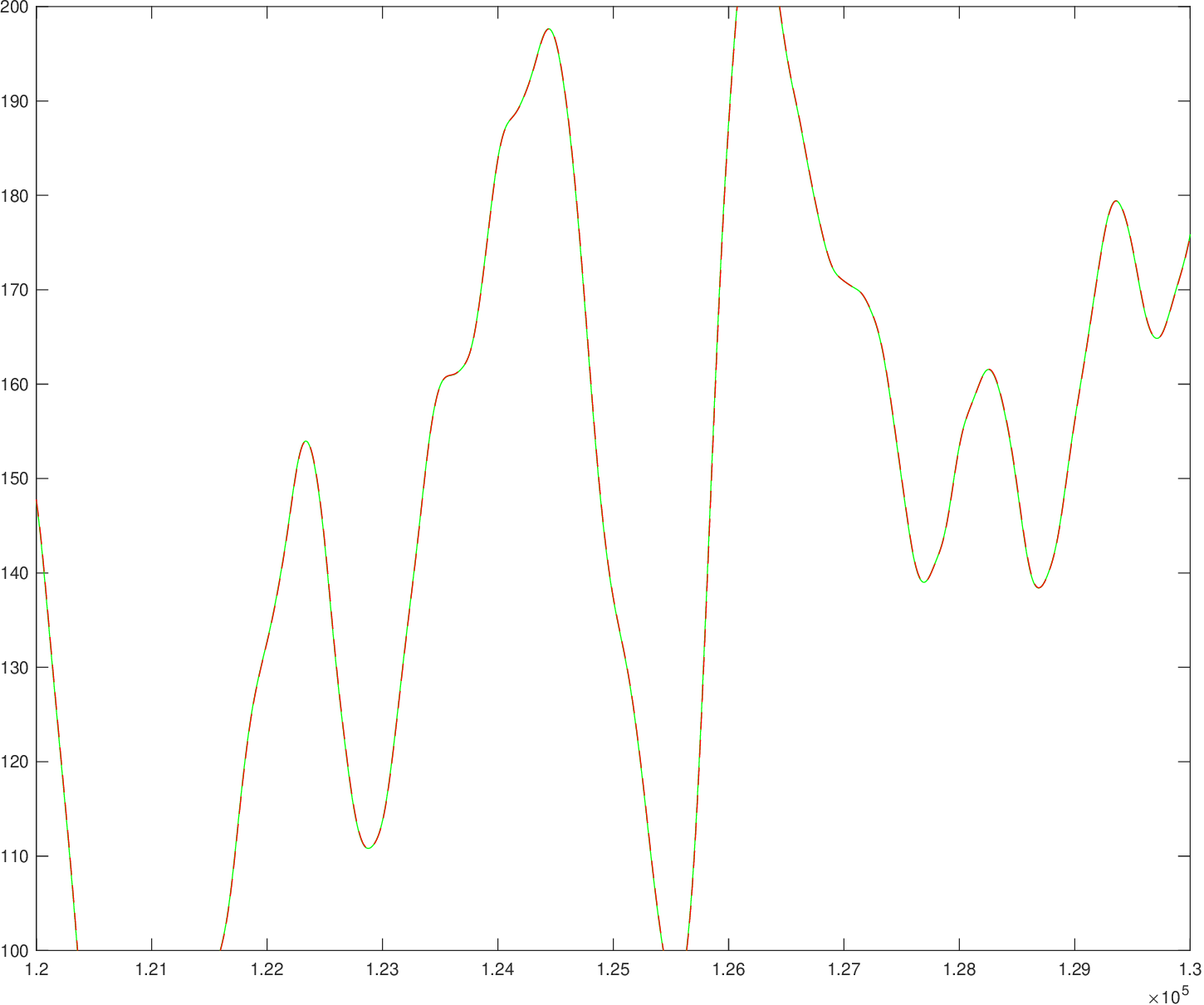}
    \end{tabular}
  \end{center}
  \caption{Comparison between the two different ways of computing the
    representation in the upper right corner in
    Figure~\ref{fig-ill-temp-sc-cov} from the corresponding
    representation in the lower left corner, using either the
    clockwise direction (marked in green) or the counterclockwise
    direction (marked in red).
    When generating this illustration, we have first essentially
    performed a rescaling of the scale-space representation of the
    signal in the left column and marked the result as solid green curve, and then
    overlayed the scale-space representation of the signal in the
    right column with a dashed red curve. (Technically, in the
    discrete implementation, we have, however, instead visualized the equivalent
    result of such a computation at a lower resolution, to avoid the
    formally ill-defined operation of interpolating the discrete signal in the
    left column to a higher resolution, and instead subsampled the
    signal in the right column, which explains the change in the
    labelling of the temporal axis.) (left) The result for the entire
    temporal interval used in the right column in
    Figure~\ref{fig-ill-temp-sc-cov}. (right) Enlargement of a central
    region of the temporal interval. As can be seen from the
    visualization, the results computed in the clockwise or
    counterclockwise directions are basically indistinguishable,
    demonstrating the scale covariance property of the temporal
    scale-space representation defined by convolution with the
    time-causal limit kernel. (The result is best viewed by zooming in
    to a digital copy of the article.) (Horizontal axes: time. Vertical axes: signal values.)}
  \label{fig-ill-temp-sc-cov-diff}
\end{figure*}

\subsubsection{Qualitative properties}

Figure~\ref{fig-trunc-exp-kernels-1D} shows graphs of this time-causal limit kernel as well
its first- and second-order temporal derivatives for a few values of
the distribution parameter $c$.
As can be seen from the graphs, the raw smoothing kernels have a skewed
shape, where the temporal delay increases with decreasing values of
the distribution parameter $c$, and with the explicit measures of the
skewness $\gamma_1$ and kurtosis
$\gamma_2$ of these kernels increasing as function of the distribution
parameter $c$ according to
(Lindeberg \citeyear[Equations~(130) and (131)]{Lin16-JMIV})
\begin{align}
  \begin{split}
     \label{eq-skewness-timecaus-kern-log-distr}
     \gamma_1 
    & = \frac{2 (c+1) \sqrt{c^2-1}}{\left(c^2+c+1\right)},
  \end{split}\\
 \begin{split}
     \label{eq-kurtosis-timecaus-kern-log-distr}
     \gamma_2 
     & = \frac{6 \left(c^2-1\right)}{c^2+1}.
  \end{split}
\end{align}

\subsubsection{Experimental results}

Figure~\ref{fig-temp-brown-smooth}
shows the result of smoothing two synthetic
temporal signals with the
time-causal limit kernel for different values of the temporal
scale parameter $\tau$.
As can be seen from the graphs, the signal is gradually smoothed from
finer to coarser temporal scales, here clearly seen in the way that
finer-scale structures are suppressed before coarser-scale structures
in the left column and that
higher frequencies are suppressed before lower frequencies in the
right column. 
In addition, the temporal delay increases from finer to coarser
temporal scales, here seen in terms of different temporal offsets
regarding the temporal moments at which the temporal peaks occur.

When using a comparably large value of the distribution parameter $c$,
as used in this figure, the temporal delay will be comparably low,
which is a preferable property when needing to respond fast in a
time-critical context. When using lower values of the distribution
parameter, the temporal delay at a given temporal scale will be
longer, which may be a preferable property if you want to use 
the temporal scale-space representations as
temporal memory buffers, with
the coarser temporal scale representations then constituting memories of
what has happened further in the past.

Figure~\ref{fig-ill-temp-sc-cov} gives an experimental illustration of
the temporal scale covariant property of the time-causal limit
kernel. Here, a synthetic signal generated from a simulated Wiener
process has been rescaled by a temporal rescaling factor $S = 2$.
From these two input signals, temporal scale-space representations have
then been computed at the matching temporal scale levels
$\sqrt{\tau} = 128$ and $\sqrt{\tau'} = 256$. Due to the temporal
scale covariance property, these temporal scale-space representations
are then also related by the same temporal scaling factor $S = 2$.

Figure~\ref{fig-ill-temp-sc-cov-diff} gives an illustration of the
equality between the two different ways of computing the
representation in the upper right corner from the signal in the lower
left corner in Figure~\ref{fig-ill-temp-sc-cov}, using either a
clockwise orientation or a counterclockwise orientation in the
corresponding commutative diagram in Figure~\ref{fig-comm-diag-temp-sc-transf}.
As can be seen from the visualization, the results are essentially
indistinguishable, showing that a good numerical approximation of
temporal scale covariance can also be achieved in a discrete
implementation (to be described further in Section~\ref{sec-comp-impl-disc-signals}).

\begin{figure}[t]
    \begin{center}
      \includegraphics[width=0.48\textwidth]{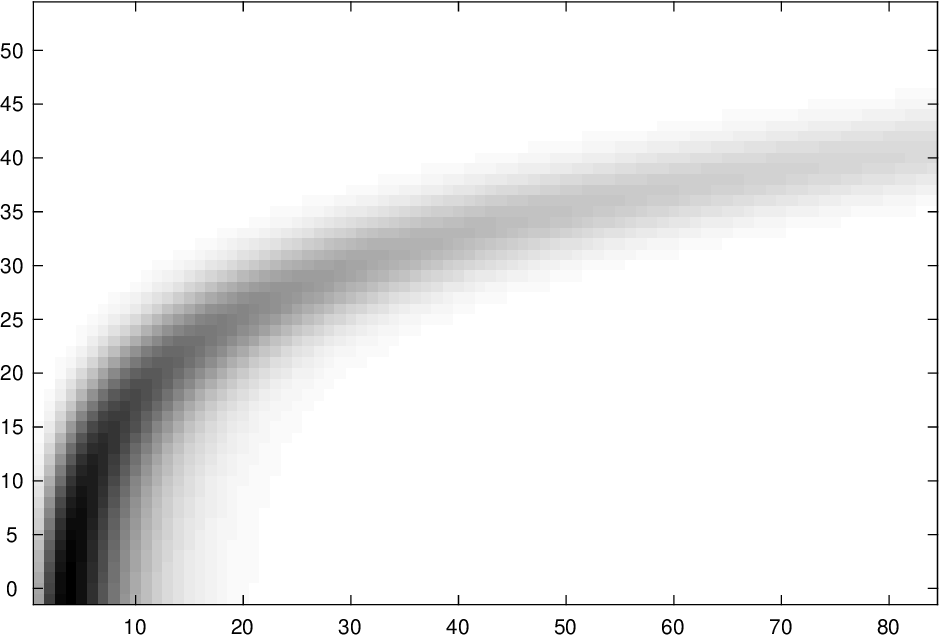}
    \end{center}
  \caption{Illustration of the temporal memory property of the
    time-causal temporal scale-space representation generated by
    convolutions with the time-causal limit kernel (and as always for
    this kernel using a
    logarithmic distribution of the temporal scale levels). The input
    signal is here a time-causal peak corresponding to a discrete
    approximation of the time-causal
  limit kernel for $\sqrt{\tau} = 4$. The time-causal temporal
  scale-space representation of this signal has then been computed by
  convolution with discrete approximations of the time-causal limit kernel for different temporal
scales, using a very low value of the distribution parameter $c =
\sqrt[8]{2}$ to enable a more clear visualization of the transition
properties between different temporal scales. As can be seen from the
illustration, the response at the onset of the temporal peak mainly occurs at
fine temporal scales, whereas the trace of the peak moves to coarser
temporal scales with increasing time, reflecting the temporal memory property
of the time-causal scale-space representation. (Horizontal axis: time,
Vertical axis: temporal scale index for temporal scale levels between
$\sqrt{\tau} \approx 0.543$ and $\sqrt{\tau} = 64$.)}
  \label{fig-ill-temp-mem}
\end{figure}

Figure~\ref{fig-ill-temp-mem} shows an illustration of the temporal memory property of this
type of temporal scale-space representation.
For a very low value of the distribution parameter, $c = \sqrt[8]{2}$
to enable a dense sampling of the temporal scale levels, and thus a
more clear visualization of the transitions from finer to coarser
temporal scales, we show the response to a temporal peak as function
of time and temporal scales in the temporal scale-space
representation. As can be seen from the figure, the trace in temporal
scale space moves from finer scales at the appearance of
the temporal peak, and then to successively coarser levels of temporal
scales as time flows. After a certain amount of time, the only
trace of the temporal peak, which originated just after $t = 0$, has
moved to only being present at coarser temporal scales. In this way,
the temporal scale channels at successively coarser levels of temporal
scales serve as a temporal memory of what has happened during different
time intervals in the past.

\subsubsection{Applications of the time-causal limit kernel}

The time-causal limit kernel and its temporal derivatives has been
used for modelling the temporal component in spatio-temporal receptive
fields in the retina, the LGN and the primary visual cortex (V1)
(Lindeberg \citeyear{Lin21-Heliyon}),
for modelling the temporal component in methods for spatio-temporal feature
detection in video data (Lindeberg \citeyear{Lin16-JMIV}),
for expressing methods for temporal scale selection in temporal
signals (Lindeberg \citeyear{Lin17-JMIV,Lin18-SIIMS}),
for modelling the temporal component of spatio-temporal smoothing in
methods for spatio-temporal scale selection (Lindeberg
\citeyear{Lin18-JMIV,Lin18-SIIMS}) and for modelling the temporal
component of smoothing in computer vision methods for video analysis
(Jansson and Lindeberg \citeyear{JanLin18-JMIV}).

In Section~\ref{sec-temp-sc-neur-sign}, we do
additionally propose to use the time-causal limit kernel for modelling
temporal phenomena at multiple temporal scales in neural signals, and
in Section~\ref{sec-spectr-temp-RFs} specifically to use this kernel
for modelling the
temporal variability in auditory receptive fields.

In Section~\ref{sec-rel-wavelet-anal} we outline how the time-causal
limit kernel can be used for defining time-causal and time-recursive
wavelet representations, and in Section~\ref{sec-rel-time-freq-anal} how the
time-causal limit kernel makes it possible to define time-causal and
time-recursive time-frequency representations (spectrograms) that
additionally obey temporal scale covariance.

\begin{figure*}[hbtp]
  \begin{center}
   \begin{tabular}{ccc}
        {\footnotesize $h(t;\; c = \sqrt{2})$} 
      & {\footnotesize $h_{t}(t;\; c = \sqrt{2})$} 
      & {\footnotesize $h_{tt}(t;\; c = \sqrt{2})$} \\
      \includegraphics[width=0.30\textwidth]{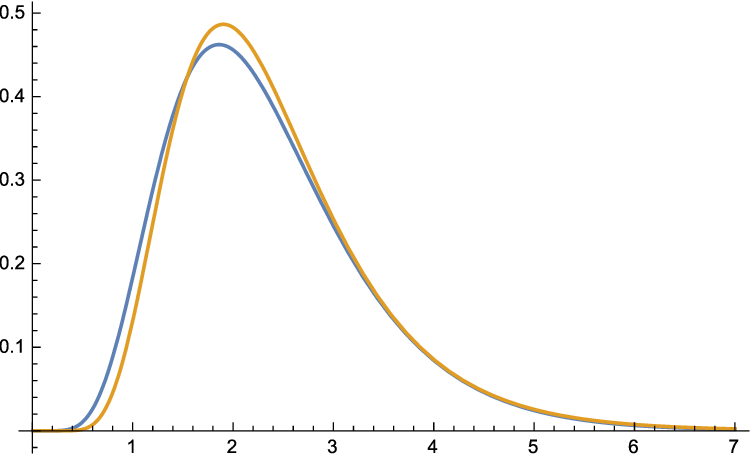} &
      \includegraphics[width=0.30\textwidth]{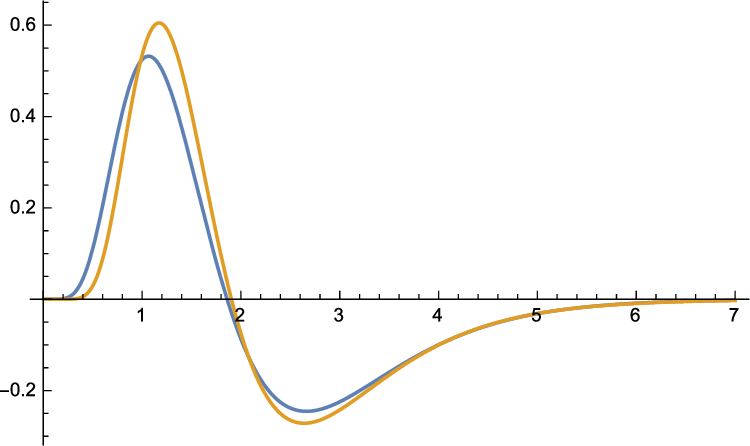} &
      \includegraphics[width=0.30\textwidth]{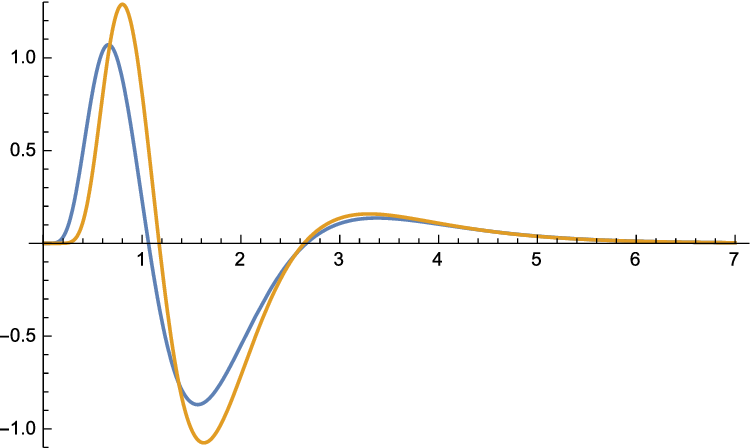} \\
\\
      {\footnotesize $h(t;\; c = 2)$} 
      & {\footnotesize $h_{t}(t;\; c = 2)$} 
      & {\footnotesize $h_{tt}(t;\; c = 2)$} \\
      \includegraphics[width=0.30\textwidth]{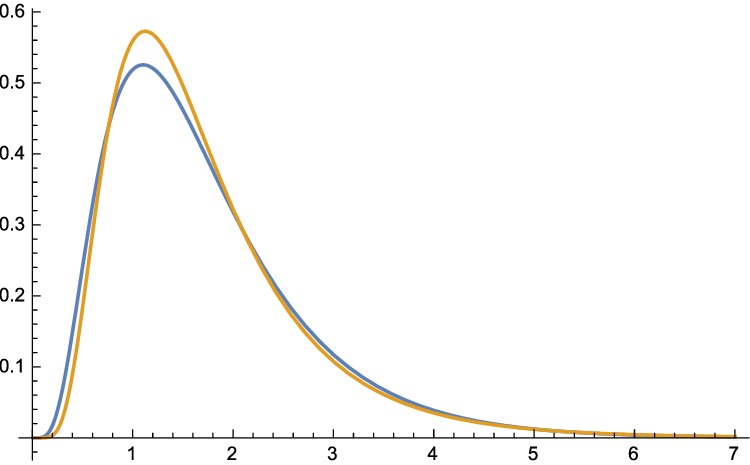} &
      \includegraphics[width=0.30\textwidth]{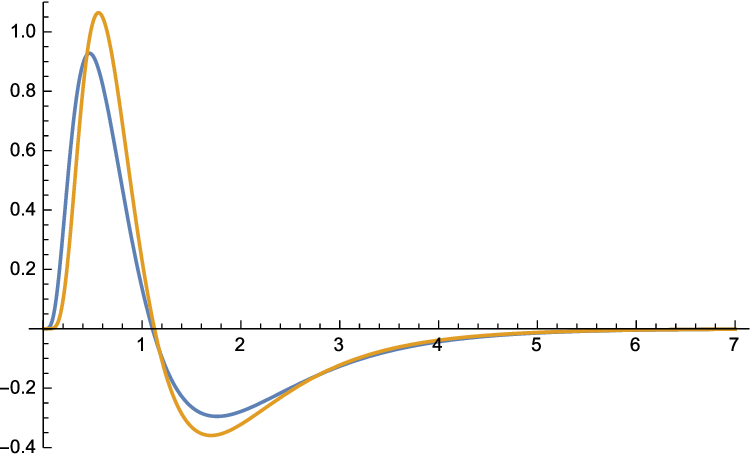} &
      \includegraphics[width=0.30\textwidth]{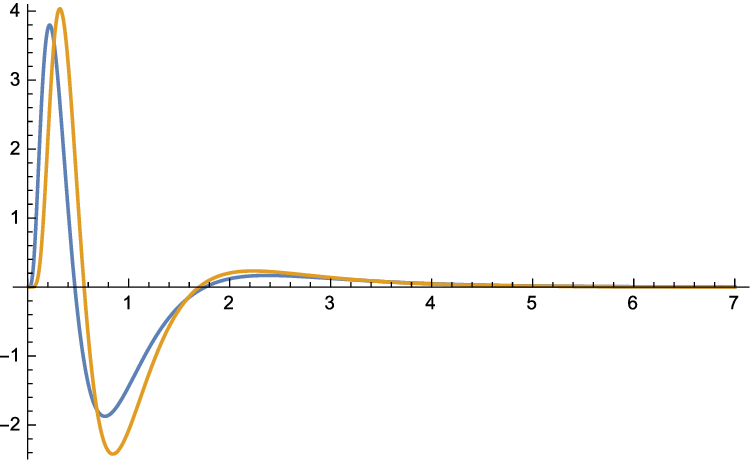} \\
    \end{tabular} 
  \end{center}
   \caption{Comparison between (blue curves) the time-causal limit kernel according
     to (\ref{eq-FT-comp-kern-log-distr-limit}) and
     approximated using the first $K = 7$ components of the infinite
     convolution of truncated exponential kernels in cascade with its
     first- and second-order temporal derivatives and 
     (brown curves) the temporal kernels in Koenderink's scale-time model
     (\ref{eq-Koe-scale-time-model})
    and their
           first- and second-order temporal derivatives.
           All kernels correspond to temporal scale (variance) $\tau = 1$ with
           the additional parameters determined such that the
           temporal mean values (the first-order temporal moments) become equal in
           the limit case when the number of temporal scale levels $K$
           tends to infinity
           (equation~(\protect\ref{eq-var-transf-par-tau-c-sigma-delta-explogdistr-scaletime-models})).
           (top row) Logarithmic distribution of the temporal scale levels
           for $c = \sqrt{2}$
         (bottom row) Corresponding results 
         for $c = 2$.
       (Horizontal axes: time. Vertical axes: function values.)}
  \label{fig-trunc-exp-kernels-1D+scaletime}
\end{figure*}

\begin{figure*}[hbtp]
  \begin{center}
   \begin{tabular}{ccc}
        {\footnotesize $h_{\mbox{\scriptsize ex-Gauss,gen}}(t)$} 
     & {\footnotesize $h_{\mbox{\scriptsize ex-Gauss,gen}}(t)$} 
     & {\footnotesize $h_{\mbox{\scriptsize ex-Gauss,gen}}(t)$} \\
      \includegraphics[width=0.30\textwidth]{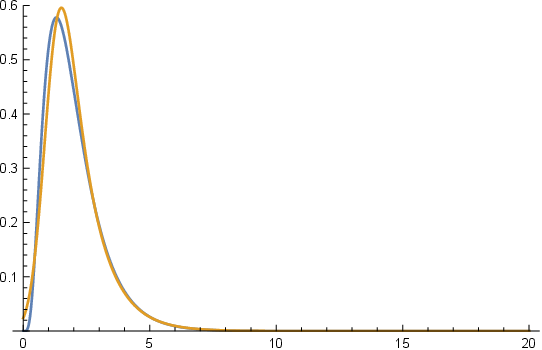} &
      \includegraphics[width=0.30\textwidth]{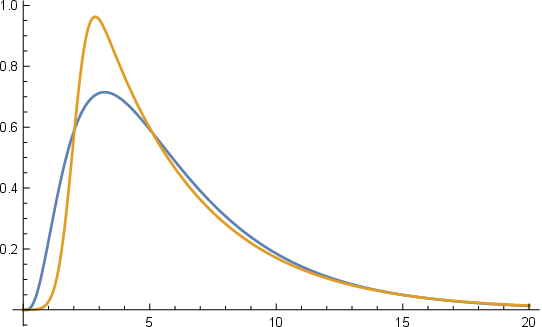} &
     \includegraphics[width=0.30\textwidth]{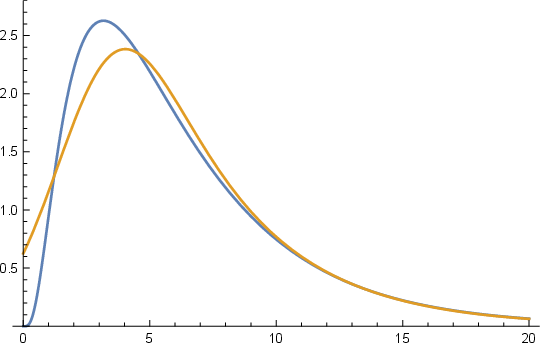} \\
  \end{tabular} 
  \end{center}
   \caption{Comparison between (brown curves) the ex-Gaussian model
     according to (\ref{eq-ex-Gaussian-gen-main}) and
     (blue curves) the time-causal limit kernel according to
     (\ref{eq-FT-comp-kern-log-distr-limit}) and
     approximated using the first $K = 7$ components of the infinite
     convolution of truncated exponential kernels in cascade.
     (left) for $\mu = 1$, $\sigma = 1/2$, $m = 1$, $a_0 = 0$ and $a_1
     = 1$ corresponding to $\tau \approx 1.24$, $c
       \approx 1.89$, $b_0 = 0$ and $b_1 \approx 1.25$,
    (middle) for $\mu = 4$, $\sigma = 1/2$, $m = 2$, $a_0 = 0$ and $a_1
     = 1$ corresponding to $\tau \approx 16.25$, $c
     \approx 2.65$, $b_0 = 0$ and $b_1 \approx 5.01$,
    (right) for $\mu = 4$, $\sigma = 2$, $m = 2$, $a_0 = 0$ and $a_1
     = 1$ corresponding to $\tau \approx 18.94$, $c
     \approx 2.89$, $b_0 = 0$ and $b_1 \approx 19.37$.
   (Horizontal axes: time. Vertical axes: function values.)}
  \label{fig-trunc-exp-kernels-1D+exGaussian}
\end{figure*}

\subsection{Alternative scale-covariant temporal models}

An alternative type of temporal model that one could also consider
from the general classification of temporal scale-space kernels is to use a set of
parallel temporal channels formed by convolution of the input signal, with a single
truncated exponential function in each channel, and with a geometric
distribution of the their time constants,
of the form (\ref{eq-temp-channel-single-exp-fcn}).
As previously explained in Section~\ref{sec-log-mem-of-past},
such temporal models 
have been previously used as models of temporal memory in neuroscience
(Howard \citeyear{How21-HandBookHumMem};
Bright {\em et al.\/} \citeyear{BriMeiCruTigBufHow20-PNAS}).

Because of the geometric distribution of the time constants in these
temporal channels, they will obey temporal scale covariance.
Temporal scale covariance will also apply to different types of
generalizations of such a model, {\em e.g.\/} by having the same
small number of truncated
exponential kernels in cascade in each temporal channel, with the time
constants between the different temporal channels coupled according to
a geometric distribution.

A fundamental difference between such
temporal models and the temporal scale-space model based on the
time-causal limit kernel, however, is that in the first class of models the temporal channels for
larger values of the scale parameter are not guaranteed to constitute
simplifications of the temporal channels for smaller values of the
scale parameter. By the temporal smoothing kernels being scale-space
kernels, each temporal channel is guaranteed to be a simplification of
the input signal. When relating different temporal scale channels to each other, however,
the number of local extrema in a temporal channel for a larger value
of the temporal scale parameter is not guaranteed to not exceed the
number of local extrema in a temporal channel for a finer smaller
value of the temporal scale parameter.

Because of the scale-recursive property (\ref{eq-recur-rel-limit-kernel})
of the time-causal limit kernel, it is on the other hand formally guaranteed that the
temporal scale-space representation at the next coarser temporal scale corresponds to the
result of applying temporal smoothing with a truncated exponential
kernel to the temporal scale-space representation at the nearest finer
temporal scale. Applied recursively, the temporal scale-space representation
at any coarser temporal scale corresponds to the result of applying a
set of truncated exponential kernels in cascade to the representation
at any finer temporal scale. In this way, for the temporal scale-space
representation generated by convolution with the time-causal limit
kernel for different values of the temporal scale parameter, every temporal scale-space
representation at a given temporal scale is guaranteed to constitute
a formal simplification of
any other temporal scale-space representation at any finer temporal scale.

The time-causal limit kernel is special in that it both obeys temporal
scale covariance and guarantees non-creation of new local extrema with
increasing temporal scales with regard to convolutions over a time-causal temporal domain.

\subsection{Relation to Koenderink's scale-time model}
\label{sec-rel-koe-scale-time}

In his scale-time model, Koenderink (\citeyear{Koe88-BC}) proposed to perform a
logarithmic mapping of the past via a temporal delay $\delta$ and then
applying Gaussian smoothing with standard deviation $\sigma$
in the transformed domain. If we
additionally normalize these kernels to unit $L_1$-norm, we obtain a
time-causal kernel of the form
(Lindeberg \citeyear[Equation~(151)]{Lin16-JMIV})
\begin{equation}
  \label{eq-Koe-scale-time-model}
  h_{\mbox{\scriptsize Koe}}(t;\; \sigma, \delta) 
  =\frac{1}{\sqrt{2 \pi } \sigma \,\delta}
  e^{-\frac{\log ^2\left(\frac{t}{\delta }\right)}{2 \sigma ^2} -\frac{\sigma^2}{2}}.
 \end{equation}
In (Lindeberg \citeyear[Appendix~2]{Lin16-JMIV}) a formal mapping between
this scale-time kernel and the time-causal limit kernel is derived, by requiring the
first- and second-order moments of these two classes of kernels to be equal:
\begin{equation}
  \label{eq-var-transf-par-tau-c-sigma-delta-explogdistr-scaletime-models}
  \left\{
    \begin{array}{l}
        \tau = \delta^2 \, e^{3 \sigma ^2} \left(e^{\sigma ^2}-1\right)  \\
      c = \frac{e^{\sigma ^2}}{2-e^{\sigma ^2}}
    \end{array}
  \right.
  \quad\quad
\left\{
    \begin{array}{l}
      \sigma = \sqrt{\log \left(\frac{2 c}{c+1}\right)}  \\
      \delta = \frac{(c+1)^2 \sqrt{\tau}}{2 \sqrt{2} \sqrt{(c-1) c^3}}
    \end{array}
  \right.
\end{equation}
which hold as long as $c > 1$ and $\sigma < \sqrt{\log 2} \approx 0.832$. 

Figure~\ref{fig-trunc-exp-kernels-1D+scaletime} shows a comparison
between the time-causal limit kernel and Koenderink's scale-time
kernels regarding the zero-order convolution kernels as well as their
first- and second-order derivatives.
As can be seen from the graphs, these two classes of kernels have
qualitatively rather similar shapes.
The time-causal limit kernel does, however, have the conceptual
advantage that it can be computed in a time-recursive manner, whereas
the scale-time kernel does not have any known time-recursive implementation,
implying that it formally requires an infinite memory of the past
(or some substantially extended temporal buffer, if the infinite
temporal convolution integral is truncated at the tail).

While we do not have any compact explicit expression for the
time-causal limit kernel over the temporal domain, if we approximate
the time-causal limit kernel by a scale-time kernel according to the
mapping
(\ref{eq-var-transf-par-tau-c-sigma-delta-explogdistr-scaletime-models}),
we obtain the following estimate for the location of the maximum point
of the time-causal limit kernel:
\begin{equation}
  \label{eq-approx-temp-pos-time-caus-log-distr}
  t_{\mbox{\scriptsize max}} \approx \frac{(c+1)^2 \, \sqrt{\tau}}{2 \sqrt{2} \sqrt{(c-1) c^3}} = \delta.
\end{equation}
This estimate can be expected to be an overestimate, and is a better
estimate of the temporal delay of the time-causal limit kernel than
the temporal mean according to (\ref{eq-mean-var-trunc-exp-filters}).

\subsection{Relation to the ex-Gaussian model used by Bright {\em et al.\/}}
\label{sec-ex-Gaussian-model}

In (Bright {\em et al.\/} \citeyear{BriMeiCruTigBufHow20-PNAS}),
a so-called ex-Gaussian model (Grushka \citeyear{Gru72-AnalChem}),
that is the convolution of
an unnormalized Gaussian function with an unnormalized truncated
exponential kernel 
\begin{equation}
    \label{eq-ex-Gaussian-gen-main}
  h_{\mbox{\scriptsize ex-Gauss,gen}}(t) = a_0 + a_1 \int_{u=0}^{\infty} e^{-\frac{(t-m-u)^2}{2 \sigma
      ^2}} e^{-\frac{u}{\mu}} \, du,
\end{equation}
is used for fitting temporal response functions of neurons to an 
analytical temporal model.
In Appendix~\ref{app-second-order-model-time-caus-limit-kern}, a relation between this
ex-Gaussian model and a corresponding model based on the time-causal
limit kernel
\begin{equation}
  \label{eq-time-caus-model-ex-Gaussian-gen-main}
 h_{\mbox{\scriptsize limit-kern,gen}}(t) = b_0 + b_1 \, \Psi(t;\; \tau, c)
 \end{equation}
is derived by requiring the
zero-, first- and second-order temporal moments of these kernels to be
equal, if the DC-offsets $a_0$ and $b_0$ are disregarded and assumed
to be equal.

This leads
to the following mapping between the parameters of the two models
\begin{align}
  \begin{split}
     b_1 = M_0,
   \end{split}\\
 \begin{split}
     c = \frac{\delta^2+V}{\delta^2-V},
   \end{split}\\
  \begin{split}
     \tau = V,
   \end{split}
\end{align}
where $\delta$ and $V$ denote the temporal mean and the temporal variance of the
ex-Gaussian model for $a_0 = 0$
\begin{align}
  \begin{split}
     \label{eq-delta-from-M1-M0}
     \delta & = \frac{M_1}{M_0},
   \end{split}\\
  \begin{split}
    \label{eq-V-from-M2-M1-M0}
     V & = \frac{M_2}{M_0} - \left( \frac{M_1}{M_0} \right)^2,
   \end{split}
\end{align}
and $M_0$, $M_1$ and $M_2$ denote the explicit expressions for the zero-, first- and second-order
moments of the ex-Gaussian model for $a_0 = 0$, according to
(\ref{eq-M0-ex-Gauss}), (\ref{eq-M1-ex-Gauss}) and (\ref{eq-M2-ex-Gauss}).

Figure~\ref{fig-trunc-exp-kernels-1D+exGaussian} shows a few examples
of ex-Gaussian temporal models approximated by models based on the time-causal limit
kernel in this way. As can be seen from the graphs, the two classes of
kernels can capture qualitative similar temporal shapes in time-causal
temporal data,%
\footnote{A certain qualifier is, however, necessary in this context,
  since the ex-Gaussian model contains one more parameter than the
  model based on the time-causal limit kernel. Hence, the above
  mapping between these models is only valid if the temporal
  delay $m$ in the ex-Gaussian model is not too large compared the
  temporal time constant $\mu$ and the amount of temporal smoothing
  $\sigma$. If a mapping is to be performed between the two models in
  the regime where this assumption does not hold, then an additional
  temporal delay parameter $t_0$ should be introduced into the model
  based on the time-causal limit kernel
 (\ref{eq-time-caus-model-ex-Gaussian-gen-with-t0}) at the cost of
 more complex analytical expressions for determining the parameters of
 the model based on the time-causal limit kernel from the temporal
 moments, now up to order 3, of the ex-Gaussian model, see
 Appendix~\ref{app-third-order-model-time-caus-limit-kern} for a further treatment
of such an extension.}
with the
conceptual differences that:
(i)~the model based on the time-causal limit kernel always tends to zero at
the temporal origin $t = 0$ when the DC-offset is zero, whereas the
ex-Gaussian model may take non-zero values for $t = 0$,
(ii)~the time-causal limit kernel does not
contain any internal non-causal temporal component as the time-shifted
Gaussian kernel in
(\ref{eq-ex-Gaussian-gen-main}) constitutes,
and (iii)~the time-causal limit kernel has a completely time-recursive
implementation, which is essential when modelling temporal phenomena
in real time as they, for example, occur in biological neurons.
The model based on the time-causal limit kernel is also specifically
possible to implement based on a cascade of first-order integrators in
cascade, which is a natural model for the information transfer in
the dendrites of neurons (Koch \protect\citeyear[Chapters~11--12]{Koch99-book}).

\subsubsection{Extension to third-order moment-based model fitting
  involving also a flexible temporal offset}

In Appendix~\ref{app-third-order-model-time-caus-limit-kern}, an
extension of the above second-order moment-based model to a
third-order moment-based model is performed, which makes it possible to
also determine a temporal offset $t_0$
\begin{equation}
 h_{\mbox{\scriptsize limit-kern,gen}}(t) = b_0 + b_1 \, \Psi(t-t_0;\; \tau, c),
\end{equation}
and which may be relevant in situations when the temporal origin of the
signal cannot be accurately determined in an experimental situation.
Since the closed-form expressions for the solutions become more
complex in this case (they are determined from the solutions of a
fourth-order algebraic equation), we restrict ourselves to a
conceptual and algorithmic description in this treatment, see
Appendix~\ref{app-third-order-model-time-caus-limit-kern} for
further theoretical details and experimental results.

\subsubsection{Extension to model fitting for other signals or
  functions}
\label{sec-gen-model-fitting-time-caus-limit-kern}

The above general procedures, whereby the parameters in the model based on
the time-causal limit kernel are
determined from the lower-order temporal moments of the data, can also be more
generally used for fitting models based on the time-causal limit
kernel to other signals and functions that: (i)~are defined for non-negative
values of time, (ii)~assume non-negative values only, (iii)~have a
roughly unimodal shape of first increasing and then decreasing and (iv)~decay
towards zero towards infinity.
The approach for fitting basically implies replacing the temporal
moments $M_0$, $M_1$, $M_2$ and optionally $M_3$ of the ex-Gaussian
model by the temporal moments of the signal or function to be fit
with a model based on the time-causal limit kernel,
see Appendix~\ref{app-gen-model-fitting-time-caus-limit-kern} for
additional details.

\section{Computational implementation of convolutions with the
  time-causal limit kernel on discrete temporal data}
\label{sec-comp-impl-disc-signals}

In the theory presented so far, we have throughout assumed that the signal
is continuous over time. When implementing this model on sampled
temporal data, the theory must be transferred to a discrete temporal
domain.

In this section, we will describe how the temporal receptive fields
can be implemented in terms of corresponding discrete temporal
scale-space kernels that possess scale-space properties over a
discrete temporal domain, and in addition are both time-causal and
fully time-recursive.

Following Lindeberg (\citeyear{Lin90-PAMI}) and in a corresponding way as the treatment in
Section~\ref{sec-time-caus-time-rec-temp-scsp-model}, let us define a
discrete kernel as a discrete scale-space kernel if for any input signal it is
guaranteed that the number of local extrema, alternatively the number
of zero-crossings, cannot increase under convolution with the discrete
scale-space kernel.

\subsection{Classification of scale-space kernels for discrete
  signals}

To characterize the class of discrete scale-space kernels, we can,
in a corresponding way as for the continuous case,
also build upon classical results by Schoenberg
(\citeyear{Sch30,Sch46,Sch47,Sch48,Sch50,Sch53,Sch88-book}),
and as further developed in the monograph by Karlin (\citeyear{Kar68}).

Making a summary of the treatment in Lindeberg (\citeyear[Section~IV]{Lin90-PAMI})
(\citeyear[Section~6.1]{Lin16-JMIV}), a
discrete smoothing kernel is a discrete scale-space kernel if and only if it
has its
generating function of the sequence of filter coefficients
$\varphi(z) = \sum_{n=-\infty}^{\infty} c_n z^n$
of the form (Schoenberg \citeyear{Sch48})
  \begin{equation}
    \label{eq-char-pf}
    \varphi(z) = c \; z^k \; e^{(q_{-1}z^{-1} + q_1z)}
                 \prod_{i=1}^{\infty} \frac{(1+\alpha_i z)(1+\delta_i
    z^{-1})} {(1-\beta_i z)(1-\gamma_i z^{-1})}
  \end{equation}
where
$c > 0$, $k \in \bbbz$, 
$q_{-1}, q_1, \alpha_i, \beta_i, \gamma_i, \delta_i \geq 0$ and
$\sum_{i=1}^{\infty}(\alpha_i + \beta_i + \gamma_i + \delta_i) < \infty$.

\subsubsection{Basic classes of primitive scale-space kernels over a discrete
  signal domain}

With regard to the original temporal domain,%
\footnote{In the general expression (\ref{eq-char-pf}) for the
  generating function of a discrete scale-space kernel, the factors
  $1+\alpha_i z$ and $1+\delta_i z^{-1}$ are the generating functions
  of generalized binomial filters of the form (\ref{eq-gen-bin-filt}),
  the factors $1-\beta_i z$ and $1-\gamma_i z^{-1}$ are the generating
  functions of recursive filters of the form
  (\ref{eq-first-order-rec-filters}), the interpretation of the factor
  $ e^{(q_{-1}z^{-1} + q_1z)}$ is explained in Footnote~\ref{footnote-inf-div-distr},
  whereas the factor $z^k$
  corresponds to a translation in the temporal domain. The product
  form of the overall expression in the domain of the generating
  functions does in turn correspond to convolutions of the
  corresponding primitives over the original temporal domain.}
this characterization
means that, besides trivial rescalings and translations,
there are three basic classes of discrete
smoothing transformations: 
  \begin{itemize}
  \item
    {two-point weighted average} or {\em generalized binomial smoothing}
    \begin{equation}
      \label{eq-gen-bin-filt}
      \begin{split}
        f_{\mbox{\scriptsize\em out}}(x)
        & =
          f_{\mbox{\scriptsize\em in}}(x) +
          \alpha_i \, f_{\mbox{\scriptsize\em in}}(x - 1)
          \quad (\alpha_i \geq 0),\\
        f_{\mbox{\scriptsize out}}(x)
        & =
          f_{\mbox{\scriptsize\em in}}(x) +
          \delta_i \, f_{\mbox{\scriptsize\em in}}(x + 1)
          \quad (\delta_i \geq 0),
      \end{split}
    \end{equation}
  \item
    moving average or {\em first-order recursive filtering\/}
    \begin{equation}
      \label{eq-first-order-rec-filters}
      \begin{split}
        f_{\mbox{\scriptsize\em out}}(x)
        & =
        f_{\mbox{\scriptsize\em in}}(x) +
          \beta_i \, f_{\mbox{\scriptsize\em out}}(x - 1) \quad
          (0 \leq \beta_i < 1), \\
        f_{\mbox{\scriptsize\em out}}(x)
        & =
          f_{\mbox{\scriptsize\em in}}(x) +
            \gamma_i \, f_{\mbox{\scriptsize\em out}}(x + 1) \quad
            (0 \leq \gamma_i < 1),
      \end{split}
    \end{equation}
  \item
   {\em infinitesimal smoothing\/}%
\footnote{\label{footnote-inf-div-distr}These kernels correspond to infinitely divisible
  distributions as can be described with the theory of L{\'e}vy processes
  (Sato \citeyear{Sat99-Book}), where specifically the case $q_{-1} = q_1$ 
corresponds to convolution with the non-causal discrete analogue of the Gaussian
kernel (Lindeberg \citeyear{Lin90-PAMI}) and the case $q_{-1} = 0$
corresponds to convolution with
time-causal temporal Poisson kernel
(Lindeberg and Fagerstr{\"o}m \citeyear{LF96-ECCV};
 Lindeberg \citeyear{Lin97-AFPAC}).}
 or diffusion as arising from the
   continuous semi-groups made possible by the factor\newline
   $e^{(q_{-1}z^{-1} + q_1z)}$.
\end{itemize}
To transfer the continuous first-order integrators derived in
Section~\ref{sec-cont-temp-scsp-kern} to a discrete implementation, 
we shall in this treatment focus on the first-order recursive
filters (\ref{eq-first-order-rec-filters}),
which by additional $l_1$-normalization constitute both the discrete
correspondence and a numerical approximation of time-causal and time-recursive first-order
temporal integration (\ref{eq-first-ord-int}).

\subsection{Discrete temporal scale-space kernels based on recursive
  filters}
\label{sec-disc-temp-scsp}

Given a signal that has been sampled by some temporal
frame rate $r$, the temporal
scale $\sigma_t$ in the continuous model in units of seconds is first 
transformed to a temporal variance $\tau$ relative to a unit time sampling
\begin{equation}
  \label{eq-transf-tau-sampl}
  \tau = r^2 \, \sigma_t^2.
\end{equation}
Then, a discrete set of intermediate temporal scale levels $\tau_k$ is defined by
(\ref{eq-distr-tau-values}) or (\ref{eq-distr-tau-values-uni}),
with the difference between successive scale levels according to 
\begin{equation}
  \label{eq-delta-tau-k}
  \Delta \tau_k = \tau_k - \tau_{k-1}
\end{equation}
with $\tau_0 = 0$.

For implementing the temporal smoothing operation between two such
adjacent scale levels (with the lower level in each pair of adjacent
scales referred to as $f_{\mbox{\scriptsize in}}$ and
the upper level as $f_{\mbox{\scriptsize out}}$), we make use of a {\em first-order
  recursive filter\/}
normalized to the form
\begin{equation}
  \label{eq-norm-update}
  f_{\mbox{\scriptsize out}}(t) - f_{\mbox{\scriptsize out}}(t-1)
  = \frac{1}{1 + \mu_k} \,
    (f_{\mbox{\scriptsize in}}(t) - f_{\mbox{\scriptsize out}}(t-1))
\end{equation}
and having a generating function of the form
\begin{equation}
  \label{eq-gen-fcn-first-order-rec-filt}
  \htransf_{\mbox{\scriptsize geom}}(z) = \frac{1}{1 - \mu_k \, (z - 1)},
\end{equation}
which is a time-causal kernel and satisfies discrete
scale-space properties of guaranteeing that the number of local extrema
or zero-crossings in the signal will not increase with increasing scale
(Lindeberg \citeyear{Lin90-PAMI}; Lindeberg and Fagerstr{\"o}m
\citeyear{LF96-ECCV}).
These recursive filters are the discrete analogue of the continuous 
first-order integrators~(\ref{eq-first-ord-int}).

Each primitive recursive filter (\ref{eq-norm-update}) has temporal mean value $m_k = \mu_k$ and temporal variance
$\Delta \tau_k = \mu_k^2 + \mu_k$, and we compute $\mu_k$ from 
$\Delta \tau_k$ in (\ref{eq-delta-tau-k}) according to
\begin{equation}
  \label{eq-disc-time-constant}
  \mu_k = \frac{\sqrt{1 + 4 \Delta \tau_k}-1}{2}.
\end{equation}
By the additive property of variances under convolution,
the discrete variances of
the discrete temporal scale-space kernels will perfectly match those
of the continuous model, whereas the temporal mean values and the temporal
delays may differ somewhat. 
If the temporal scale $\tau_k$ is large relative to the
temporal sampling distance, the discrete model should be a good approximation in this respect.

By the time-recursive formulation of this temporal scale-space
concept, the computations can be performed based on a
compact temporal buffer over time, which contains the temporal
scale-space representations at temporal scales $\tau_k$, and with
no need for storing any additional temporal buffer of what
has occurred in the past, to perform the corresponding temporal
smoothing operations.

For practical implementations, we often approximate the time-causal
limit kernel using 4 to 8 layers of recursive filters coupled in
cascade using either $c = \sqrt{2}$ or $c = 2$.

A summarizing algorithmic description of how to implement these temporal filtering
operations in
practice is given in Appendix~\ref{app-disc-impl-time-caus-filt}.

\section{Computation of temporal scale-space derivatives}
\label{sec-temp-scsp-ders}

So far, we have been concerned with the problem of how to smooth a temporal signal in
such a way that the smoothing transformation is guaranteed to not
increase the number of local extrema in the signal, or equivalently
the number of zero-crossings.
In many applications, one is, however, more interested in studying the
{\em change\/} in the signal over time, as can be modelled by temporal
derivatives.

For a purely time-dependent signal, the first-order temporal derivative will
lead to strong responses in the signal when the temporal slope is high,
corresponding to {\em e.g.\/} onsets or offsets of a sound in auditory
processing, or motion in the world, alternatively changes in the illumination, for video
processing. Regarding visual processing over a purely spatial
domain, first-order spatial derivatives will respond to edges in
the image domain, which in turn may correspond to discontinuities in
either depth, surface orientation, reflectance or illumination in the world.

For a purely time-dependent signal, the second-order derivatives may on the
other hand often lead to strong responses near local maxima or minima
over time, if the sign of the first-order temporal derivative changes
rapidly at those points. Concerning audio processing, a second-order temporal
derivative applied to a spectrogram representation may give a strong
response to {\em e.g.\/} a beep or some other brief temporal sound,
provided that the temporal scale is sufficiently near the temporal
duration of the sound. Applying second-order derivatives with respect
to logarithmic frequencies to a spectrogram will, in
turn, enhance spectral bands and formants, provided that the
logspectral scales are appropriately selected. Regarding visual processing, a second-order
temporal derivative applied to a video stream may give a strong response to a flashing light,
again assuming that the temporal scale is sufficiently near the temporal
duration of the flash. Assuming that the visual observer does
not fixate a moving object, second-order temporal derivatives may
also give strong responses to image patterns that move relative to
the viewing direction. For visual processing on a purely spatial domain,
second-order spatial derivative operators can be specially designed to give
strong responses to blob-like or corner-like image structures, which can
be detected by interest point detectors.

Beyond such pointwise or regionwise responses over time, as described
above, temporal derivatives can also be interpreted and used densely, for every
time moment, and, for example, be combined according to a local Taylor expansion
around any temporal moment $t_0$:
\begin{align}
  \begin{split}
     L(t_0+ \Delta t;\; \tau) & = 
    L(t_0;\; \tau) + \Delta t \, L_t(t_0;\; \tau)
  \end{split}\nonumber\\
  \begin{split}
      & + \frac{(\Delta t)^2}{2} \, L_{tt}(t_0;\; \tau) + {\cal O}((\Delta t)^3),
    \end{split}
\end{align}
to characterize the local temporal structures in the temporal signal
at any scale $\tau$.
Such a representation involving temporal derivatives up to order $N$
is referred to as a {\em temporal $N$-jet representation\/}.%
\footnote{This temporal $N$-jet concept is a transfer of the corresponding notion of
  spatial $N$-jet representation for image data, initially proposed by
  Koenderink and van Doorn (\citeyear{KoeDoo87-BC,KoeDoo92-PAMI}). A
  conceptual difference between the temporal $N$-jet and the spatial
  $N$-jet concepts, however, is that the temporal derivatives in the temporal
  $N$-jet are associated with temporal delays, and that these temporal
  delays, in
  addition, also differ between different temporal scales.}

A practical complication that, however, arises, when computing
temporal derivatives at multiple scales concerns how to compare the
responses between different levels of scale. Due to the temporal smoothing
operation, the amplitude of the temporal derivatives can be expected
to decrease monotonically with
increasing amount of temporal smoothing, provided that the temporal smoothing
operation is sufficiently well-designed. This does, for example, hold
for temporal smoothing with the truncated exponential kernels, which
arise as the only possible temporal smoothing primitives in the
time-causal scale-space kernels, including the time-causal limit kernel.

In this section, we will describe a
way to reduce the problem of decreasing amplitude of temporal
derivatives with increasing values of the temporal scale parameter,
by instead using scale-normalized temporal derivatives.
The intention is that by using appropriately designed scale-normalized
derivative operators, it should be possible to judge if a temporal
derivative response of a certain order at a certain temporal scale should be
regarded as stronger or weaker than a corresponding temporal
derivative response at some other temporal scale.
We will also describe how temporal scale covariance can be obtained
for temporal derivative operators that are combined with the time-causal limit kernel.

\begin{figure}[hbtp]
  \begin{center}
    \begin{tabular}{c}
       {\footnotesize\em $L_{\zeta\zeta}$ at temporal scale $\sqrt{\tau} = 64$}\\
      \includegraphics[width=0.34\textwidth]{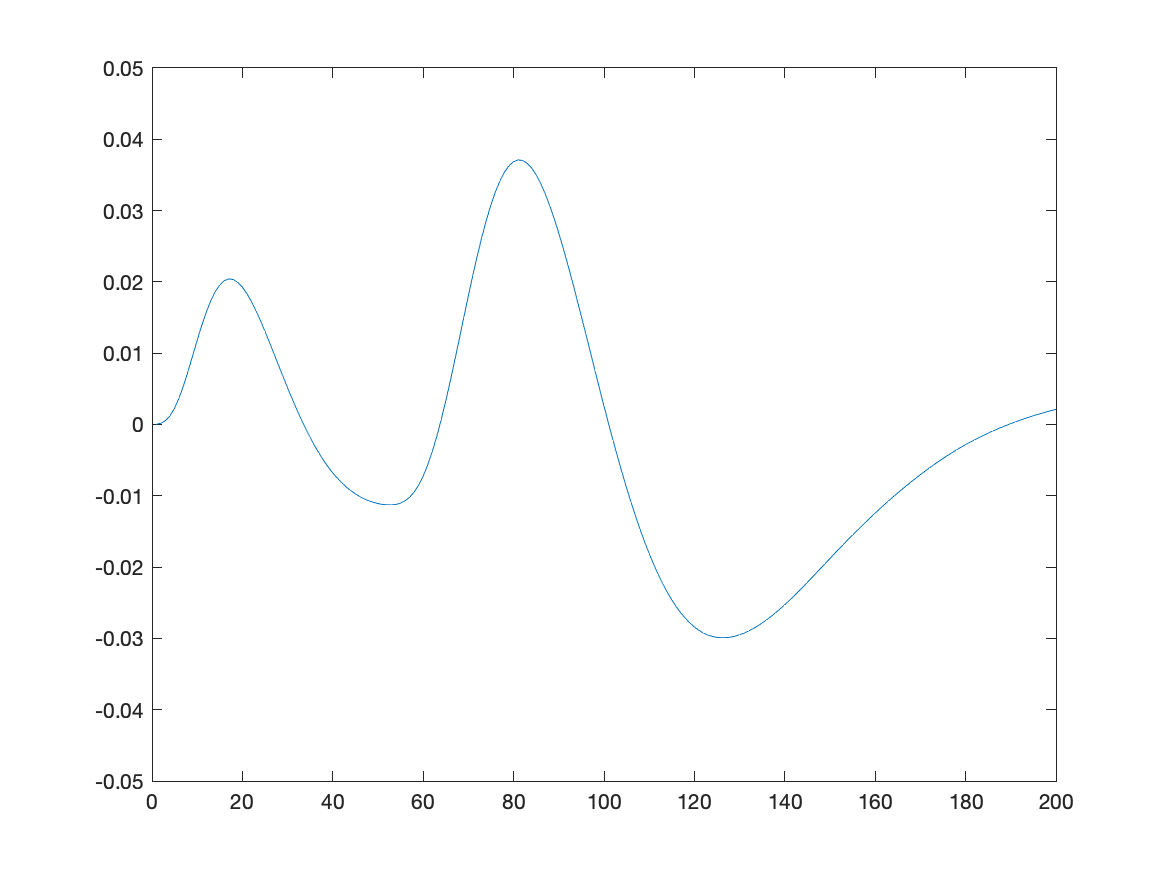}\\
       {\footnotesize\em $L_{\zeta\zeta}$ at temporal scale $\sqrt{\tau} = 16$}\\
      \includegraphics[width=0.34\textwidth]{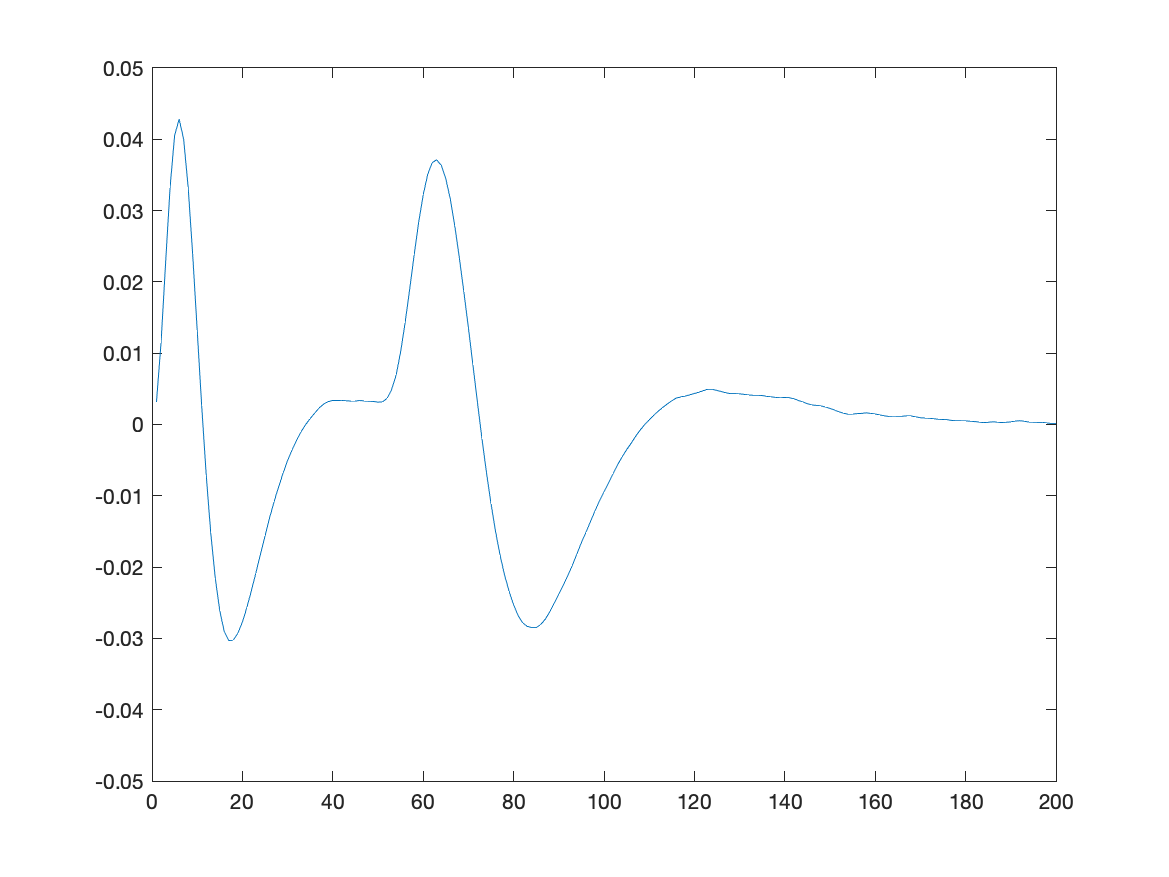}\\
       {\footnotesize\em $L_{\zeta\zeta}$ at temporal scale $\sqrt{\tau} = 4$}\\
      \includegraphics[width=0.34\textwidth]{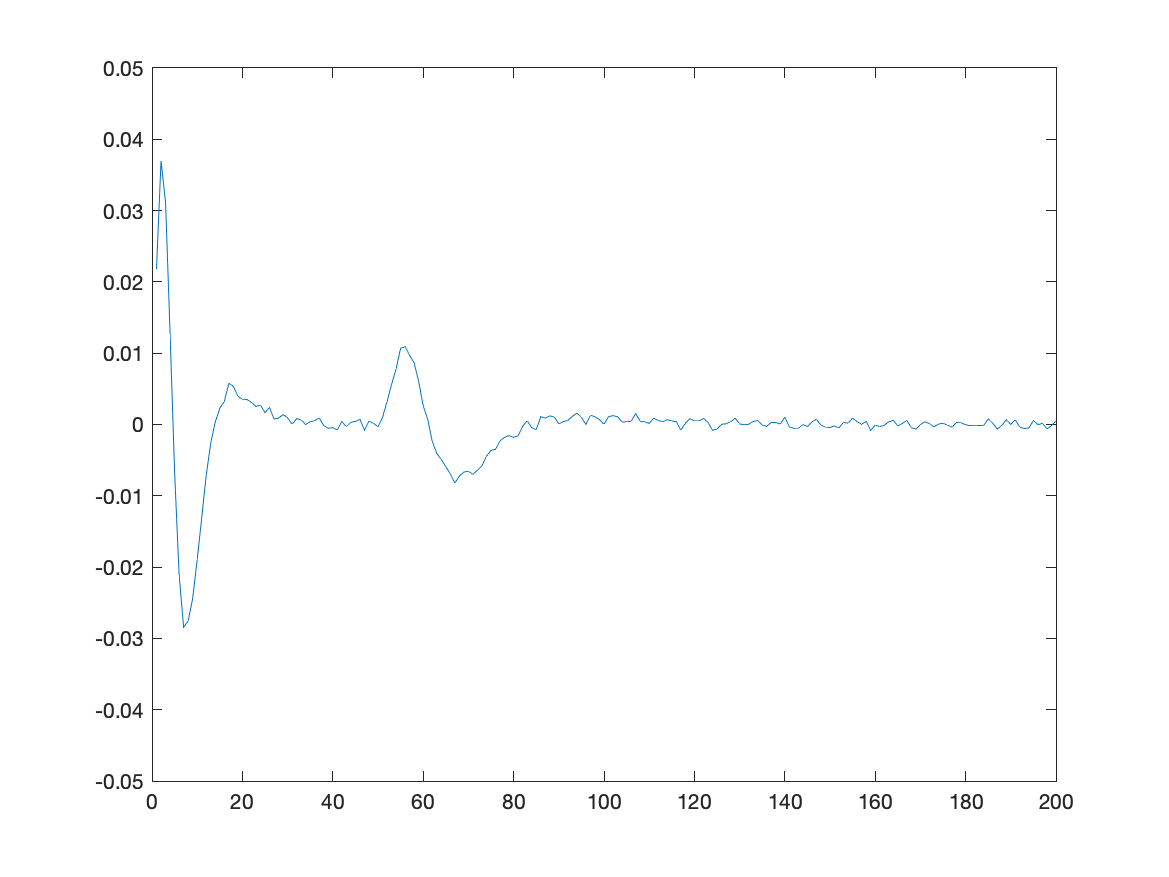}\\
      {\footnotesize\em Input signal $f$}\\
      \includegraphics[width=0.34\textwidth]{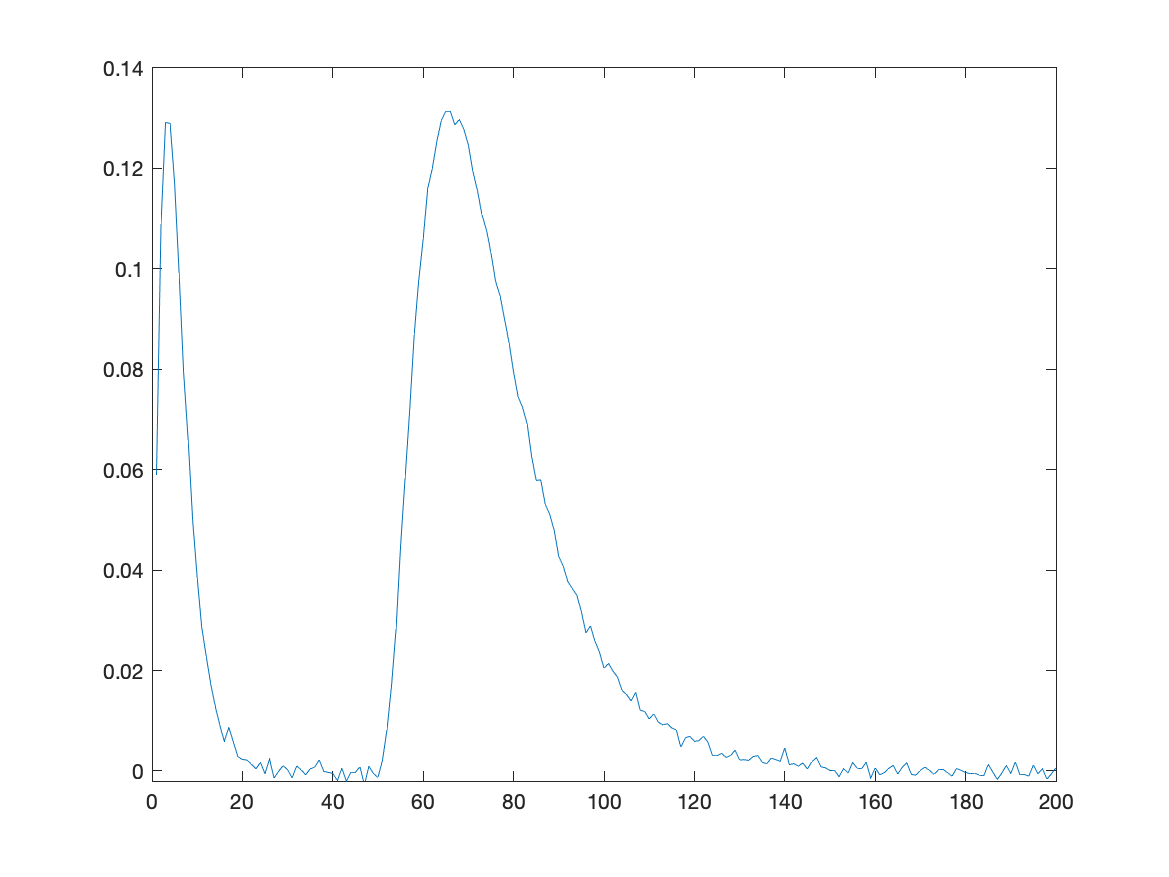}\\
    \end{tabular}
  \end{center}
  \caption{Illustration of the result of computing discrete
    approximations of second-order
    scale-normalized temporal derivatives
    $L_{\zeta\zeta}$ from the time-causal temporal scale-space representation
    $L$ at different scales (using distribution parameter $c = 2$ and
    scale normalization power $\gamma = 1$),
    here for a synthetic input signal $f$
    consisting of two temporal peaks generated as discrete
    approximations to time-causal limit kernels for temporal scales
    $\tau = 16$ and $\tau = 256$ with a certain amount of relative
    temporal delay to separate the responses as well as a
    small amount of added white Gaussian noise.
    (Horizontal axes: time. Vertical axes: Signal values.)}
  \label{fig-tempders2-ill}
\end{figure}
  
\subsection{The scale-normalized derivative concept}
\label{sec-sc-norm-ders}

For the non-causal Gaussian scale-space concept defined over a purely spatial
domain, and corresponding to Gaussian smoothing at all scales,
it can be shown that the canonical way of defining scale-normalized
derivatives at different spatial scales $s$ is according to 
(Lindeberg \citeyear{Lin97-IJCV,Lin98-IJCV,Lin21-EncCompVis})
\begin{equation}
  \partial_{\xi} = s^{\gamma/2} \, \partial_{x},
 \end{equation}
where $\gamma$ is a free parameter. Specifically, it can be shown 
(Lindeberg \citeyear[Section~9.1]{Lin97-IJCV}) that
this notion of $\gamma$-normalized derivatives corresponds to 
normalizing the $m$:th order Gaussian derivatives 
$g_{\xi^m}$ over $N$-dimensional image space to constant $L_p$-norms over scale 
\begin{equation}
  \| g_{\xi^m}(\cdot;\; s) \|_p 
  = \left( 
        \int_{t \in \bbbr} |g_{\xi^m}(x;\; s)|^p \, dt
      \right)^{1/p} 
   = G_{m,\gamma}
\end{equation}
with the power $p$ in the $L_p$-norm depending on the scale
normalization power $\gamma$, the order of differentiation $m$ and the
spatial dimensionality $N$ of the signal according to
\begin{equation}
 \label{eq-sc-norm-p-from-gamma}
  p = \frac{1}{1 + \frac{m}{N} \, (1 - \gamma)},
\end{equation}
where the perfectly scale-invariant case $\gamma = 1$ corresponds to
$L_1$-normalization for all orders $m$.

\subsection{Scale normalization for time-causal temporal derivatives}

For temporal derivatives%
\footnote{For notational convenience,
  and as is common in the field of scale-space theory, we write
  derivatives with subscripts, such that $L_t$ denotes the first-order
  derivative of the scale-space representation $L$
  with respect to time $t$, otherwise often written as
  $\frac{\partial L}{\partial t}$, and $L_{tt}$ denotes the
  second-order derivative,
  which can also be written as
  $\frac{\partial^2 L}{\partial t^2}$. In a corresponding manner,
  $L_{t^n}$ denotes the $n$:th order derivative, elsewhere often
  written as $\frac{\partial^n L}{\partial t^n}$. The operator
  $\partial_{t^n}$ denotes the $n$:th order temporal derivative operator, such
  that $L_{t^n} = \partial_{t^n} L$. The operator
  $\partial_{\zeta^n}$ does in turn denote the $n$:th order
  scale-normalized temporal derivative operator, such
  that $L_{\zeta^n} = \partial_{\zeta^n} L$.}
defined from the time-causal scale-space
concept corresponding to convolution with truncated exponential
kernels coupled in cascade, it can be shown to be meaningful to define time-causal
scale-space derivatives in a corresponding manner
(Lindeberg \citeyear{Lin16-JMIV,Lin17-JMIV}):
\begin{itemize}
\item
By {\em variance-based scale normalization\/}, we define scale-normalized
temporal derivatives according to
\begin{equation}
  \label{eq-sc-norm-der-var-norm}
  \partial_{\zeta^n} = \tau^{n \gamma/2} \, \partial_{t^n},
\end{equation}
where $\tau$ denotes the variance of the temporal smoothing kernel.
\item
By {\em $L_p$-norm-based scale normalization\/}, we determine a
temporal scale normalization factor $\alpha_{n,\gamma}(\tau)$
\begin{equation}
  \label{eq-sc-norm-der-Lp-norm-1}
  \partial_{\zeta^n} = \alpha_{n,\gamma}(\tau) \, \partial_{t^n}
\end{equation}
such that the $L_p$-norm (with $p$ determined as function of $\gamma$ according to
(\ref{eq-sc-norm-p-from-gamma})) of the corresponding composed scale-normalized
temporal derivative computation kernel $\alpha_{n,\gamma}(\tau) \, h_{t^n}$
equals the $L_p$-norm of some other reference kernel, where we
may initially take the $L_p$-norm of the corresponding Gaussian
derivative kernels (Lindeberg \citeyear[Section~7.3]{Lin16-JMIV})
\begin{align}
  \begin{split}
     \| \alpha_{n,\gamma}(\tau) \, h_{t^n}(\cdot;\; \tau) \|_p 
     & = \alpha_{n,\gamma}(\tau) \, \| h_{t^n}(\cdot;\; \tau) \|_p 
  \end{split}\nonumber\\
  \begin{split}
     \label{eq-sc-norm-der-Lp-norm-2}
     & = \| g_{\xi^n}(\cdot;\; \tau) \|_p = G_{n,\gamma}.
  \end{split}
\end{align}
\end{itemize}

\subsection{Scale covariance property of scale-normalized temporal derivatives}
\label{sec-sc-cov-temp-ders}

In the special case when the temporal scale-space representation is
defined by convolution with the scale-covariant time-causal limit
kernel according to (\ref{eq-temp-scsp-conv-limit-kernel}) and
(\ref{eq-FT-comp-kern-log-distr-limit}), it is shown in
(Lindeberg \citeyear[Appendix~3]{Lin16-JMIV}) that the
corresponding scale-normalized derivatives become truly scale covariant under temporal
scaling transformations $t' = c^j t$ with scaling factors $S = c^j$ that are
integer powers of the distribution parameter $c$
\begin{align}
  \begin{split}
     L'_{\zeta'^n}(t';\, \tau', c) 
     & = c^{j m (\gamma-1)} \, L_{\zeta^n}(t;\, \tau, c) 
  \end{split}\nonumber\\
  \begin{split}
    \label{eq-transf-prop-time-caus-ders}
     & = c^{j (1 - 1/p)} \, L_{\zeta^n}(t;\, \tau, c)
  \end{split}
\end{align}
between matching temporal scale levels $\tau' = c^{2j} \tau$.
Specifically, for $\gamma = 1$ corresponding to $p = 1$ the
magnitude values of the scale-normalized temporal derivatives at
matching scales become fully scale invariant
\begin{equation}
  L'_{\zeta'^n}(t';\, \tau', c) = L_{\zeta^n}(t;\, \tau, c),
\end{equation}
allowing for well-defined comparisons between the magnitude values of
different types of temporal structures in a signal at different
temporal scales.

\subsection{A canonical class of time-causal, time-recursive and scale-covariant
  temporal basis functions}

The above scale covariance property implies that the scale-normalized temporal derivatives
of the time-causal limit kernel constitute a canonical class of
temporal basis functions over a time-causal temporal domain.

These kernels have been used as temporal basis functions for
spatio-temporal receptive fields
(Lindeberg \citeyear{Lin16-JMIV,Lin21-Heliyon};
Jansson and Lindeberg \citeyear{JanLin18-JMIV})
and for expressing methods for
temporal scale selection (Lindeberg \citeyear{Lin17-JMIV,Lin18-SIIMS})
and spatio-temporal scale selection (Lindeberg \citeyear{Lin18-JMIV,Lin18-SIIMS})
that detect and compare temporal structures at different temporal
scales in a completely scale-invariant manner.

In this treatment, we additionally propose to use this family of
temporal basis functions to model the temporal variability of neurons
over multiple scales (Section~\ref{sec-temp-sc-neur-sign})
and specifically the temporal variability in computational models of auditory receptive fields
(Section~\ref{sec-spectr-temp-RFs}).

\subsection{Discrete approximations of scale-normalized temporal scale-space
  derivatives}
\label{sec-disc-temp-scsp-ders}

For the discrete temporal scale-space concept over discrete time
described in Section~\ref{sec-disc-temp-scsp},
discrete approximations of temporal derivatives are obtained by
applying temporal difference operators
\begin{equation}
  \label{eq-temp-der-approx-molecules}
     \delta_t = (-1, +1), \quad\quad
     \delta_{tt} = (1, -2, 1)
\end{equation}
to the discrete temporal scale-space representation at any temporal scale, which in turn is constructed from
a cascade of first-order recursive filters of the form
(\ref{eq-norm-update}), with the time constants $\mu_k$ given by
(\ref{eq-disc-time-constant}) from the differences in temporal scale
levels $\Delta \tau_k = \tau_k - \tau_{k-1}$ with $\tau_k$ according
to (\ref{eq-distr-tau-values}).

Scale normalization factors for discrete $l_p$-normal\-ization are
then defined in an analogous way as for continuous signals,
(\ref{eq-sc-norm-der-var-norm}) or (\ref{eq-sc-norm-der-Lp-norm-1}),
with the only difference that the 
continuous $L_p$-norm is replaced by a discrete $l_p$-norm.

\subsubsection{Experimental results}

Figure~\ref{fig-tempders2-ill} shows an illustration of computing
discrete approximations of second-order scale-normalized temporal derivatives in this
way,%
\footnote{Here, using distribution parameter $c = 2$ for the time-causal limit
kernel and
scale normalization power $\gamma = 1$ for the scale-normalized
temporal derivative operator.}
for a synthetic input signal consisting of two temporal peaks
generated from discrete approximations of the time-causal limit kernel
for $\tau = 16$ and $\tau = 256$, respectively, and with some amount
of relative temporal delay to separate the responses as well as a
small amount of added white Gaussian noise.

Observe how the dominant responses to the finer-scale structures in
the input signal are obtained at finer levels of scale in the temporal
scale-space representation, whereas the dominant responses to the
coarser-scale structures in the input signal are obtained at coarser
levels of scale in the temporal scale-space.

Do also observe how the responses at coarser temporal scales are
associated with longer temporal delays, manifesting themselves as temporal
peaks corresponding to the underlying signal structures appearing at
later time moments at coarser levels of scale.

Do furthermore note
that the range of values on the vertical axis in these graphs is the
same for all the scale values, demonstrating the ability to make
relative comparisons between the magnitudes of the derivative
responses at different scales, due to the notion of scale
normalization of the temporal derivatives,
here with regard to the $l_1$-norm.

\section{Relations to wavelet analysis and time-frequency
  analysis}
\label{sec-rel-wavelets-time-frequency-repr}

For analyzing temporal signals at multiple temporal scales,
wavelet analysis
(Grossmann and Morlet \citeyear{GroMor84-SIAM};
Mallat \citeyear{Mal89-PAMI}, \citeyear{Mal99-book};
Heil and Walnut \citeyear{HeiWal89-SIAM};
Meyer \citeyear{Mey92-book};
Daubechies \citeyear{Dau92-book};
Chui \citeyear{Chu92-book};
Rioul and Duhamel \citeyear{RioDuh92-TIT};
Graps \citeyear{Gra95-CompSciEng};
Debnath and Shah \citeyear{DebSha02-book})
and time-frequency analysis
(Gabor \citeyear{Gab46}; Cohen \citeyear{Coh95-book};
Feichtinger and Strohmer \citeyear{FeiStr98-book};
Qian and Chen \citeyear{QiaChe99-SignProcMag};
Gr{\"o}chenig \citeyear{Gro01-book}; Flandrin \citeyear{Fla18-book})
constitute two other main classes
of conceptual tools. In this treatment, we do, however, not follow
those notions as prototype models, instead adhering to the scale-space
paradigm because of its special properties.
Nevertheless, the presented temporal scale-space theory can
be related to wavelet analysis and time-frequency analysis in the
following ways:

\subsection{Relations to wavelet analysis}
\label{sec-rel-wavelet-anal}

By construction, the temporal derivatives of the time-causal limit
kernel $\Psi(t;\; \tau, c)$ defined from
(\ref{eq-FT-comp-kern-log-distr-limit}) have integral equal to zero
\begin{equation}
   \int_{t = -\infty}^{\infty} (\partial_{t^n} \Psi)(t;\; \tau, c) \, dt = 0.
\end{equation}
In this respect, the temporal derivatives of the time-causal kernel,
complemented by normalization with respect
to a suitably chosen norm, can serve%
\footnote{Additionally, one usually states a requirement that the
  wavelet function should decrease sufficiently fast at the tails, such
  that $\int_{t=-\infty}^{\infty} (1 + |t|^{\alpha}) |\partial_{t^n}
  \Psi| \, dt < \infty$ for some $\alpha > 0$. As will be shown
  later, for the temporal derivatives of the time-causal limit kernel,
  fulfilment of this condition follows from an exponential decrease towards
zero at the infinity, see Equation~(\ref{eq-ser-decomp-der-time-caus-limit-kern}).}
as a mother wavelet
over a continuous time-causal temporal domain,
\begin{equation}
  \label{eq-time-caus-mother-wavelet}
   W(t;\; \tau, c) = \frac{(\partial_{t^n} \Psi)(t;\; \tau, c)}{\| (\partial_{t^n} \Psi)(t;\; \tau, c) \|},
\end{equation}
in a similar way as Gaussian derivative kernels of a certain order
\begin{equation}
   W(t;\; \sigma) = \frac{(\partial_{t^n} g)(t;\; \sigma)}{\| (\partial_{t^n} g)(t;\; \sigma) \|}
\quad\mbox{with}\quad
   g(t;\; \sigma) = \frac{1}{\sqrt{2\pi} \sigma} e^{-t^2/2\sigma^2},
 \end{equation}
such as the Mexican hat wavelet (Marr \citeyear{Mar76,Mar82}),
also known as a Ricker wavelet
(Ricker \citeyear{Ric44-GeoPhys}; Hosken \citeyear{Hos88-FirstBreak}), and
corresponding to the second-order derivative of
the Gaussian,
can serve as a mother wavelet over a continuous non-causal
temporal domain.

In wavelet analysis, one usually normalizes both the mother wavelet
and the child wavelets to unit $L_2$-norm, leading to translated and rescaled
child wavelets of the form
\begin{equation}
  \label{eq-child-wavelet-L2}
   \psi_{a,b}(t) = \frac{1}{\sqrt{a}} \, W\left(\frac{t-b}{a}\right).
\end{equation}
In scale-space theory, the most common way of normalizing the Gaussian derivative
kernels as well as temporal derivatives of the time-causal limit
kernel is to constant $L_1$-norm over scales (and corresponding to
scale-normalized derivatives for $\gamma = 1$ according to
Section~\ref{sec-sc-norm-ders}), although other scale
normalizations, including $L_2$-normalization,
are also possible, as further described in
Section~\ref{sec-sc-norm-ders}. Such $L_1$-normalization then leads to translated and
rescaled child wavelets of the form
\begin{equation}
  \label{eq-child-wavelet-L1}
   \psi_{a,b}(t) = \frac{1}{a} \, W\left(\frac{t-b}{a}\right).
 \end{equation}
In the following, we will describe how the corresponding wavelet
representations obtained my mapping a signal $f$ onto the child
wavelets can be computed if the mother wavelet is chosen as a
temporal derivative of the time-causal limit kernel.

\subsubsection{Handling the transformation properties of the child
  wavelets within the algebra of the time-causal temporal
  scale-space representation}

By using the transformation properties of scale-normalized derivatives
of the time-causal scale-space representation of the time-causal limit
kernel (\ref{eq-transf-prop-time-caus-ders}), it follows that under a
scaling transformation of time $t' = c^j t$ for some integer $j$ with
$c$ being the distribution parameter of the time-causal limit kernel,
and with a corresponding
transformation of the temporal scale parameter $\tau' = c^{2j} \tau$, similar
transformation properties hold for the scale-normalized temporal
derivatives of the time-causal limit kernel (let the input signal be
the continuous delta function $f(t) = \delta(t)$ in
(\ref{eq-transf-prop-time-caus-ders}))
\begin{align}
  \begin{split}
     \Psi'_{\zeta'^n}(t';\, \tau', c) 
     & = c^{j m (\gamma-1)} \, \Psi_{\zeta^n}(t;\, \tau, c) 
  \end{split}\nonumber\\
  \begin{split}
    \label{eq-transf-prop-time-caus-ders-limit-kern}
     & = c^{j (1 - 1/p)} \, \Psi_{\zeta^n}(t;\, \tau, c),
  \end{split}
\end{align}
where $\gamma$ is the power in the temporal scale-normalized
derivative concept and $p$ is the power in the corresponding
$L_p$-norm that is kept constant over scale by the scale-normalized derivatives.

This implies that if we choose the mother wavelet as a temporal
derivative of the time-causal limit kernel according to
(\ref{eq-time-caus-mother-wavelet}), then the temporal scaling and
translation operations of the child wavelets in
(\ref{eq-child-wavelet-L2}) and (\ref{eq-child-wavelet-L1}) can be
expressed fully within the algebra of the time-causal scale-space
representation, provided that the temporal scaling factors $a$ are
chosen as integer powers of the distribution parameter $c$ in the
time-causal limit kernel according to $a = c^j$. This does in turn imply that
the result of expanding a temporal test signal onto the child wavelets
can be {\em directly extracted\/} as the corresponding temporal
derivatives of the time-causal temporal scale-space
representation of the temporal test signal at the different temporal
scales,
possibly complemented by a scale-dependent scaling of the magnitude
values, depending on the choice of $L_p$-norm in the wavelet
representation and the choice of scale normalization power $\gamma$ in
the scale-normalized derivative concept.


\subsubsection{Finite $L_p$-norms for the temporal derivatives of the
  time-causal limit kernel}

A regularity requirement that one usually imposes on wavelet functions
is that they should be in both $L_1(\bbbr)$ and $L_2(\bbbr)$.
This property can be easily shown for the temporal derivatives of the
time-causal limit kernel, as follows:

Consider a partial fraction decomposition of the Laplace transform
(\ref{eq-expr-comp-kern-trunc-exp-filters})
of the infinite convolution
of truncated exponential kernels that defines the time-causal limit
kernel according to (\ref{eq-FT-comp-kern-log-distr-limit}):
\begin{equation}
  H_{\mbox{\scriptsize $\Psi$}}(q;\; \tau, c)
  = \prod_{k=1}^{\infty} \frac{1}{1 + \mu_k q}
  = \sum_{k=1}^{\infty} \frac{A_k}{1 + \mu_k q},
\end{equation}
with $\mu_k$ as functions of $\tau$ and $c$ according to (\ref{eq-mu1-log-distr}) and
(\ref{eq-muk-log-distr}), and where the coefficients $A_k$ can be
determined by first multiplying both sides of the equation by $(1 + \mu_k q)$
and then setting $q = -1/\mu_k$, leading to
\begin{equation}
  A_k = \prod_{i=1, i \neq k}^{\infty} \frac{1}{1 - \frac{\mu_i}{\mu_k}}.
\end{equation}
Interpreted over the original temporal domain, this means that the
time-causal limit kernel can be written in terms of the following
decomposition as a sum of truncated exponential functions:
\begin{equation}
  \Psi(t;\; \tau, c)
  = \sum_{k = 1}^{\infty} A_k \, h_{\mbox{\scriptsize  exp}}(t;\; \mu_k)
  = \sum_{k = 1}^{\infty} \frac{A_k}{\mu_k} \, e^{-t/\mu_k} \quad (t \geq 0).
\end{equation}
Thus, the $n$:th order temporal derivative of the time-causal limit
kernel will have the following series representation:
\begin{equation}
  \label{eq-ser-decomp-der-time-caus-limit-kern}
  (\partial_{t^n} \Psi)(t;\; \tau, c)
  = \sum_{k = 1}^{\infty} \left( \frac{-1}{\mu_k} \right)^n
  \frac{A_k}{\mu_k} \,  e^{-t/\mu_k} \quad (t \geq 0).
\end{equation}
When time $t$ tends to infinity, this function will in the limit tend
towards zero, and as fast as exponentially with respect to he slowest time constant $\mu_1$.
Since $(\partial_{t^n} \Psi)(t;\; \tau, c)$ is additionally finite for finite values of
$t$, it follows that both the $L_1$- and the $L_2$-norms of $\partial_{t^n} \Psi$
will be finite, implying that $\partial_{t^n} \Psi \in L_1(R) \cap
L_2(R)$, thus proving the result.

\subsubsection{Time-causal and time-recursive wavelets for real-time and
  time-critical applications}

These resulting wavelets described in this section,
consisting of temporal derivatives of
the time-causal limit kernel, will be completely time-causal. The
convolutions%
\footnote{In wavelet analysis, the expansion of a test function onto a
  a set of wavelet functions is usually computed in terms of inner products,
  corresponding to correlations. The reason why we instead use convolutions
  here is to avoid the additional step of reversing the time direction
  for the temporal derivatives of the time-causal limit kernel in
  relation to how they are used in the other parts of this article, in
  terms of convolutions.}
between these wavelet kernels and a temporal measurement function can
also be computed in a completely time-recursive way, thus eliminating
the need for additional temporal buffering and in turn allowing for
minimal temporal response times in a time-critical context.
In these respects, the temporal derivatives of the time-causal limit
kernel may thus have interesting
potential use for wavelet analysis with regard to applications that are
to be performed over time-causal and
time-recursive temporal domains, such as for real-time signal analysis
systems, or when modelling physical or biological systems for which
access to the relative future in relation to any time moment is not possible.

Another type of time-causal wavelet representation has been proposed and
studied by Szu {\em et al.\/} (\citeyear{SzuTelLoh92-OptEng}),
based on linear combinations
of sine and cosine waves multiplied by a truncated exponential function.
In this context, the wavelets based on temporal derivatives of the time-causal limit
kernel have the conceptual advantage that they are solely based on truncated exponential
kernels coupled in cascade, and can therefore be implemented in a
fully time-recursive manner.%
\footnote{In their work, Szu {\em et al.\/}
  (\citeyear{SzuTelLoh92-OptEng}) propose optical computations to
  achieve real-time performance for their time-causal wavelets,
  whereas discrete approximations of
  the time-causal limit kernel can be expressed in terms of
  recursive filters, which in turn can be implemented in real time on standard
  digital signal processing hardware.}
Additionally, with regard to
the discrete implementation of such temporal receptive fields in terms
of recursive filters coupled in cascade (according to Section~\ref{sec-disc-temp-scsp}),
the computation of wavelets based on temporal derivatives of the time-causal limit
kernel, an additional temporal scale level can be
computed with just the addition of a single recursive filter,
complemented with a discrete temporal difference operator
(according to Section~\ref{sec-disc-temp-scsp-ders}).

\begin{figure*}[hbtp]
  \begin{center}
    \begin{tabular}{cc}
       {\footnotesize $\operatorname{Re} \chi(t, \omega;\; \tau, c) =
      \Psi(t;\; \tau, c) \, \cos \omega t$}
       & {\footnotesize $\operatorname{Im} \chi(t, \omega;\; \tau, c) =
      \Psi(t;\; \tau, c) \, \sin \omega t \, $} \\
      \includegraphics[width=0.40\textwidth]{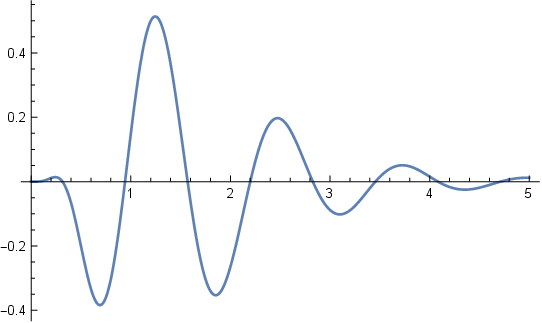} &
      \includegraphics[width=0.40\textwidth]{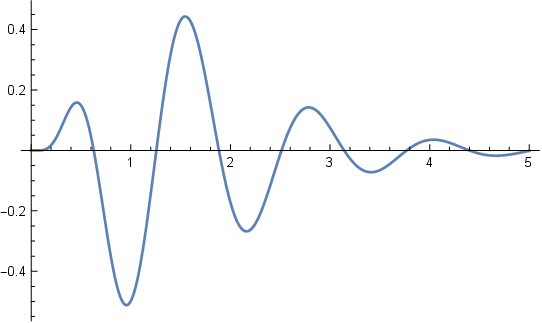}
      \\
    \includegraphics[width=0.40\textwidth]{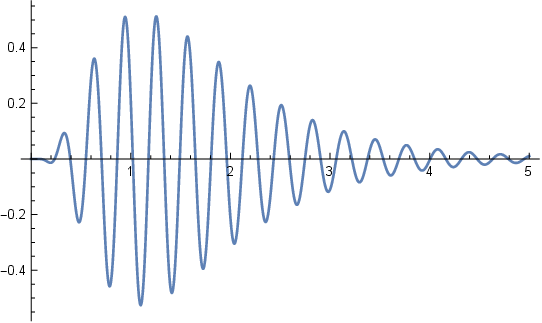} &
      \includegraphics[width=0.40\textwidth]{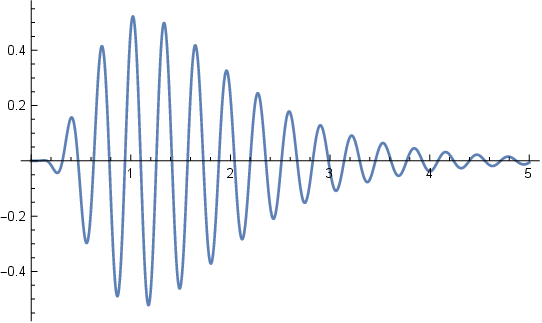}
      \\
   \end{tabular}
  \end{center}
  \caption{Graphs of the complex-valued extension $\chi(t, \omega;\;
    \tau, c) = \Psi(t;\; \tau, c) \, e^{i \omega t}$ of the
    time-causal limit kernel $\Psi(t;\; \tau, c)$ for temporal scale
    $\tau = 1$ with distribution parameter $c = 2$ and
    different values of the angular frequency $\omega$.
  (left column) The real component, corresponding to the time-causal
  limit kernel multiplied by a cosine wave. (right column) The
  imaginary component, corresponding to the time-causal
  limit kernel multiplied by a sine wave.
  (top row) Angular frequency $\omega = 5$.
  (bottom row) Angular frequency $\omega = 20$.
  (Horizontal axes: time. Vertical axes: kernel values.)}
  \label{fig-time-freq-kernels}
\end{figure*}

\subsection{Relations to time-frequency analysis}
\label{sec-rel-time-freq-anal}

If we combine the time-causal limit kernel $\Psi(t;\; \tau, c)$
defined according to (\ref{eq-FT-comp-kern-log-distr-limit})
with pointwise multiplication by a complex exponential function
$e^{i\omega t}$, then we obtain a straightforward way of defining
a time-causal time-frequency representation
of a temporal signal $f(t)$ according to
\begin{equation}
  \label{eq-time-caus-limit-kern-time-freq-repr}
  S(\omega;\; \tau, c)
  = \int_{u=0}^{\infty} f(t-u) \, \Psi(u;\; \tau, c) \, e^{i \omega u} \, du,
\end{equation}
where the complex-valued extension of the time-causal limit kernel
\begin{equation}
  \label{eq-compl-time-caus-limit-kern}
  \chi(t, \omega;\; \tau, c) = \Psi(t;\; \tau, c) \, e^{i \omega t} 
\end{equation}
can be seen as a time-causal analogue%
\footnote{In this context, do specifically note that according to the
  classification of continuous scale-space kernels according to
  Section~\ref{sec-class-cont-scsp-kernels}, the only possible continuous
  scale-space kernels are Gaussian kernels and truncated exponential kernels
  coupled in cascade. The Gaussian kernel is the canonical choice over
  a non-causal temporal domain, whereas composed convolutions of
  truncated exponential kernels are the only possible temporal
  scale-space kernels over a time-causal temporal domain. Out of the
  latter family of options, the time-causal limit kernel is a special
  choice that additionally allows for temporal scale covariance over a time-causal
  temporal domain, in a corresponding way as the regular Gaussian
  kernel allows for temporal scale covariance over a non-causal
  temporal domain.}
  of the Gabor function
(Gabor \citeyear{Gab46}), with the role of the Gaussian kernel
$g(t;\; \sigma)$ in the
Gabor function
\begin{equation}
  G(t, \omega;\; \sigma)
  = g(t;\; \sigma) \, e^{i \omega t}
  = \frac{1}{\sqrt{2 \pi} \, \sigma} e^{-t^2/2\sigma^2} \, e^{i \omega t} 
\end{equation}
now replaced the by the time-causal limit kernel $\Psi(t;\; \tau, c)$
for $\tau = \sigma^2$.
Figure~\ref{fig-time-freq-kernels} shows graphs of a few examples of
such complex-valued extensions of the time-causal limit kernel
for different values of the angular frequency $\omega$ in
relation to a given temporal scale $\tau$.

In this context, the time-causal limit kernel serves as a temporal
window function for computing a windowed Fourier transform, to give
better localization properties in the temporal domain compared to a
regular Fourier transform, and where the window function in this case,
in contrast to the more common choice of a Gaussian window function, is
fully time-causal, to allow for real-time processing as well as realistic
modelling of real-world physical and biological processes, where access
to the relative future in relation to any time moment is simply not possible.

\subsubsection{Relations to the Gammatone filter}

The complex-valued extension of the time-causal limit kernel in
(\ref{eq-compl-time-caus-limit-kern}) is specifically closely related
to the Gammatone filter (Johannesma \citeyear{Joh72-HearTheory}; 
Patterson {\em et al.\/} \citeyear{PatNimHolRic87-GammaTone}, \citeyear{PatAllGig95-JASA}; 
Hewitt and Meddis \citeyear{HewMed94-JASA}) in auditory processing
\begin{equation}
   \gamma(t) = a \, t^{n-1} e^{-2\pi b t} \cos(2\pi \phi \, t + \alpha),
 \end{equation}
with the main difference being that the truncated exponential kernels used
in this auditory filter have equal time constants, and can thus under
a convolution operation be composed into a single monomial multiplied
by the complex exponential, in analogy with
Equation~(\ref{eq-temp-scsp-kernel-uni-distr}),
and thereby corresponding
to a uniform distribution of the temporal scale levels according to
Section~\ref{sec-uni-distr-temp-sc-levels}, whereas the temporal scale levels in the
complex-valued extension of the time-causal limit kernel are
constructed according to a geometric distribution of the temporal
scale levels according to Section~\ref{sec-log-distr-temp-sc-levels},
thus, in turn, allowing for different and more rapid temporal dynamics.

Another minor difference is that the phase of the Gammatone filter is
represented as a phase angle $\alpha$ of a cosine function, whereas
the phase of the complex-valued extension of the time-causal limit
kernel is represented as the phase value of a complex exponential.

\subsubsection{Relations to the Heisenberg group}

The time-frequency representation defined according to
(\ref{eq-time-caus-limit-kern-time-freq-repr}) has the theoretically
attractive property that it is closed under (i)~translations over
time, (ii)~multiplicative shifts in the frequency of periodic or
repetitive temporal signals and
(iii)~uniform scaling transformations of the temporal axis
with discrete scaling factors $S$ that are integer powers of the
distribution parameter $c$. Hence, except for the necessary
discretization of the temporal scale parameter according to a
geometric distribution, which implies
closedness over a discrete set of scaling factors as opposed to as
over a continuum, this time-frequency representation has
the ability to capture similar types of transformations of the signal
as the Gabor family, and as can be modelled by the Heisenberg group, see
(Feichtinger and Gr{\"o}chenig \citeyear{FeiGro92-Wavelets}).
In this way, the complex-valued time-causal limit kernel provides a way to define a
scale-covariant time-frequency representation also over a time-causal
temporal domain.

\subsubsection{Extension to an additionally time-recursive time-frequency transform}

If one additionally wants these time-frequency representations to also
be time recursive,
then it is possible to modify this construction slightly, by instead
multiplying the input signal by a set of complex exponentials
and then filtering the resulting complex-valued signal with
the time-causal limit kernel
(according to Equation~(\ref{eq-time-caus-spectrogram})), thus
implying that this time-frequency transform can be implemented
discretely in
terms of a set of recursive filters that operate over time on the
pointwise multiplication of the input signal with a set of complex
exponential functions.
The difference will then be that the phase values will have to be
compensated {\em a posteriori\/}, whereas the magnitude values of the
corresponding spectrogram will be preserved.
An earlier version%
\footnote{Without taking the number of temporal scale levels $K$ to
  infinity to enable true temporal scale covariance.}
of this type of theoretical model has been successfully used for
computing auditory receptive fields
(Lindeberg and Friberg \citeyear{LinFri15-PONE,LinFri15-SSVM}),
as will be further described in Section~\ref{sec-spectr-temp-RFs}.

\section{Applications to modelling temporal variations in biological systems}
\label{sec-temp-model-biol-perc-neur}

In this section, we will describe different application domains of
using the theory for temporal scale-space representation, specifically
the time-causal limit kernel, to model temporal variations in
biological signals.

\subsection{Temporal basis functions in spatio-temporal receptive
  field models}
\label{sec-spat-temp-RFs}

In (Lindeberg \citeyear{Lin10-JMIV,Lin13-BICY}), a general model for
spatio-temporal receptive fields is derived of the form
\begin{equation}
      \label{eq-spat-temp-RF-model}
       T(x_1, x_2, t;\; s, \tau;\; v, \Sigma)  
       = g(x_1 - v_1 t, x_2 - v_2 t;\; s, \Sigma) \, h(t;\; \tau)
\end{equation}
  where
  \begin{itemize}
  \item
     $x = (x_1, x_2)^T$ denotes the image coordinates,
  \item
     $t$ denotes time,
  \item
    $s$ denotes the spatial scale,
\item
    $\tau$ denotes the temporal scale,
  \item
   $v = (v_1, v_2)^T$ denotes a local image velocity,
  \item
    $\Sigma$ denotes a spatial covariance matrix determining the
    spatial shape of an affine Gaussian kernel
   $g(x;\; s, \Sigma)  = \frac{1}{2 \pi s \sqrt{\det\Sigma}} e^{-x^T \Sigma^{-1} x/2s}$,
  \item
     $g(x_1 - v_1 t, x_2 - v_2 t;\; s, \Sigma)$ denotes a spatial affine Gaussian kernel
     that moves with image velocity $v = (v_1, v_2)$ in space-time and
\item
   $h(t;\; \tau)$ is a temporal smoothing kernel over time.
 \end{itemize}
This model for zero-order spatio-temporal receptive fields should, in
turn, be complemented by spatial and temporal differentiation to lead
to spatio-temporal receptive fields with positive and negative lobes
that are balanced in the sense of the integral of the filter weights being equal to zero.

In (Lindeberg \citeyear{Lin16-JMIV,Lin18-JMIV,Lin21-Heliyon}), it is
described how the time-causal limit kernel can be successfully
be used as the temporal smoothing kernel in this context,
{\em i.e.\/}, $h(t;\; \tau) = \Psi(t;\; \tau)$ with $\Psi$ defined
from its Fourier transform according to
(\ref{eq-FT-comp-kern-log-distr-limit}), and allowing for truly time-causal
and time-recursive model of spatio-temporal receptive fields, which in
turn enable provable scale covariance and scale invariance properties
over the temporal domain.

By comparisons with biological visual receptive fields measured by
electrophysiological cell recordings by
DeAngelis {\em et al.\/}
(\citeyear{DeAngOhzFre95-TINS,deAngAnz04-VisNeuroSci}), it is shown in
(Lindeberg \citeyear{Lin16-JMIV,Lin21-Heliyon}) that this spatio-temporal
receptive field model very well captures the qualitative shape of
lagged and non-lagged LGN neurons as well as simple cells in the primary
visual cortex (V1).

\subsection{Temporal basis functions in spectro-temporal receptive
  field models}
\label{sec-spectr-temp-RFs}

In (Lindeberg and Friberg \citeyear{LinFri15-PONE,LinFri15-SSVM}), a
theoretical framework for idealized models of auditory receptive
fields is presented, based on a two-stage model consisting of
time-causal spectrograms followed by time-causal spectro-temporal receptive fields
applied on these, and which comprises covariance and invariance
properties under natural sound transformations, such as 
frequency shifts and glissando transformations.

The time-causal spectrograms in  this model are defined according to
\begin{equation}
  \label{eq-time-caus-spectrogram}
     S_h(t, \omega;\; \mu) 
       = \int_{t'=-\infty}^{\infty}
                h_{composed}(t - t';\; \mu) \, f(t') \,
                e^{-i\omega t'} \, dt',
\end{equation}
where the temporal integration kernel $h_{composed}$ is from theoretical
arguments constrained to be the convolution of a set of truncated
exponential kernels coupled in cascade.
Following the arguments in this paper, and further restricting this
kernel to be a time-causal limit kernel $\Psi$, we can extend the previous
theoretical framework for multi-scale spectrograms to also comprise
temporal scale covariance.

In the second-stage model of spectro-temporal receptive fields in this
theory, the
idealized form of auditory receptive fields are from theoretical
arguments constrained to be of the form
\begin{equation}
  \label{eq-spectr-temp-recfields-gen-form}
   A(t, \nu;\; \Sigma) 
  = \partial_{t^{\alpha}} \partial_{\nu^{\beta}} 
      \left( 
         g(\nu - v t;\; s) \,
         T(t;\; \tau_a) 
      \right)
\end{equation}
where
\begin{itemize}
\item
  $\partial_{t^{\alpha}}$ represents a {\em temporal derivative operator\/} of order
  $\alpha$ with
  respect to time $t$ which could alternatively be replaced by a glissando-adapted
  temporal derivative of the form $\partial_{\overline{t}} = \partial_t + v \, \partial_{\nu}$,
\item
  $\partial_{\nu^{\beta}}$ represents a {\em logspectral derivative operator\/} of order
  $\beta$ with
  respect to logarithmic frequency $\nu$,
\item
  $T(t;\; \tau_a)$ represents a {\em temporal smoothing kernel\/} with
  temporal scale parameter $\tau_a$, which should in the time-causal
  case be a set of truncated exponential kernels
  coupled 
 in cascade,
\item
  $g(\nu - v t;\; s)$ represents a Gaussian {\em spectral smoothing kernel\/} over logarithmic
  frequencies $v$ with logspectral scale parameter $s$ and $v$
  representing a glissando parameter making it possible to adapt the
  receptive fields to variations in frequency $\nu' = \nu + v t$ over time.
\end{itemize}
By comparison with biological auditory receptive fields measured by
electrophysiological cell recordings by
Qiu {\em et al.\/} \protect(\citeyear{QiuSchEsc03-JNeuroPhys}),
Andoni {\em et al.\/}  \protect(\citeyear{AndLiPol07-JNeuroSci}),
Machens {\em et al.\/}  \protect(\citeyear{MacWehZad04-JNeuroSci}),
Elhilali {\em et al.\/}  \protect(\citeyear{ElhFriChiSha07-JNeuroSci}) and
Atencio and Schreiner \protect(\citeyear{AteSch12-PONE}),
it is shown in (Lindeberg and Friberg \citeyear{LinFri15-PONE})
that the idealized receptive fields from this model agree
qualitatively very well with biological auditory receptive fields
measured in the inferior colliculus (ICC) and
primary auditory cortex (A1) of mammals.

By following the arguments regarding temporal smoothing in this paper,
and constraining the temporal
kernel in the above model to be a time-causal limit kernel,
$T(t;\; \tau_a) = \Psi(t;\; \tau_a)$, it follows that the auditory
covariance properties in the spectro-temporal receptive field model can be extended to
also comprise temporal scale covariance.

\subsection{Temporal scales in neural signals}
\label{sec-temp-sc-neur-sign}

In this section, we describe previous evidence and use of multiple
temporal scales in neural signals, with relations to the theory for
processing temporal signals at multiple scales presented in this paper.

Concerning the use of multiple temporal scales for processing neural signals,
Goldman (\citeyear{Gol09-Neur}) shows how neural responses can be
maintained by a purely feedforward mechanism, which thus implements a
temporal memory. In his model, a set of first-order integrators with
equal time constants is used. By instead using different time
constants of the first-order integrators, as used for the
implementation of the time-causal limit kernel, we can get a more
compact model for the memory buffers, requiring less wetware or
computational modules, with the additional benefit that
the time constants obey a self-similar logarithmic distribution.

Tsao {\em et al.\/} (\citeyear{TsaSugLuWanKniMosMos18-Nature}) show how
temporal information in the lateral entorhinal cortex is robustly
encoded over a wide range of temporal scales, from time scales of seconds to hours,
where specifically the brain handles multiple scales in parallel,
consistent with the underlying construction of a multi-scale
representation over the temporal domain, and specifically using a
multi-scale temporal representation as a temporal memory.
In a further study of the primate entorhinal cortex,
Bright {\em et al.\/} (\citeyear{BriMeiCruTigBufHow20-PNAS})
experimentally model time cells in this brain area as single truncated
exponentials, in line with theoretical model
in Equation~(\ref{eq-temp-channel-single-exp-fcn}), although also
complemented with a Gaussian smoothing step that leads to the
ex-Gaussian model, and conclude that the time cells in
the entorhinal cortex use a spectrum of time constants to construct a
temporal record of the past in support of episodic memory.
In a study of cerebellar unipolar brush cells,
Guo {\em et al.\/} (\citeyear{GuoHusMacReg21-Nature}) show that the
population of neurons generates a continuum of multi-scale temporal
representations, with essentially a logarithmic distribution of the
temporal scale levels, consistent with the distribution of temporal
scale levels used for the temporal scale-space representation and its
associated temporal memory model based on the time-causal limit kernel.

In their computational model, of temporal memory, Howard and Hasselmo
(\citeyear{HowHas20-arXiv}) propose that  time cells in the hippocampus can
be understood as a compressed estimate of events as a function of the
past, and that temporal context cells in the entorhinal cortex can be
understood as the (real-valued) Laplace transform of that function, respectively,
where the Laplace transform in turn arises from the integration with 
truncated exponential kernels with different time constants,
as are used as the unique primitive time-causal temporal smoothing
kernel that are guaranteed to not
increase the number of local extrema or zero-crossings in the signal.
Howard (\citeyear{How21-HandBookHumMem}) gives a more general overview
of mechanisms for temporal memory, including the use of 
multiple first-order temporal integrators as arising from this theory.

In an fMRI study of memory recall in human subjects over large
variations in the time elapsed after the event,
Monsa {\em et al.\/} (\citeyear{MonPeeArz20-JCognNeuroSci})
conclude that scale-selective activity characterizes autobiographical
memory processing and may provide a basis for understanding how the
human brain processes and integrates experiences across temporal
scales in a hierarchical manner.

Holcombe (\citeyear{Hol09-TICS}) gives a general overview of different
temporal scale limits in visual perception, in particular describing a
distinction into slow and fast temporal processes, which are
hypothesized to originate from neural processes over different ranges
of temporal scales.
In an fMRI study of the human ventral stream,
Gauthier {\em et al.\/} (\citeyear{GauEgeHesGIrKle12-JNeuroSci})
show that the widths of temporal integration windows increase
at higher hierarchical levels in the visual hierarchy.

Regarding the use of multiple temporal scales in auditory perception,
Atencio and Schreiner (\citeyear{AteSch12-PONE}) show examples of
spectro-temporal receptive fields in the primary auditory cortex
(A1) with different spectro-temporal scale characteristics; broadly
tuned receptive fields with short temporal duration and narrowly tuned
receptive fields with longer temporal duration.
Chait {\em et al.\/} (\citeyear{ChaGreAraSimPoe15-FrontNeurSci}) investigate
how different temporal scales interact in speech perception and suggest
that human speech perception uses multi-time resolution processing.
Teng {\em et al.\/} (\citeyear{TenTiaPoe16-SciRep}) provide evidence
that the auditory system extracts fine-detail acoustic information
using short temporal windows and uses long temporal windows
to abstract global acoustic patterns.
Concerning the specific area of birdsong, Gentner
(\citeyear{Gen08-JASA}) shows how the use of
multiple temporal scales within the acoustic pattern hierarchy conveys
information about the individual identity of the singer.
Osman {\em et al.\/} (\citeyear{OsmLeeEscRea18-JNeuroSci}) also propose a
hierarchy of temporal scales for discriminating and classifying the
temporal shapes of sound in different auditory cortical areas.

In a wider study regarding the visual, somatosensory and auditory cortices,
Latimer {\em et al.\/} (\citeyear{LatBarSokAwwKatNelLamFaiPri19-NeuroSci})
found that the behaviour of the adaptive responses that they observe can
be accounted for by fixed filters that operate over multiple time scales.
By developing a method for estimating temporal scales in neuronal
dynamics, Spitmaan {\em et al.\/} (\citeyear{SpiSeoLeeSol20-PNAS}) found
that most neurons exhibited multiple temporal scales in their response,
which consistently increased from parietal to prefrontal and cingulate
cortex.
Miri {\em et al.\/} (\citeyear{MirBhaAksTanGol22-JPhys}) in turn
suggest that gaze control requires integration over distributed
temporal scales.

We propose that if the aim is to build mathematical models of such
neural, perceptual or memory processes, then the mathematical theory
for time-causal scale-space kernels presented
in this paper should be ideally suited for building such models that
are both time-causal and time-recursive. Specifically, if the aim is
to build such temporal models that can handle multiple temporal scales in a way
that respects temporal scale covariance, and under a architectural
setting that corresponds to multiple primitive temporal smoothing stages
coupled in cascade, then the time-causal limit
kernel (described in Section~\ref{sec-time-caus-limit-kern}) with its
temporal derivatives (described in Section~\ref{sec-temp-scsp-ders}) 
constitutes a canonical class of temporal basis functions to be used in such models.

As a
consequence of the temporal delay of such time-causal kernels
(Equations~(\ref{eq-mean-var-trunc-exp-filters}) and
(\ref{eq-approx-temp-pos-time-caus-log-distr})), any time-causal
perceptual process will be
associated with an inherent temporal delay (complemented with the
processing time of the neural processes that implement the
corresponding computations), implying that the 
representation of the present (White \citeyear{Whi20-PsychBullRev})
will, in practice, be a representation of some (temporally extended)
temporal moment(s)%
\footnote{For a time-causal temporal filtering process that operates
  over multiple temporal scales, there will, in general, be a different temporal
  delay for each temporal scale, in the sense that the temporal delay
  will be shorter for temporal filtering over a short temporal scale
  and longer for temporal filtering over a long temporal scale.
This raises an interesting theoretical problem concerning how to
maintain an internally consistent representation of the time-delayed
present, given that different components in such a representation may
be associated with different temporal delays.}
in the past, unless
complemented with extrapolation/prediction (White \citeyear{Whi18-VisCogn})
over a time period corresponding to
the temporal delay(s) of the perceptual process that lead to that
percept.
Still, however, a representation of the present, with or without
temporal prediction implying without or with an inherent
temporal delay, will by necessity be a representation of a temporally ``fuzzy'' present.

In their review of the use of multiple temporal scales in the brain,
Cavanagh {\em et al.\/} (\citeyear{CavHunKen20-FronNeurCirc})
state that 
short temporal windows facilitate adaptive responding in dynamic
environments, whereas longer temporal windows promote the gradual
integration of information across time, and specifically concerning
the notion of multiple temporal scales they conclude a heterogeneity of temporal receptive
fields at the level of single neurons within a cortical region,
consistent with the aims behind
the theory for temporal scale-space representation described in this article.

\section{Implications of the presented theory with regard to the
  philosophy of time and perceptual agents}
\label{sec-impl-phil-time-perc-agent}

The subject of this paper has been to describe a theoretical framework for handling the
notions of time and temporal scales for a perceptual system or a neural system,
in a both principled and theoretically well-founded manner.
Since this subject has implications regarding how we consider the
notion of time for a perceptual agent, we will in this section
describe relations to the philosophy of time
(M{\"o}lder {\em et al.\/} \citeyear{MolArsOhr16-book};
Callender (\citeyear{Cal17-book}),
which is still an open topic in the area of philosophy.

The notion of time is something that we usually take for
granted. Still there is no fully established definition for this concept.
Already St.\ Augustine (354-430) stated (Outler, transl.\ \citeyear[Book~11, page~193]{AugOut95-book}):
\begin{quote}
  ``What, then, is time? If no one asks me, I know what it is. If I wish to explain it to him who asks me, I do not know. Yet I say with confidence that I know that if nothing passed away, there would be no past time; and if nothing were still coming, there would be no future time; and if there were nothing at all, there would be no present time.''
\end{quote}
According to Newtonian or Galilean space-time, we can treat time as flowing
continuously and define a universally valid notion of global time. According to
Einstein's relativity theory (\citeyear{Ein05-AnnPhy,Ein16-book}), different observers can measure
time differently, being affected by the relative velocity between the
observers. Thus, measurement of time is a
local property (attached to the path that an observer or a clock 
follows in space-time), and (at very high relative velocities) different observers may not even be able to agree on the
temporal ordering between different temporal events in the world.%
\footnote{To understand how relative temporal ordering can be
  different for two observers with different relative velocities,
  consider two observers Alice and Bob in relation to a moving train
  with constant velocity.
  Alice is in a waggon of the train, whereas Bob is standing on the
  ground outside. Let us assume that Alice is positioned in the middle
of the waggon, and uses a special flashlight to emit two photons
simultaneously from exactly the middle of the waggon, one photon is
emitted in the forward direction of the train and the other one is
emitted in the backward direction. Since light
always travels at the speed of light in relation to any Galilean
frame according to Einstein, the two photons in the forward direction and the backward
direction will hit the walls in the forward and the backward
directions of the train simultaneously, from the viewpoint of Alice. From the viewpoint of Bob,
the situation will, however, be different. Since the train moves in the
forward direction, the photon emitted in forward direction will have to travel
a longer distance from the temporal moment of emission to the temporal
moment of arrival than the photon emitted in the backward direction,
since the train is moving and changing the positions of the walls
during the time it takes for the light to travel between the two
positions. Therefore, the photon emitted in the
forward direction will hit the wall in the forward direction after the
photon emitted in the backward direction hits the wall in the backward
direction. Thus, because of the way that space-time is transformed by
high velocities, Alice and Bob will arrive at different conclusions
regarding the relative temporal ordering of the two events.}%
\footnote{A minor
note concerning this thought experiment: If you find the situation
artificial in the respect that the two photons arrive at the two walls
exactly simultaneously from the viewpoint of Alice, you could modify
the thought experiment slightly:  Move the flashlight just a tiny bit in
the forward direction, so that the photon emitted in the forward
direction arrives at the forward wall slightly before the photon
emitted in the backward wall from the viewpoint of Alice,
but not too much so that that the photon emitted in the forward
direction arrives at the forward wall before the photon emitted in the
backward direction arrives at the backward wall from the viewpoint of
Bob. Then, we have a complete reversal of the temporal ordering of the
two temporal events.}%
\footnote{For a biological perceptual agent, the relativistic time
  corrections that he or she may encounter due to relative velocities
  between two observers who observe everyday phenomena in the world with their own
  perceptual systems only will, however, be much shorter than the
  inner time scales of their perceptual systems, implying that
  relativistic time effects can be ignored in a treatment of how to handle
  the notion of time for a perceptual agent that observes everyday phenomena.}
This treatment deals with the handling of time for a single perceptual
agent that observes a dynamic world using time-causal receptive fields
as temporal primitives in its perceptual system.

Originating from a paper by McTaggart (\citeyear{McTag08-Mind}), there
are two main theories regarding time in the area of philosophy:
According to the A-theory, A-series events are ordered by which are
present, which are past, and which are future (tensed propositions), whereas according to
B-theory, B-series events are ordered by which come before and which come
after (tenseless propositions) (Zalta (ed.), Stanford Encyclopedia of Philosophy
\citeyear{TimeStanfEncPhil20}).
Thus, A-theory is closer to how we perceive time as humans (and similar to
St. Augustine's view above), whereas B-theory is closer to how we describe
temporal phenomena in physical theories of the world.

In a treatment about the notion of temporal presence, Power
(\citeyear{Pow16-PhilPsyTime}) discusses how we are able to 
maintain a perception of changes in the world in our representation of
the present. Essentially using the argument that the temporal present
is a instantaneous property (valid at a single time moment only), while
arguing that the perception of changes requires access to properties
of the world over an extended temporal interval, he concludes that
A-theory is false, since extended temporal properties cannot exist in
a representation of the temporal presence at a single time moment.%
\footnote{The view that the present is an instantaneous property does also go
back to St. Augustine (Outler, transl.\ \citeyear[Book~11, page~194]{AugOut95-book}):
  ``But the present has no extension whatever.''}

From the viewpoint of a temporal
multi-scale analysis as developed in this paper,
where each measurement of properties in the
world requires integration over a non-infinitesimal temporal interval,
it does, however, follow that any perceptual measurement of the world
will have to be
performed at some non-infinitesimal inner temporal scale, and thus
correspond to integration over a non-infinitesimal duration over time.
From such a viewpoint there is no contradiction relative to a {\em perceptual
representation of the present\/}, since a multi-scale representation of
the present will always occur over multiple temporal scales, and will
thus have the possibility to collect information about how properties
in the world change over time over extended temporal intervals.

Additionally, in human perception,
there are dedicated perceptual mechanisms for registering changes or motion over
time;%
\footnote{In computational models of vision, such temporal changes can be
                measured in a {\em direct\/} way, by receptive fields in terms
                of temporal or spatio-temporal derivatives, in other
                words not by first perceiving the underlying spatial
                structures at each time moment and then inferring
                temporal relations as a secondary process, but in
                instead directly in the sense of using specific change
                detectors or motion
              detectors that operate directly on the spatio-temporal
              image structure caused by a dynamic scene.}
compare, for example, with the illusion of the motion after effect
(Wohlgemuth \citeyear{Woh11-PsyMonSuppl}),
implying that if you look out of a window of a moving train for a
long time, and if the train suddenly stops, you may for a while
perceive a (physically non-existent) motion in the opposite
direction. Alternatively, you may encounter a similar illusion if
looking at the motion of streaming water for a sufficiently long time,
and then perceive motion in the opposite direction if you change your
viewing direction to focus on a static object. There are also static
stimuli that give rise to perception of motion (see {\em e.g.\/}
Conway {\em et al.\/} (\citeyear{ConKitYazPacLiv05-JNeuroSci})).

\begin{figure}[t]
  \begin{center}
      \includegraphics[width=0.48\textwidth]{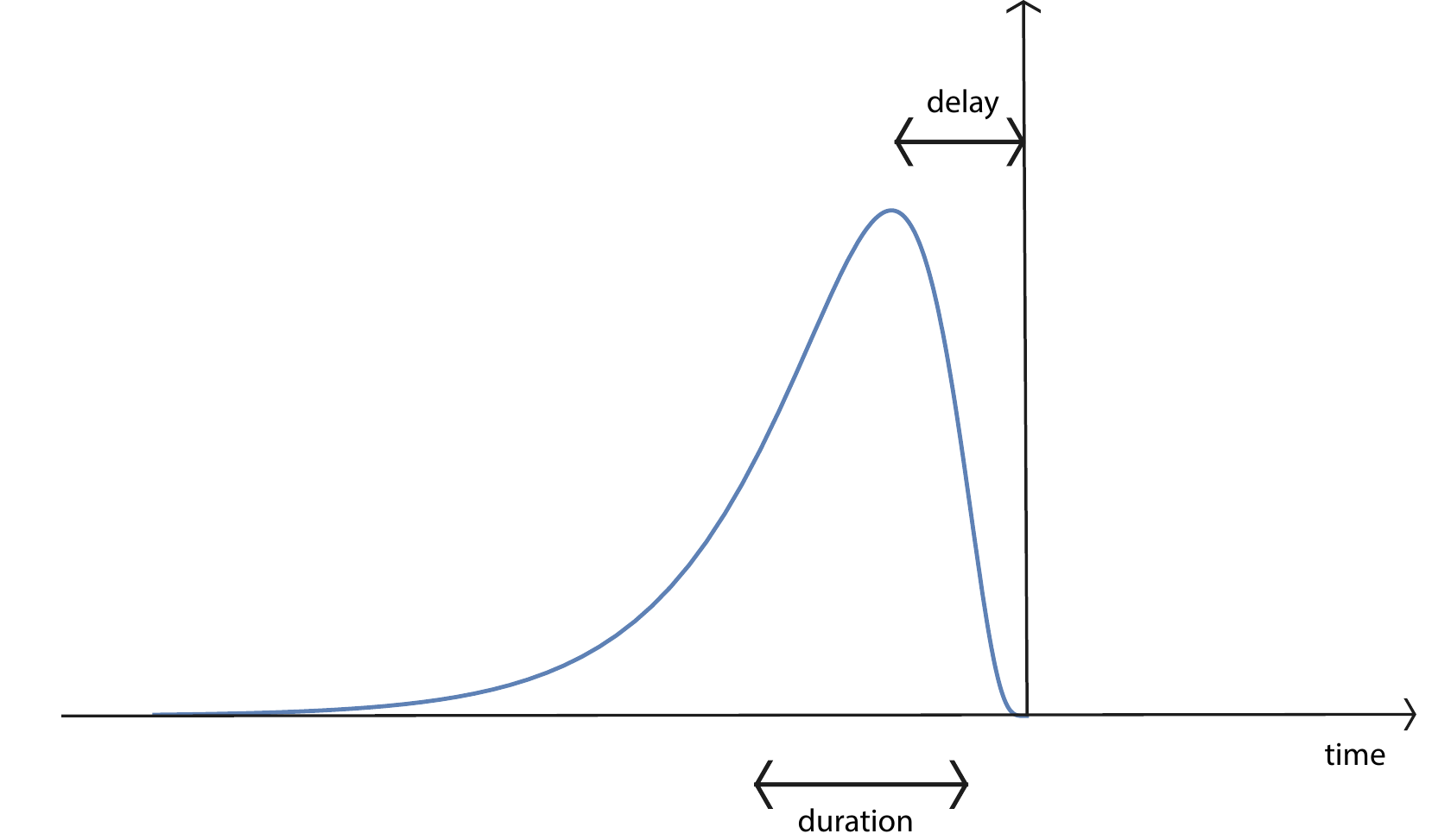}
    \end{center}
    \caption{Illustration of non-infinitesimal temporal duration of any
      physical measurement that arises as a consequence of a
      non-infinitesimal inner temporal scale in a physical temporal
      measurement device, specifically for any biological sensory or
      perceptual system,
      as well as the non-zero temporal delay of any time-causal
      temporal receptive field, which implies that the representation
      at any present moment will {\em de facto\/} instead be a
      representation of what happened some amount of time ago in the
      past. For the scale-covariant time-causal limit kernel proposed
      as the most idealized model of a temporal receptive field
      in this article, the temporal delay will specifically be
      proportional to the temporal scale measured in units of
      $[\mbox{time}]$, thus implying longer temporal delays at coarser
      temporal scales. (For a physical or biological implementation of these
      notions, there will also be another complementary
      temporal delay, not treated further here, caused
      by the time it takes to carry out the actual computations in the
      perceptual system.) (The vertical arrow in this illustration is
      intended to represent the present moment. The blue curve, in
      turn, reflects how different information from different temporal moments
      in the past contribute to the representation of the present at that
      present moment. To represent the temporal duration of the
      time-causal temporal smoothing kernel, we have in this
      illustration drawn the ``full width half maximum'' (FWHM), which
    is proportional to the temporal standard deviation of the
    temporal scale-space kernel, in other words proportional to the square root of
    the temporal scale parameter $\tau$.)}
    \label{fig-temp-delay-duration}
\end{figure}

The model for temporal multi-scale processing developed in this paper
does thus make the following assumptions concerning the handling
of the notion of time for a perceiving agent: The perceptual system of the
perceiving agent has a lowest layer of biophysical sensors, which performs temporal
integration of the underlying physical signal with some shortest time
constant corresponding to the smallest possible inner temporal scale
of the perceiving agent. Then, successive layers of such operations
are coupled in cascade in a hierarchical manner over that first layer, leading to a layered
architecture in the perception system, with successively longer
effective time constants at higher layers corresponding to coarser
temporal scales. Each such representation in any layer of the
hierarchy operates on input
information acquired in the present, possibly complemented with access
to memory buffers of the past. Thus, from the perspective of the
perceiving agent, he or she
cannot have any access to the actual physical present in the external world
(``das Ding an sich''; Kant 1783, \citeyear{Kan1902-book}), but instead
just access to a
temporally blurred representation of the present, which from the
perspective of the perceiving agent is the {\em only\/} available
representation of the present%
\footnote{Note, however, that this notion of an inner temporal scale
  for {\em any\/} representation of the present is, however, not the
  same notion as the notion of ``the specious present'' in the
  area of philosophy and psychology of time, a terminology forwarded by
  James (\citeyear[pages 609--610]{Jam1890-book}), who
  stated that:
  ``In short, the practically cognized present is no knife-edge, but a saddle-back, with a certain breadth of its own on which we sit perched, and from which we look in two directions into time. The unit of composition of our perception of time is a duration, with a bow and a stern, as it were---a rearward- and a forward-looking end. It is only as parts of this duration-block that the relation of succession of one end to the other is perceived.''
  A main difference
  between these concepts is that the notion of an inner temporal scale will be associated with
  any primitive that can be represented in the specious present, for
  example, in short-term temporal buffers of the immediate past, each
  with a different temporal delay.}
(see Figure~\ref{fig-temp-delay-duration}).


From the representation of the (temporally blurred fuzzy) present, the
internal perceiving system of the agent may also compute
representations at coarser temporal scales, which by the temporal
delays inherent to the time-causal temporal processes will also serve
as a temporal memories of the past. The perceiving agent has no access
to a video or audio recording of the past. Instead, the only possible
representation of the past is what is stored in the temporal memories
of the perceiving agent.%
\footnote{Here, we disregard representations of the past that can be
  acquired by other external means, such as by communication with other individuals,
  by reading written records, by observing explicit video or audio
  recordings of the past, or by finding traces in the world of past events,
  as done in archeology or forensics.}
Some of these memories may be of a short term
nature and soon be overwritten by more recent information,
while other memories may be stored for further longer term access.

A more technical problem in relation to temporal memory concerns
making estimates of the duration of a temporal event. According to the
standard methodology in physics, one would use a clock, register the
times of the beginning and the end of the temporal event and compute
the duration from the difference between these temporal moments
(a B-series type of measurement). A biological perceiving agent does, however,
not have access to any explicit clock, and there is no evidence for an
accurate inner
clock in the human brain that a human perceiving agent could relate
to for directly measuring the duration of temporal events
(Wittman \citeyear{Wit09-RoySoc}). 

From the viewpoint of a temporal multi-scale 
analysis, it is, however, in principle possible to estimate the
duration of a temporal event by operating on representations at multiple
temporal scales and comparing the relative strengths of their responses, thus using A-type
measurements in the (time-delayed) present as opposed to quantitative B-type
temporal relations for estimating temporal duration. In (Lindeberg
\citeyear{Lin18-JMIV}), it is shown how it is possible to define
multi-scale spatio-temporal visual operations that respond
by their strongest response over temporal scales at a temporal scale
corresponding to the temporal duration of the temporal event, thus
estimating the duration of a temporal event based on measurements at a
single temporal moment only, although a very special temporal moment
at which the response assumes extrema over both time and temporal scales. This is
an extension of spatial scale selection
(Lindeberg \citeyear{Lin97-IJCV}, \citeyear{Lin21-EncCompVis}), which
makes it possible to estimate spatial scales without need for
explicitly laying out a ruler.%
\footnote{The spatial scale selection methodology involves estimating the spatial size
  of image structures by detecting the spatial scale levels at which
  multi-scale spatial image operations assume their maxima over
  spatial scales (Lindeberg \citeyear{Lin97-IJCV}, \citeyear{Lin21-EncCompVis}).}

Due to the temporal delays of the time-causal receptive fields that
drive this perceptual engine over time, any representation of the
present will not be a representation of the actual present moment, but
instead of what had occurred at some temporal moments
(or rather temporal intervals) in the past. Furthermore,
representations at coarser temporal scales will harbour the traces of events that
occurred further in the past compared to representations at finer
scales, thus providing basic mechanisms for temporal memory buffers.

\begin{figure}[t]
  \begin{center}
      \includegraphics[width=0.48\textwidth]{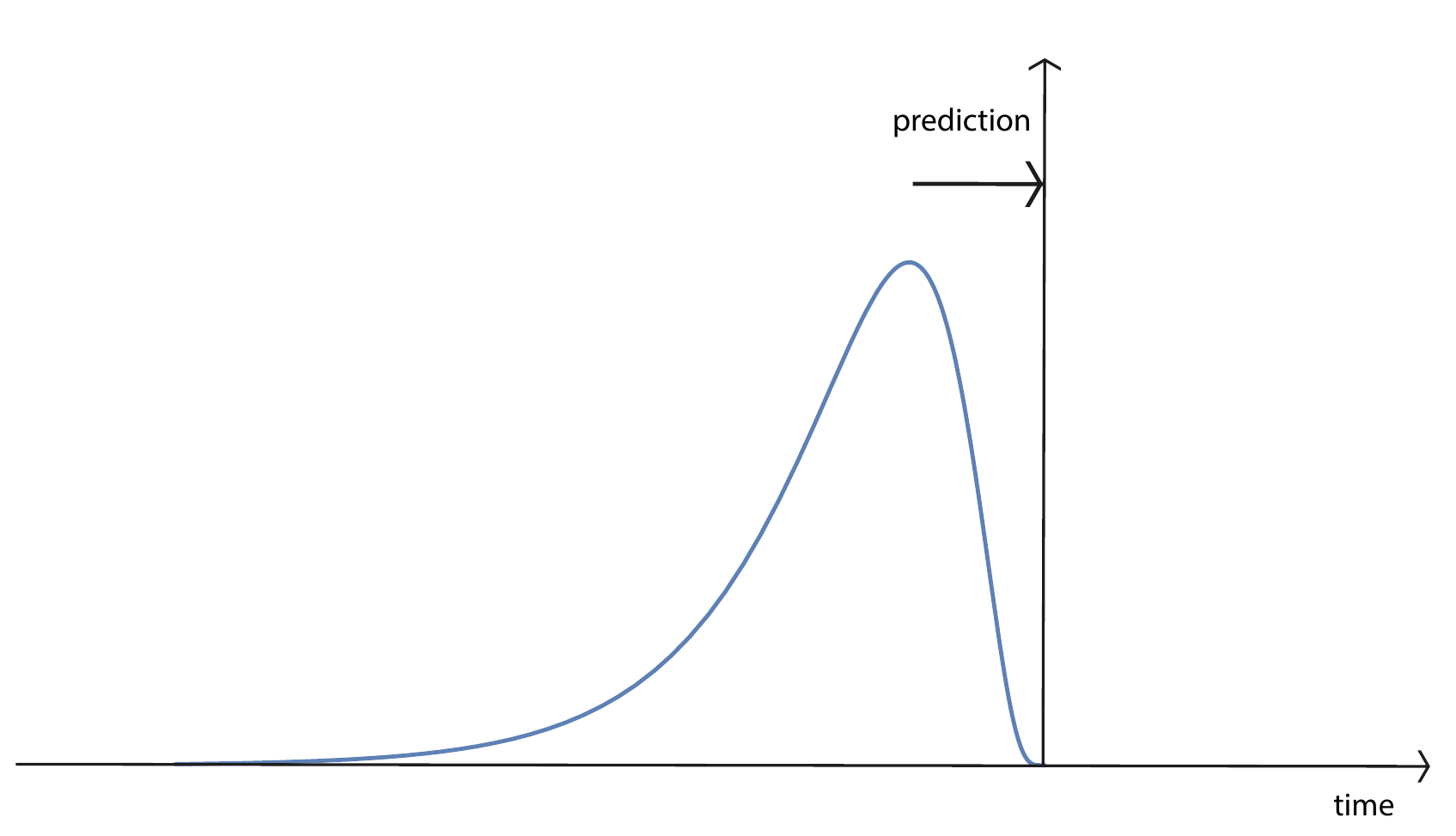}
    \end{center}
    \caption{The temporal delays of the time-causal receptive fields
      resulting from the presented theory
      call for a mechanism for performing temporal prediction to extrapolate
      the {\em de facto\/} time-delayed representation of the present
      (here represented as the temporal peak of the temporal receptive
      field marked in blue) to a better representation of the actual
      present (here represented by the vertical line on the time axis), to enable better
      temporal dynamics for a perceptual agent that interacts with a
      dynamic world. (Additionally, it is, of course, for other purposes also preferably to
      also attempt to perform predictions into the actual future in relation to any
      time moment, to enable temporal planning and to compensate for the time
      it will take to execute the actions called for by the perceptual
      agent. The latter types of temporal predictions are, however,
      not assumed to influence the representation of the present
      in this treatment.)}
    \label{fig-temp-prediction}
  \end{figure}
  
To make it possible for
the perceiving agent to handle fast occurring temporal events in a
dynamic world, it is therefore extremely valuable for a perceiving agent to be
able to perform predictions from the time-delayed perceptual present to
at least the actual physical present, so as to be able to coordinate his
or her actions with fast occurring temporal phenomena
(Figure~\ref{fig-temp-prediction}). Given that it will additionally
take time to plan and execute an action in practice, it is in a
similar way essential that the perceptual agent can perform
predictions into the actual future in relation to the actual present
moment when planning and executing an event.
Even further predictions to the future may of course also be
valuable for longer term planning, and to be able to make such longer term predictions, it is
very valuable to have an explicit memory of the past over
longer temporal scales. Thus, the notion of multiple temporal scales
is also important for making predictions into the future, for
different time scales into the future.

For the brain of a perceiving agent, its ability to predict what will
happen in the future may therefore be one of the most critical factors that
determine its ability to survive and reproduce in a competition
between individuals and species in the survival of the fittest
(Darwin 1859, \citeyear{Dar04}; Spencer 1864, \citeyear{Spe20-book}).
Minimizing the prediction error, has been proposed as main principle
underlying brain function (Friston \citeyear{Fri10-NatRevNeuro};
McCrone \citeyear{McCro22-Lancet}).
It has also been argued that the sensory cortex is optimized for
prediction of future input
(Singer {\em et al.\/} \citeyear{SinTerWilSchKinHar18-Elife}),
and furthermore been demonstrated that it is possible to
learn the receptive fields of deep neural
networks by training the networks to predict the relative future from
pre-recorded video sequences of natural scenes
(Singer {\em et al.\/} \citeyear{SinTerWilSchKinHar18-Elife};
Kwan and Park \citeyear{KwoPar19-ICCV};
Lotter {\em et al.\/} \citeyear{LotKreCox20-NatMachIntell}).
Low-level neuronal learning mechanisms have also been proposed in terms of
predicting future activity
(Luczak {\em et al.\/}  \citeyear{LucMcNauKub22-NatMachIntell}).

To conclude, we argue that in a A-theory type treatment of time for a
perceptual agent, it is
essential to complement previous such treatments with explicit notions
of (i)~non-infinitesimal temporal scales for {\em any\/} representations of the
present, and also to incorporate (ii)~the unavoidable temporal delays of
time-causal receptive fields that determine the functional properties
of perceptual systems. In a corresponding manner, given the extended
temporal delays of even the fastest temporal processes in {\em e.g.\/}
human vision, it is essential to complement the perceptual process
with (iii)~mechanisms for temporal predictions, since otherwise the actions
of the perceiving agent will be too slow to be able to handle and cope
with rapid temporal phenomena in the environment.%
\footnote{Considering, for example, the sport of playing tennis. For a
  professional tennis player, the speed of the tennis ball after a
  serve may be up to the order 200~km/h, corresponding to more than
  50~m/s. For among the faster temporal processes in human vision, the temporal
  limit of visual judgement is of the order of 20~ms, while for among
  some slower processes the limit of visual judgement is of the order
  of 100~ms (Holcombe \citeyear{Hol09-TICS}).
  Regarding spatio-temporal receptive fields in the primary
  visual cortex (V1), explicit modelling of examples of such receptive
  fields measured by DeAngelis {\em et al.\/}\
 (\citeyear{DeAngOhzFre95-TINS,deAngAnz04-VisNeuroSci}) lead to temporal
 scale values in the range from 50 to 80~ms (Lindeberg
 \citeyear{Lin16-JMIV}).
 The composed temporal scale level for the entire visual hierarchy has
 to be longer, whereas the
 fastest possible visual reaction times are of the order of 200~ms
 (Jain {\em et al.\/} \citeyear{JaiBanKumSin15-AppBasMedRes}).
 During temporal intervals in the range 20 to 100~ms,
 the tennis ball will be able to move
 by the order of 1~m to 5~m, or during 200~ms even as far as 10~m (if
 we neglect the loss of speed because of the air resistance).
  As a tennis player, you quickly learn
  that you have to fixate on the ball in order to be able to hit
  it properly, which compensates for some amount of the relative
  motion. Nevertheless, it seems unlikely that we would
  be able to judge the position and the timing of the ball properly, unless our
  conscious perception of it involves at least some component of temporal
  extra\-polation or prediction. Specifically, the process of fixating
  on a rapidly moving target also needs an explicit
  temporal prediction mechanism.}%
\footnote{For further support of the working hypothesis that our
  perception of the present likely
  involves essential components of prediction or extrapolation in the
  forward direction of time,
  see Nijhawan (\citeyear{Nij94-Nature,Nij08-BehBrainSci}), Grush
  (\citeyear{Gru07-PhilZeit}, \citeyear{Gru08-NewIdeaPsych}),
  Changizi {\em et al.\/} (\citeyear{ChaHsiNijKanShi08-CognSci})
  and White (\citeyear{Whi18-VisCogn}).}
These three notions are immediate consequences of treating temporal
perception as a consequence of a temporal measurement problem, where
information in physical stimuli has to be integrated over
non-infinitesimal durations over time (a main assumption underlying
the formulation of the presented temporal scale-space theory),
and making a notion such as the
instantaneous present {\em de facto\/} impossible for a perceptual agent.

Given the working hypothesis that perception has to involve some mechanisms
for temporal prediction to compensate for the non-avoidable temporal
delays associated with time-causal temporal integration over
non-infinitesimal neighbourhoods over time,
our conscious experience of the present in the world, thus has to
{\em synthesize\/} a view of the world, created by our brain, and truly
corresponding to ``controlled hallucination''
(Koenderink \citeyear{Koe11-PercBeyInf}; Clark \citeyear{Cla15-book};
 Paolucci \citeyear{Pao21-CognSemi}; Seth \citeyear{Set21-book}).
It is a ``hallucination''
in the sense that the view of the present is not actually a view of
how the world is or was at the moment that it was first registered and
then passed on to further processing. It is on the other hand ``controlled'' in the sense
that it is grounded on biophysical measurements of properties in the
world, and processed by a biological system that has been refined over
evolution over a very large number of generations.

Let us finally emphasize that this treatment does not make any claim
of being able to judge about the properties of time itself, which can only be made by
physical experiments, possibly complemented by theoretical modelling
and analysis, as done in the area of theoretical physics.
Instead, the treatment in this section concerns how the
notion of time is handled by a perceptual agent, specifically how the notion of
multiple temporal scales with their associated temporal delays have
to be considered in such a context, with a set of immediate implications thereof.

Let us also stress that the model used as basis for this treatment is
continuous in time, whereas for a biological neural system that communicates
with spikes between its neurons, the underlying communication channels
are in reality discrete, however, here assumed to be operating at a temporal scale
below the inner temporal scale of the
functional processes in the perceptual system.

\section{Historical developments of temporal scale-space theory}
\label{sec-hist-devel-temp-scsp-theory}

For the reader interested in a historical overview of previous
developments of temporal scale-space theory, this section gives an
overview of some the main previous contributions in this area that this
paper is based on, follows and extends.

Koenderink (\citeyear{Koe88-BC}) pioneered the area of the temporal
scale-space representation by proposing his scale-time model based on
applying Gaussian smoothing over a logarithmically transformed
temporal domain.

A complete classification of the general class of continuous scale-space kernels was
given in (Lindeberg \citeyear{Lin93-Dis}). While this
classification also included the truncated exponential kernels used as
main temporal primitives in
this paper, the main topic of that book was spatial computer vision, and
the specific detailed structure of time-causal
scale-space kernels was at first developed further in the more dedicated treatment in
(Lindeberg and Fagerstr{\"o}m \citeyear{LF96-ECCV}) aimed at video
processing, specifically including the logarithmic distribution of the temporal
scale parameter in the set of temporal scale channels.

The topic of temporal scale selection was first addressed in
(Lindeberg \citeyear{Lin97-AFPAC}), including detailed investigations of the
response properties of time-causal receptive fields
over temporal scales and time, and illustrating
how a closely related temporal model based on the time-causal Poisson
kernel, in turn assuming a semi-group property over temporal scales,
can also serve as a temporal memory of the past.

In (Lindeberg \citeyear{Lin97-ICSSTCV,CVAP257}) the time-causal model
based on the temporal Poisson
kernel, specifically the temporal derivatives of this kernel,
was used for modelling the temporal variability in biological
spatio-temporal receptive fields.
In ter~Haar Romeny {\em et al.\/} (\citeyear{RomFloNie01-SCSP})
the temporal variability in biological spatio-temporal
receptive fields was modelled using temporal derivatives of
Koenderink's scale-time kernels.

Other temporal scale-space models based on a semi-group property over temporal scales were then
studied in (Fagerstr{\"o}m \citeyear{Fag05-IJCV,Fag07-ScSp}) and
(Lindeberg \citeyear{Lin10-JMIV}).

In (Lindeberg \citeyear{Lin16-JMIV})
a substantial theoretical extension was made of the temporal
model based on truncated exponential kernels coupled in cascade,
by deriving the time-causal limit kernel, which allows for temporal
scale covariance.
In (Lindeberg \citeyear{Lin17-JMIV}) this model was
extended to temporal scale selection, including detailed studies of
the temporal response properties and scale selection properties for
the cases of a uniform sampling {\em vs.\/} a logarithmic sampling
of the temporal scale parameter. A general proof was also presented,
explaining how previous temporal models based on the
assumption of a semi-group property
over temporal scales lead to poor temporal dynamics, specifically
undesirably long temporal delays.

In (Lindeberg \citeyear{Lin16-JMIV}) the developments of the time-causal limit
kernel were performed in the context of video processing, and were
used for deriving theoretical models of spatio-temporal receptive
fields with close relations to biological receptive fields in the
lateral geniculate nucleus (LGN) and the primary visual cortex
(V1). In (Lindeberg \citeyear{Lin18-JMIV}) this theoretical
framework for spatio-temporal receptive fields was extended to scale-covariant
spatio-temporal feature detection with integrated spatio-temporal
scale selection. In (Lindeberg \citeyear{Lin18-SIIMS}) corresponding
extensions were made for dense temporal scale selection as well as
dense spatio-temporal scale selection.
In (Jansson and Lindeberg \citeyear{JanLin18-JMIV}) a specific
application to video analysis was developed to analyze dynamic
textures in a temporally scale-covariant manner.
In (Lindeberg \citeyear{Lin21-Heliyon}) the same theoretical model for
spatio-temporal receptive fields based on using the time-causal limit
kernel and its temporal derivatives as temporal basis functions was used for modelling biological
vision in an axiomatic normative theory of visual receptive fields

In (Lindeberg and Friberg \citeyear{LinFri15-SSVM,LinFri15-PONE})
parallel developments were made for auditory signals, showing how main
classes of
time-frequency transforms (spectrograms) can be derived in an axiomatic
manner, as well as how auditory receptive fields at a higher level can
also be axiomatically derived with very close similarities to
biological auditory receptive fields.

Most of the previous developments of the temporal scale-space theory relevant
for the treatment in this paper have, however, been performed with regard to visual
processing, and in the context of models for spatio-temporal receptive
fields. Some parallel developments have on the other hand been
performed with regard to auditory processing.

Anticipating that this could be a cause to problems for a
reader from a background in biology or signal processing, who is interested
in analysing or modelling purely temporal phenomena using a
corresponding theory, and wanting to get reasonably quickly into the
associated concepts, 
a first main purpose of this article has therefore been to give a dedicated
and self-contained
treatment that develops the relevant temporal scale-space theory for
the specific domain of purely temporal signals, without having the
theory intertwined with concepts regarding spatial or frequency domains, as is the case
in the previously available literature, dealing with visual or auditory processing.

We do additionally outline extensions of this temporal scale-space theory to forming
time-causal and time-recursive wavelet representations as well as
time-causal and scale-covariant time-frequency representations, which
do both provide novel contributions with regard to these areas.

With regard to modelling of
temporal phenomena in biology, we develop detailed comparisons to
other purely temporal models that can be used for such purposes, including
ways of translating results from those models to models based on the
time-causal limit kernel studied in this paper. With regard to
such purposes, we do also extensively relate to previous work
on modelling temporal scales in neural signals, for which we proposed that
the presented temporal scale-space model could provide a both theoretically and practically
valuable tool. Specifically, we
present a general procedure for fitting the time-causal limit kernel
to non-negative data, without any need for making use of an explicit
expression of the time-causal limit kernel over the temporal domain.

We do finally present implications of the presented theory to
fundamental concept formation in the area of the philosophy of time and regarding
non-infinitesimal inner temporal scales for any temporal sensor
measurement in a perceptual agent, including the resulting inevitable non-zero
temporal delays implied by that, in turn implying a need for making
predictions into the real present moment, to be able to handle rapid
temporal phenomena in the environment.

\section{Summary and conclusions}
\label{sec-summ-concl}

We have presented a theory for how temporal smoothing of temporal
signals can be performed in such a way that it guarantees that the
smoothing process does not create new artificial structure in the
signal, in the sense that the number of local extrema in the signal, or
equivalently the number of zero-crossings, is guaranteed to not
increase from finer to coarser temporal scales.
Additional critical components of this theory are temporal causality,
implying that we are not allowed to access information from the future
in relation to any time moment, and temporal recursivity, implying
that the temporal smoothing process should not require any other
temporal memory of the past than the resulting temporal scale-space
representations themselves.

A complete classification of the linear and shift-invariant
convolution kernels that obey these properties has been given, based on an
earlier treatment in (Lindeberg and Fagerstr{\"o}m \citeyear{LF96-ECCV}), in
turn based on earlier classical results by
Schoenberg (\citeyear{Sch48,Sch50}).
For continuous signals, the corresponding temporal scale-space kernels
consist of truncated exponential kernels coupled in cascade, corresponding to first-order
integrators coupled in cascade, and for discrete signals, first-order
recursive filters coupled in cascade (Section~\ref{sec-cont-temp-scsp-kern}).

As a conceptual extension of this general approach, we have 
described a specific subset of choosing these kernels in such a way
that temporal scale covariance is obtained. The corresponding
time-causal limit kernel that permits scale covariance, which is a
novel construction in (Lindeberg \citeyear{Lin16-JMIV}), is the limit case
of an infinite number of truncated exponential kernels coupled in
cascade, with specific choices of the temporal time constants
(Section~\ref{sec-time-caus-limit-kern}).

Temporal scale covariance in this context means that if the input
signal is rescaled by some uniform temporal scaling factor $S = c^i$,
where $c$ is the distribution parameter of the time-causal limit
kernel and $i$ is some integer, then the result of performing temporal
smoothing on the rescaled temporal signal is the same as
performing temporal smoothing on the input signal, followed by a
corresponding rescaling of the processed original signal, and complemented
by a shift of $i$ units along the scale dimension
(Section~\ref{sec-prov-temp-sc-cov}).

These temporal kernels, optionally combined with their temporal
derivatives, do in this way constitute a canonical class of temporal
basis functions for numerous purposes of temporal modelling, in
situations when the temporal operations have to be time-causal and
time-recursive, and in addition have the ability to handle temporal
information over multiple temporal scales in a theoretically
well-founded manner.
With appropriate scale normalization of the temporal derivatives,
the temporal derivatives of the time-causal limit kernel are also
truly scale covariant, with preserved magnitude values of temporal
derivatives at matching temporal scale levels under scaling
transformations, in turn allowing for truly scale-invariant processing
under temporal scaling transformations of the input signal
(Section~\ref{sec-sc-cov-temp-ders}).

We have given an explicit expression for the time-causal limit kernel
in the Fourier domain (\ref{eq-FT-comp-kern-log-distr-limit}) and
although the kernel lacks a compact closed-form expression over the temporal
domain, we have shown how it can be related to other temporal models,
such as Koenderink's scale-time kernels (Section~\ref{sec-rel-koe-scale-time})
and the ex-Gaussian model,
which is the convolution with an exponential kernel with a single
truncated exponential function (Section~\ref{sec-ex-Gaussian-model}).
We have also presented a general methodology for how the parameters in a model
based on a (temporally either unshifted or time-shifted) time-causal
limit kernel can be determined from lower-order temporal moments of
some other temporal function or temporal signal
(Section~\ref{sec-gen-model-fitting-time-caus-limit-kern} and
Appendix~\ref{app-gen-model-fitting-time-caus-limit-kern}).

We have described how these kernels can be implemented on discrete data, based on a set
of first-order recursive filters coupled in cascade, where also the
discrete implementation guarantees that new local extrema, or
equivalently new zero-crossings, cannot be created from finer to
coarser levels of scale (Section~\ref{sec-comp-impl-disc-signals}).
The discrete implementation of temporal
derivatives is straightforward, in terms of small support finite
difference operators applied to the discrete temporal scale-space
representation (Section~\ref{sec-disc-temp-scsp-ders}).
Thus, the discrete implementation is highly efficient
and lends itself to real-time applications.

We propose that the presented theory, serving as a {\em normative theory
of purely temporal receptive fields\/}, provides a canonical way of
defining multi-scale representations of temporal signals in
situations where the signal operations have to be truly time-causal,
because of lack to access of future information in real-time
scenarios, and time-recursive, because of a need to keep memory
buffers of the past to a minimum in terms of memory requirements.
Specifically, we propose that the time-causal limit kernel with its
temporal derivates constitutes a canonical class of temporal basis
functions in situations when the temporal scales may vary, especially
when temporal scale covariance and temporal scale
invariance are desirable properties.

We have also related the theory to other approaches for processing
temporal signals at multiple temporal scales, specifically wavelet
analysis and time-frequency analysis. We have outlined how the
temporal derivatives of the time-causal limit kernel can serve as
time-causal and time-recursive wavelet bases
(Section~\ref{sec-rel-wavelet-anal}) and how a complex-valued
extension of the time-causal limit kernel can be seen as time-causal
analogue of Gabor functions, in turn enabling truly scale-covariant
time-frequency analysis also over time-causal and time-recursive
temporal domains (Section~\ref{sec-rel-time-freq-anal}).

Concerning applications of the presented theory,
we have described how these time-causal kernels constitute a canonical
class of temporal kernels for modelling spatio-temporal and
spectro-temporal receptive fields in biological perception
(Sections~\ref{sec-spat-temp-RFs}-\ref{sec-spectr-temp-RFs}).
We have also given a more general overview of the applicability of multiple temporal
scale levels in perceptual, memory and cognitive processes in biological
nervous systems, as well as given arguments proposing that the time-causal kernels
treated in this paper should constitute a corresponding canonical class of temporal
kernels when modelling neural signals as well as more general perceptual and temporal
memory processes by explicit mathematical models (Section~\ref{sec-temp-sc-neur-sign}).

Finally, we have presented general arguments for the need for
incorporating the notion of non-infinitesimal temporal scales with
their associated non-zero temporal delays when considering a perceptual
representation of the present (not the same concept as the
instantaneous actual present, which a perceptual agent has no possible
access to), which then also leads to a direct need for
temporal extrapolation or prediction in order to compensate for the temporal delays
associated with the time-causal temporal filtering operations in
a time-causal perceptual system (Section~\ref{sec-impl-phil-time-perc-agent}).
We propose that these arguments should have essential implications for the logical
reasoning in A-type theories of time in the philosophy of time,
as well as when modelling perceptual agents.

\section*{Acknowledgements}

I would like to thank the reviewers for valuable comments that
improved the presentation.

\appendix

\section{Appendix A: Relation between the time-causal limit kernel
  and the ex-Gaussian model used by Bright {\em et al.\/}}
\label{sec-app-ex-Gauss-model}

In (Bright {\em et al.\/} \citeyear[see Equations~(2) and (3)]{BriMeiCruTigBufHow20-PNAS}),
the authors fit a so-called ex-Gaussian model, which is the convolution of
an unnormalized Gaussian function with an unnormalized truncated
exponential kernel, to the temporal response functions of neurons.
With slightly different naming of the variables to avoid notational
clashes with the notation used elsewhere in this article, let us consider a temporal
response function of the form
\begin{equation}
    \label{eq-ex-Gaussian-gen}
  h_{\mbox{\scriptsize ex-Gauss,gen}}(t) = a_0 + a_1 \int_{u=0}^{\infty} e^{-\frac{(t-m-u)^2}{2 \sigma
      ^2}} e^{-\frac{u}{\mu}} \, du,
\end{equation}
which after explicit computation of the convolution integral in
Mathematica assumes the form
\begin{multline}
 h_{\mbox{\scriptsize ex-Gauss,gen}}(t) = \\ = a_0 + a_1 \sqrt{\frac{\pi }{2}} \, \sigma  \, e^{\frac{2 m \mu -2 \mu  t+\sigma ^2}{2 \mu ^2}}
   \operatorname{erfc}\left(\frac{m \mu -\mu  t+\sigma ^2}{\sqrt{2} \mu  \sigma }\right).
 \end{multline}

\subsection{Second-order moment-based method without flexible
  temporal offset parameter}
\label{app-second-order-model-time-caus-limit-kern}

In this appendix, we will derive a relation between the above ex-Gaussian model and a
corresponding model based on the time-causal limit kernel
\begin{equation}
  \label{eq-time-caus-model-ex-Gaussian-gen}
 h_{\mbox{\scriptsize limit-kern,gen}}(t) = b_0 + b_1 \, \Psi(t;\; \tau, c) ,
 \end{equation}
with the time-causal limit kernel $\Psi(t;\; \tau, c)$ in
(\ref{eq-temp-scsp-conv-limit-kernel}) defined from
its Fourier transform according to
(\ref{eq-FT-comp-kern-log-distr-limit}).

For simplicity, let us first assume that we are in range of the parameter
space of the ex-Gaussian model where the temporal
delay is small relative to the standard deviation and the time
constant $\mu$, such that we do not need to introduce
an additional temporal delay in the model
(\ref{eq-time-caus-model-ex-Gaussian-gen}) based on the time-causal limit
kernel. Let us also assume that we can assume that the DC levels in
the two models should be equal, such that we can throughout assume that $b_0 = a_0$.
Then, our task is to derive a mapping to compute the parameters $b_1$,
$\tau$ and $c$ in the model based on the time-causal limit kernel from
the parameters $a_1$, $m$, $\sigma$ and $\mu$ in the ex-Gaussian model.

The approach that we shall follow is to compute the zero-, first- and
second-order temporal moments of the two models with the DC-offsets
$a_0$ and $b_0$ suppressed
\begin{equation}
    \label{eq-ex-Gaussian}
 h_{\mbox{\scriptsize ex-Gauss}}(t) = a_1 \sqrt{\frac{\pi }{2}} \, \sigma  \, e^{\frac{2 m \mu -2 \mu  t+\sigma ^2}{2 \mu ^2}}
   \operatorname{erfc}\left(\frac{m \mu -\mu  t+\sigma ^2}{\sqrt{2} \mu  \sigma }\right)
 \end{equation}
 and
\begin{equation}
  \label{eq-time-caus-model-ex-Gaussian}
 h_{\mbox{\scriptsize limit-kern}}(t) = b_1 \, \Psi(t;\; \tau, c) ,
 \end{equation}
and determine the mapping between the parameters of the two models
from the requirement that the integral, the temporal mean and the
temporal variance should be equal.

Computing the (uncentered) temporal moments up to order two of the
ex-Gaussian model (\ref{eq-ex-Gaussian}) in Mathematica gives
\begin{align}
  \begin{split}
    \label{eq-M0-ex-Gauss}
   M_0 & = \int_{t = 0}^{\infty}  h_{\mbox{\scriptsize ex-Gauss}}(t) \, dt
 \end{split}\nonumber\\
  \begin{split}
    & = a_1\sqrt{\frac{\pi }{2}} \mu  \sigma  \left(\text{erf}\left(\frac{m}{\sqrt{2}
          \sigma }\right)
      \right.
 \end{split}\nonumber\\
  \begin{split}
    & \phantom{= a_1\sqrt{\frac{\pi }{2}} \mu  \sigma} \left.
      \quad +e^{\frac{2 m \mu +\sigma ^2}{2 \mu ^2}} \operatorname{erfc}\left(\frac{m \mu
   +\sigma ^2}{\sqrt{2} \mu  \sigma }\right)+1\right),
  \end{split}\\
  \begin{split}
     \label{eq-M1-ex-Gauss}   
     M_1 & = \int_{t = 0}^{\infty}  t \, h_{\mbox{\scriptsize ex-Gauss}}(t) \, dt
   \end{split}\nonumber\\
  \begin{split}
    & = a_1 \sqrt{\frac{\pi }{2}} \mu  \sigma  \left(\mu  e^{\frac{2 m \mu +\sigma ^2}{2
   \mu ^2}} \operatorname{erfc}\left(\frac{m \mu +\sigma ^2}{\sqrt{2}
   \mu  \sigma }\right)
     \right.
  \end{split}\nonumber\\
  \begin{split}
     \phantom{= a_1 \sqrt{\frac{\pi }{2}} \mu  \sigma} \left.
+(m+\mu
   ) \left(-\operatorname{erfc}\left(\frac{m}{\sqrt{2} \sigma
       }\right)\right)
        \right.
  \end{split}\nonumber\\
  \begin{split}
      \phantom{= a_1 \sqrt{\frac{\pi }{2}} \mu  \sigma} \left.
   +\sqrt{\frac{2}{\pi
   }} \sigma  e^{-\frac{m^2}{2 \sigma ^2}}+2 (m+\mu )\right),
  \end{split}\\
  \begin{split}
    \label{eq-M2-ex-Gauss}   
     M_2 & = \int_{t = 0}^{\infty}  t^2 \, h_{\mbox{\scriptsize ex-Gauss}}(t) \, dt
   \end{split}\nonumber\\
  \begin{split}
    & = a_1 \sqrt{\frac{\pi }{2}} \mu  \sigma  \left(-\operatorname{erfc}\left(\frac{m}{\sqrt{2}
          \sigma }\right) \left(m^2+2 m \mu +2 \mu ^2+\sigma ^2\right)
     \right.
  \end{split}\nonumber\\
  \begin{split}
     \phantom{= a_1 \sqrt{\frac{\pi }{2}} \mu  \sigma} \left.
      +2 \mu ^2 e^{\frac{2 m \mu
   +\sigma ^2}{2 \mu ^2}} \operatorname{erfc}\left(\frac{m \mu +\sigma ^2}{\sqrt{2} \mu  \sigma
 }\right)
    \right.
  \end{split}\nonumber\\
  \begin{split}
     \phantom{= a_1 \sqrt{\frac{\pi }{2}} \mu  \sigma} \left.
       +2 \left(m^2+2 m \mu +2 \mu ^2+\sigma ^2\right)
    \right.
  \end{split}\nonumber\\
  \begin{split}
     \phantom{= a_1 \sqrt{\frac{\pi }{2}} \mu  \sigma} \left.
        +\sqrt{\frac{2}{\pi }} \sigma 
   (m+2 \mu ) e^{-\frac{m^2}{2 \sigma ^2}}\right),
  \end{split}
\end{align}
from which we in turn obtain the temporal mean $\delta$ and
the temporal variance $V$ according to
\begin{align}
  \begin{split}
     \label{eq-delta-from-M1-M0}
     \delta & = \frac{M_1}{M_0},
   \end{split}\\
  \begin{split}
    \label{eq-V-from-M2-M1-M0}
     V & = \frac{M_2}{M_0} - \left( \frac{M_1}{M_0} \right)^2.
   \end{split}
\end{align}

\subsubsection{Method for second-order moment-based model fitting}

Using the fact that the temporal mean and the temporal variance of
the time-causal limit kernel are given by
(Lindeberg \citeyear[Equations~(34) and (35)]{Lin16-JMIV})
\begin{align}
  \begin{split}
     \label{eq-delta-time-caus-limit-kern}
     \delta & = \sqrt{\frac{c+1}{c-1}} \sqrt{\tau },
   \end{split}\\
  \begin{split}
     \label{eq-V-time-caus-limit-kern}
     V & = \tau,
   \end{split}
\end{align}
identifying these expressions and solving for $b_1$, $c$ and $\tau$ in
the model (\ref{eq-time-caus-model-ex-Gaussian}) based on the
time-causal limit kernel gives
\begin{align}
  \begin{split}
    \label{eq-b1-from-M0}
     b_1 = M_0,
   \end{split}\\
 \begin{split}
     c = \frac{\delta^2+V}{\delta^2-V},
   \end{split}\\
  \begin{split}
     \tau = V,
   \end{split}
\end{align}
which with $\delta$ and $V$ according to (\ref{eq-delta-from-M1-M0})
and (\ref{eq-V-from-M2-M1-M0}) as well as $M_0$, $M_1$ and $M_2$
according to (\ref{eq-M0-ex-Gauss}), (\ref{eq-M1-ex-Gauss}) and
(\ref{eq-M2-ex-Gauss}) gives the desired mapping between the
ex-Gaussian model (\ref{eq-ex-Gaussian-gen}) and the model
(\ref{eq-time-caus-model-ex-Gaussian-gen}) based on the time-causal limit kernel.

\subsubsection{Experimental results}

Figure~\ref{fig-trunc-exp-kernels-1D+exGaussian}
shows examples of ex-Gaussian temporal models
approximated by time-causal limit kernels in this way.
A conceptual advantage of the time-causal limit kernel in this
context, is that we do not need to use or modify a Gaussian kernel to
model the initial transient phenomena in a time-causal temporal response function that decays towards zero in an exponential
manner towards the tail.
In this way, a neural response modelled by the model based
time-causal limit kernel would also correspond to a biologically plausible
implementation corresponding to temporal integration of
the form illustrated in Figure~\ref{fig-first-order-integrators-electric}.

\begin{figure*}[hbtp]
  \begin{center}
   \begin{tabular}{ccc}
        {\footnotesize\em second-order model fitting to $h_{\mbox{\scriptsize ex-Gauss,gen}}(t)$} 
     & {\footnotesize\em third-order model fitting to $h_{\mbox{\scriptsize ex-Gauss,gen}}(t)$} \\
      \includegraphics[width=0.30\textwidth]{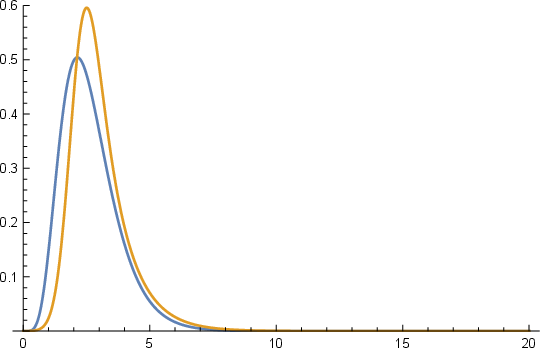}
     &
     \includegraphics[width=0.30\textwidth]{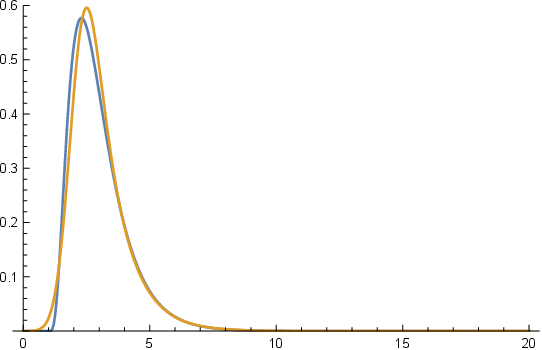} \\
  \end{tabular} 
\end{center}
   \caption{Comparison between (brown curves) the ex-Gaussian model
     according to (\ref{eq-ex-Gaussian-gen-main}) and
     (blue curves) the time-causal limit kernel according
     (\ref{eq-FT-comp-kern-log-distr-limit}) 
     approximated using the first $K = 7$ components of the infinite
     convolution of truncated exponential kernels in cascade.
     (left) A model with an unshifted time-causal limit kernel fitted using 
     the second-order moment-based method to an
     ex-Gaussian model with parameters  $\mu = 1$, $\sigma = 1/2$, $m = 2$, $a_0 = 0$ and $a_1
     = 1$ corresponding to $\tau \approx 1.25$, $c
       \approx 1.32$, $b_0 = 0$ and $b_1 \approx 1.25$.
    (right) A model with a time-shifted time-causal limit kernel fitted using using
    the third-order moment-based method to the same ex-Gaussian model with parameters
    $\mu = 4$, $\sigma = 2$, $m = 2$, $a_0 = 0$ and $a_1
     = 1$ corresponding to $\tau \approx 1.25$, $c
     \approx 1.88$, $t_0 \approx 0.98$, $b_0 = 0$ and $b_1 \approx 1.25$.
   (Horizontal axes: time. Vertical axes: function values.)}
  \label{fig-trunc-exp-kernels-1D+exGaussian-2nd-3rd-order-models}
\end{figure*}

\subsection{Extension to a third-order moment-based method involving an additional
  temporal offset parameter}
\label{app-third-order-model-time-caus-limit-kern}

As a remark concerning extensions, if the ex-Gaussian model is in a
range of the parameter space where the temporal delay is large
relative relative to temporal duration of temporal onset of the
composed kernel, then an additional
temporal offset $t_0$ can be added to the model
(\ref{eq-time-caus-model-ex-Gaussian-gen}) based on the time-causal
limit kernel
\begin{equation}
  \label{eq-time-caus-model-ex-Gaussian-gen-with-t0}
 h_{\mbox{\scriptsize limit-kern,gen}}(t) = b_0 + b_1 \, \Psi(t-t_0;\; \tau, c)
 \end{equation}
and an additional computation and identification of the third-order
central moments be performed to determine also this parameter in the mapping
between the two types of temporal models.

The explicit expression for the unnormalized and uncentered third-order temporal moment
of the ex-Gaussian model for $a_0 = 0$ is
\begin{align}
   \begin{split}
    \label{eq-M3-ex-Gauss}   
     M_3 & = \int_{t = 0}^{\infty}  t^3 \, h_{\mbox{\scriptsize ex-Gauss}}(t) \, dt
   \end{split}\nonumber\\
  \begin{split}
    & = a_1 \sqrt{\frac{\pi }{2}} \sigma  \left(\mu  \operatorname{erf}\left(\frac{m}{\sqrt{2}
          \sigma }\right) \left(m^3+3 m^2 \mu +6 m \mu ^2+3 m \sigma ^2
      \right.
      \right.
      \end{split}\nonumber\\
  \begin{split}
    & \phantom{a_1 \sqrt{\frac{\pi }{2}} \sigma \left( \mu  \operatorname{erf}\left(\frac{m}{\sqrt{2}
          \sigma }\right)  \right.}
    \left.
      \left.
        \quad +6 \mu ^3+3 \mu  \sigma
 ^2\right)
      \right.
  \end{split}\nonumber\\
  \begin{split}
     & \phantom{a_1 \sqrt{\frac{\pi }{2}} \sigma} \left.
+\mu  e^{-\frac{m^2}{2 \sigma ^2}} \left(-6 \mu ^3 e^{\frac{\left(m \mu
   +\sigma ^2\right)^2}{2 \mu ^2 \sigma ^2}} \operatorname{erf}\left(\frac{m \mu +\sigma
 ^2}{\sqrt{2} \mu  \sigma }\right)
\right.
\right.
 \end{split}\nonumber\\
  \begin{split}
    & \left.
      \left.
    \phantom{a_1 \sqrt{\frac{\pi }{2}} \sigma \mu  e^{-\frac{m^2}{2 \sigma ^2}} }
\quad\quad +\sqrt{\frac{2}{\pi }} \sigma  \left(m^2+3 m \mu +6
  \mu ^2+2 \sigma ^2\right)
\right.
\right.
 \end{split}\nonumber\\
  \begin{split}
   & \left.
      \left.
    \phantom{a_1 \sqrt{\frac{\pi }{2}} \sigma \mu  e^{-\frac{m^2}{2 \sigma ^2}} }
 \quad\quad +e^{\frac{m^2}{2 \sigma ^2}} \left(m^3+3 m^2 \mu +6 m \mu
  ^2+3 m \sigma ^2
\right.
\right.
\right.
 \end{split}\nonumber\\
  \begin{split}
    & \left.
      \left.
        \left.
   \phantom{a_1 \sqrt{\frac{\pi }{2}} \sigma \mu  e^{-\frac{m^2}{2 \sigma ^2}} }
  \quad\quad\quad\quad\quad +6 \mu ^3+3 \mu  \sigma ^2\right)
\right.
\right.
 \end{split}\nonumber\\
  \begin{split}
    & \phantom{a_1 \sqrt{\frac{\pi }{2}} \sigma \mu  e^{-\frac{m^2}{2 \sigma ^2}} }
    \left.
      \left.
        \quad\quad
       +6 \mu ^3 e^{\frac{\left(m \mu
             +\sigma ^2\right)^2}{2 \mu ^2 \sigma ^2}}\right)\right),
  \end{split}
\end{align}
whereas the expression for the normalized and centered third-order moment of the time-causal limit
kernel is (Lindeberg \citeyear[Equation~(36)]{Lin16-JMIV})
\begin{equation}
  \label{eq-kappa3-time-caus-limit-kern}
  \kappa_3 = \frac{2 (c+1) \sqrt{c^2-1} \, \tau^{3/2}}{\left(c^2+c+1\right)}.
\end{equation}
Using the relationship between the centered and uncentered third-order moments
\begin{align}
   \begin{split}
    \label{eq-M3-ex-Gauss}   
     C_3 & = \int_{t = 0}^{\infty}  (t - \delta)^3 \, h(t) \, dt
   \end{split}\nonumber\\
   \begin{split}
     & = \int_{t = 0}^{\infty}  t^3 \, h(t) \, dt
     - 3 \delta \int_{t = 0}^{\infty}  t^2 \, h(t) \, dt
    \end{split}\nonumber\\
   \begin{split}
     & \phantom{=} + 3 \delta^2 \int_{t = 0}^{\infty}  t \, h(t) \, dt
     - 3 \delta^3 \int_{t = 0}^{\infty} h(t) \, dt
   \end{split}\nonumber\\
   \begin{split}
     & =
     M_3 - \frac{3 M_1}{M_0} M_2 + \frac{3 M_1^2}{M_0^2} M_1 - \frac{M_1^3}{M_0^3} M_0,
   \end{split}
\end{align}
we obtain the following expression for the normalized and centered
third-order moment of the ex-Gaussian model
\begin{equation}
  \label{eq-kappa3-from-M2-M1-M0}
  \kappa_3 = \frac{C_3}{M_0}
  = \frac{M_3}{M_0} - \frac{3 M_1 M_2}{M_0^2} + \frac{2 M_1^3}{M_0^3}.
\end{equation}

\subsubsection{Method for third-order moment-based model fitting}

To determine the parameters in the model based on the time-shifted
time-causal limit kernel
(\ref{eq-time-caus-model-ex-Gaussian-gen-with-t0}) with the DC-offset
disregarded ($b_0 = 0$), we can hence
proceed as follows:
\begin{enumerate}
\item
   Compute the unnormalized and uncentered moments $M_0$, $M_1$, $M_2$
   and $M_3$ of the ex-Gaussian
   model according to (\ref{eq-M0-ex-Gauss}),
   (\ref{eq-M1-ex-Gauss}), (\ref{eq-M2-ex-Gauss}) and
   (\ref{eq-M3-ex-Gauss}).
 \item
   Compute the variance $V$ of the ex-Gaussian model according to
   (\ref{eq-V-from-M2-M1-M0}) and let the variance $\tau$  of the time-causal
   limit kernel be equal to this value according to
   (\ref{eq-V-time-caus-limit-kern}).
 \item
   Identify the normalized and centered third-order moments of the
   ex-Gaussian model and the model based on the time-causal limit kernel according
   to (\ref{eq-kappa3-from-M2-M1-M0}) and
   (\ref{eq-kappa3-time-caus-limit-kern}).
 \item
   With the third-order moment $\kappa_3$ of the ex-Gaussian model
   computed according to (\ref{eq-kappa3-from-M2-M1-M0}) and the
   variance $\tau$ of the time-causal limit kernel according to
   (\ref{eq-V-time-caus-limit-kern}), square the
   expression (\ref{eq-kappa3-time-caus-limit-kern}) and solve the
   resulting fourth-order algebraic equation in terms of the
   distribution parameter $c$ of the time-causal limit kernel.
   This will give four roots for $c$, out of which only two of the roots can
   be expected to satisfy the original unsquared equation, because of
   the squaring operation that may introduce new false roots.
 \item
   Select%
   \footnote{The skewness measure $\gamma_1 = \kappa_3/V^{3/2} = 2
     (c+1) \sqrt{c^2-1}/ (c^2+c+1)$, which is used for determining the
     distribution parameter $c$ in the model based on the time-causal
     limit kernel with a flexible temporal offset $t_0$, increases
     monotonically with $c$ for $c > 1$
   and assumes values in the range $]0, 2[$. Hence, provided that the
   skewness measure determined from $\kappa_3$ and $V$ is in this
   range, there will a unique real root for $c$ that satisfies $c > 1$.}
   the real root of the original equation (\ref{eq-kappa3-time-caus-limit-kern}) that additionally
   satisfies $c > 1$. Then, determine the temporal offset
   $t_0$ of the time-shifted time-causal limit kernel in
   (\ref{eq-time-caus-model-ex-Gaussian-gen-with-t0}) from the
   normalized and centered first-order moment of the time-causal limit kernel
   \begin{equation}
    \label{eq-delta-time-caus-limit-kern-with-t0}
     \delta = \sqrt{\frac{c+1}{c-1}} \sqrt{\tau } + t_0,
   \end{equation}
   with $\delta$ identified with the normalized first-order moment of
   the ex-Gaussian model according to (\ref{eq-delta-from-M1-M0}).
 \item
   Compute the amplitude $b_1$ of the time-shifted time-causal limit
   kernel in (\ref{eq-time-caus-model-ex-Gaussian-gen-with-t0})
   according to (\ref{eq-b1-from-M0}).
 \end{enumerate}
This procedure can either be carried out purely numerically or in a package
for symbolic computation, such as Mathematica.

\subsubsection{Experimental results}

Figure~\ref{fig-trunc-exp-kernels-1D+exGaussian-2nd-3rd-order-models} shows
the result of applying this procedure for fitting a time-shifted
time-causal limit kernel to an ex-Gaussian model that does not obey
the assumptions for fitting a model based on the non-shifted time-causal limit kernel
to it according to the previous second-order moment-based method.
In the left figure, the result of the second-order moment-based method is shown,
demonstrating a substantial difference because of the fixed zero
offset of the original time-causal limit kernel.
The right figure shows corresponding results for the third-order moment-based
method, demonstrating a much better agreement between the two models,
when an additional degree of flexibility is introduced into the model
based on the time-causal limit kernel by adding the temporal offset parameter.

\subsection{Fitting models with the time-causal limit kernel to other
  functions or signals}
\label{app-gen-model-fitting-time-caus-limit-kern}

Note that with replacement of the moments $M_0$, $M_1$, $M_2$ and
optionally $M_3$ with the moments of some other non-negative function or signal,
the same overall procedures
can more generally be used for fitting models based on the time-causal
limit kernel to other one-dimensional signals or functions that: (i)~are
defined for positive values of time, (ii)~assume non-negative values
only, (iii)~have a roughly unimodal shape of first increasing and then decreasing
and (iv)~tend to zero towards infinity.
The second-order moment-based fitting approach is in this context intended for situations
when the temporal origin of the signal or function is known in
advance and in some sense 
intended to be minimal, whereas the third-order moment-based fitting approach
is intended for situations when the temporal origin in the data is
unknown and hence needs to be adapted to each situation.

\section{Appendix B: Implementing temporal filtering with a discrete
  approximation of the time-causal limit kernel}
\label{app-disc-impl-time-caus-filt}

This appendix gives a brief explicit description about how to
implement temporal filtering of a sampled discrete signal
with a discrete approximation of the
time-causal limit kernel.

For simplicity, assume%
\footnote{If the input signal has been sampled with a frame rate
  $r$ not equal to one, then first transform the temporal standard deviation
  $\sigma_t$ relative to the original temporal axis to a standard
  deviation relative to a temporal axis with unit frame rate according
to $\sigma = r \, \sigma_t$, in analogy with (\ref{eq-transf-tau-sampl}).}
that the input signal has been sampled with a unit time increment
$\Delta t = 1$.
Then, given a temporal standard deviation of the kernel $\sigma$ in
such units of time, compute
   its variance $\tau = \sigma^2$ and choose a suitable value of the
   distribution parameter $c > 1$ that determines the sampling density in
   the temporal scale direction.
\begin{enumerate}
\item
   Compute a set of temporal scale levels $\tau_k$ according to a geometric
   distribution (\ref{eq-distr-tau-values}):
   \begin{equation}
     \tau_k = c^{2(k-K)} \tau \quad\quad (1 \leq k \leq K).
   \end{equation}
\item
   Compute a corresponding set of scale increments:
   \begin{equation}
     \Delta \tau_k = \tau_k - \tau_{k-1} \quad\quad (1 \leq k \leq K)
   \end{equation}
   with the additional definition $\tau_0 = 0$.
\item
    Compute the time constants $\mu_k$ for a set of temporal recursive
    filters with generating
    functions of the form (\ref{eq-gen-fcn-first-order-rec-filt})
    according to (\ref{eq-disc-time-constant}):
    \begin{equation}
      \mu_k = \frac{\sqrt{1 + 4 \Delta \tau_k}-1}{2} \quad\quad (1 \leq k \leq K).
    \end{equation}
\item
    Couple the following sets of first-order recursive filters in
    cascade (\ref{eq-norm-update}):
    \begin{equation}
       f_{\mbox{\scriptsize out}}(t) - f_{\mbox{\scriptsize out}}(t-1)
       = \frac{1}{1 + \mu_k} \,
            (f_{\mbox{\scriptsize in}}(t) - f_{\mbox{\scriptsize out}}(t-1)).
          \end{equation}
     Note that, in a real-time scenario or an offline scenario
     where memory efficiency is important, if the task is to compute a
     single temporal scale level only, such as the first temporal scale
     level in a cascade, this operation can be
     performed without explicitly
     storing the representations at the intermediate temporal scale
     levels, except for at the current and the previous temporal
     frames.

     Furthermore, when computing multiple temporal scale levels in
     parallel, the temporal scale-space representation at the next coarser temporal scale
     is most efficiently computed by applying a single recursive
     filter to the temporal scale-space representation at the nearest
     finer temporal scale (if we assume a dense representation over
     temporal scale levels, where all the temporal scale levels are
     assumed to be used in the later processing stages).
 \item
    Optionally, compute discrete approximations of scale-normalized
    temporal derivatives for some $\gamma > 0$ (where $\gamma = 1$
    is a standard default value) by applying the following discrete derivative
    approximation operators (according to Equations~(\ref{eq-sc-norm-der-var-norm})
    and (\ref{eq-temp-der-approx-molecules}))
    \begin{equation}
        \delta_{\mbox{\scriptsize t,norm}} = \sigma^{\gamma} \, (1, -1) \quad\quad
       \delta_{\mbox{\scriptsize tt,norm}} = \sigma^{2 \gamma} \, (1, -2, 1)
    \end{equation}
    to the temporally smoothed signal, alternatively instead using
    $L_p$-normalization according to (\ref{eq-sc-norm-der-Lp-norm-1})
    as opposed to variance-based normalization according to (\ref{eq-sc-norm-der-var-norm}).
\end{enumerate}

\bibliographystyle{abbrvnat}

{\footnotesize
\bibliography{bib/defs,bib/tlmac}
}

\end{document}